\documentclass[10pt]{elsarticle}
\usepackage[utf8]{inputenc}
\usepackage[margin=3cm]{geometry}
\usepackage{amsmath}
\usepackage{graphicx}
\usepackage{framed} 
\usepackage[labelfont=bf]{caption}
\usepackage[title]{appendix}
\usepackage{wrapfig}
\usepackage{multicol}
\usepackage{floatrow}
\usepackage{caption}
\DeclareCaptionType{equ}[][]
\usepackage{xcolor}
\usepackage{siunitx}
\usepackage{subcaption}
\usepackage{gensymb}
\usepackage[parfill]{parskip}
\usepackage{textcomp}
\usepackage{mathtools}
\usepackage{color}
\usepackage{longtable}
\usepackage{listings}
\usepackage{enumitem}
\let\vec\mathbf
\usepackage{indentfirst}
\usepackage{booktabs}
\usepackage{tabularx}
\definecolor{codegreen}{rgb}{0,0.6,0}
\definecolor{codegray}{rgb}{0.5,0.5,0.5}
\definecolor{codepurple}{rgb}{0.58,0,0.82}
\definecolor{backcolour}{rgb}{0.95,0.95,0.92}
\lstdefinestyle{mystyle}{
    backgroundcolor=\color{backcolour},   
    commentstyle=\color{codegreen},
    keywordstyle=\color{magenta},
    numberstyle=\tiny\color{codegray},
    stringstyle=\color{codepurple},
    basicstyle=\ttfamily\footnotesize,
    breakatwhitespace=false,         
    breaklines=true,                 
    captionpos=b,                    
    keepspaces=true,                                     
    numbersep=5pt,                  
    showspaces=false,                
    showstringspaces=false,
    showtabs=false,                  
    tabsize=4
}
\newcommand{\subsectionbf}[1]{\subsection{\textbf{#1}}}
\newcolumntype{P}[1]{>{\centering\arraybackslash}p{#1}}
\graphicspath{{fig/}}
\usepackage[breaklinks=true,colorlinks=true,linkcolor=blue,urlcolor=blue,citecolor=blue]{hyperref}
\biboptions{sort&compress}
\usepackage{cleveref}
\usepackage{physics}
 
\newcommand{\blue}[1]{{\color{blue}#1}} 

\makeatletter
\def\ps@pprintTitle{%
  \let\@oddhead\@empty
  \let\@evenhead\@empty
  \def\@oddfoot{\reset@font\hfil\thepage\hfil}
  \let\@evenfoot\@oddfoot
}
\newcommand{\myfootnote}[1]{
    \renewcommand{\thefootnote}{}
    \footnotetext{\scriptsize#1}
    \renewcommand{\thefootnote}{\arabic{footnote}}
}
\makeatother
\usepackage{setspace}
\DeclareMathOperator*{\argmin}{argmin}
\usepackage{fancyhdr}
\usepackage{etoolbox}
\newbool{censorflag}

\newcommand{\censor}[2][]{\ifbool{censorflag}{#1}{#2}}  

\booltrue{censorflag}  

\begin{document}

\begin{frontmatter}

\title{A tutorial on inversion-based shape control with design \censor{applications for NSTX-U and SPARC}{application to NSTX-U}}
\author[Princeton]{J.T. Wai$^{*,}$}
\author[PPPL]{M.D. Boyer$^{*,}$}
\author[CFS]{D.J. Battaglia}
\author[EPFL]{F. Carpanese}
\author[GDM]{F. Felici}
\author[GA]{W.P. Wehner}
\author[GA]{A.S. Welander}
\author[Princeton,PPPL]{E. Kolemen} 

\address[Princeton]{Princeton University, Princeton, New Jersey, USA}
\address[PPPL]{Princeton Plasma Physics Laboratory, Princeton, New Jersey, USA}
\address[CFS]{Commonwealth Fusion Systems, Devens, MA, USA}
\address[EPFL]{École Polytechnique Fédérale de Lausanne, Lausanne, Switzerland}
\address[GDM]{Google Deepmind, London, UK}
\address[GA]{General Atomics, San Diego, California, USA}

\begin{abstract}

One of the most common designs for magnetic control in tokamaks is to ``linearize an equilibrium'' to obtain a sensitivity mapping, then invert this mapping in order to determine the feedback control currents or voltages. For this work, we refer to this broad class of methods as inversion-based shape control (IBSC). In this work we describe the IBSC framework in a comprehensive manner and show how variations such as using dynamic voltage mappings and quadratic-program constrained control fit naturally into the framework. Despite the prevalence of IBSC, some of these extensions and the challenges associated with specific variations of IBSC are less widely known. We pay special attention to the challenge of decoupling the interaction between shape control and vertical control, which was a source of degraded vertical control performance on NSTX-U. This work is intended to provide a systematic overview of IBSC, and to that end, we have provided background material, proposed design procedures, and tutorials on the magnetic control design process within the appendices. Applying the systematic design procedure to NSTX-U, we find that the vertical control bobble on NSTX-U can be removed via decoupling, and that the vertical control phase margin can be improved by 6$^\degree$ just by including PF1A and PF2 as vertical actuators. \censor{The design procedure is also applied to SPARC and we report on progress and plans for the SPARC magnetic control design. SPARC will use both feedforward control, with trajectories obtained via the GSPulse code, and an isoflux constrained-control IBSC feedback controller. Using the FGE code, the controller will be tested against a library of nonlinear flight simulations before most pulses. Simulations indicate robust phase margins ($>90^\degree$) for vertical control and reasonable timescales ($\sim 150$ms) for rejecting shape disturbances. The baseline controller for SPARC has been implemented within the SPARC PCS; however we also scope an upgraded IBSC control design based on voltage sensitivity maps computed with the frequency-decoupling algorithm ALIGN, and demonstrate superior decoupling. This design may replace the baseline controller if deemed necessary.}{}

\end{abstract}
\end{frontmatter}
\myfootnote{Corresponding authors: jwai@cfs.energy, ekolemen@princeton.edu}
\myfootnote{$^*$ Finalization of this work was conducted while affiliated with Commonwealth Fusion Systems.}
\crefname{appendix}{}{}


\section{Introduction and outline}

One of the most prevalent and practical methods for performing magnetic shape control in tokamaks is to linearize an equilibrium to obtain a sensitivity mapping between the shape parameters and shaping actuators \cite{Yuan2013,Wang2019,Albanese2016,Walker1997,Humphreys2000,Boyer2018,Treutterer1997,Anand2024,Ariola2005,Hofmann1990,Corona2019,Hofmann1994,Marchiori2016,Ambrosino2003,Lennholm2000,Albanese2003,Albanese2005,Pesamosca2022,Anand2019,Mele2024,Pangione2013,Hahn2009} . Inverting this mapping gives a direct relationship from shape errors to actuator commands. During feedback control, the actuator commands are computed from this inverse mapping, in combination with some additional controller weights or PID gains. For conciseness, in this work we will refer to the broad class of methods that use this principle as ``inversion-based shape control'' (IBSC). 

In this work we present a general framework for classifying IBSC designs. While the main principle is straightforward, there are many specific design choices to consider such as: the underlying model used to compute the maps; whether the map is a static linearization or an approximation to the dynamical model; and the choice of inversion method. The specific choices have downstream implications on the controller implementation and performance, and can motivate additional consideration such as the choice of weights or use of decoupling filters. A rough outline of this paper is as follows: 

\begin{itemize}
    \item Section 2 describes the IBSC framework and variations in detail. 
    \item Section 3 discusses the use of static linearization maps and implications for vertical decoupling. 
    \item This work is intended to provide a systematic overview of IBSC, and instructive supplementary material is provided in the appendices. Appendix D provides background material and intuition on shape control, and Appendices \censor{E-H}{E-F} contain design procedures and tutorials for NSTX-U\censor{ and SPARC}{}. 
    \item Section 4 summarizes the main findings from applying the systematic design procedure to NSTX-U. 
    \censor{\item Section 5 summarizes the main findings from applying the design procedure to SPARC, and discusses detailed plans for the SPARC magnetic controller. }{}
\end{itemize}

\section{Inversion-based shape control (IBSC) framework}

\subsectionbf{Principles of IBSC}

Shape control is a coupled multi-input multi-output (MIMO) control problem and can typically have around 10-20 coil actuators and a similar or greater number of controlled shaping parameters (for background material, see Appendix D). Each coil will usually strongly affect a number of shaping parameters. First-principles physics models can be used to obtain a state-space representation of the circuit dynamics of the power supplies, currents in the vessel conducting structures, and how these affect the plasma magnetic equilibrium. While there are many relevant MIMO control techniques that could be applied to this type of problem (LQR/LQG \cite{Rui2024,Belyakov1999}, MPC \cite{Gerki2018,Wai2020,Gerksic2016,Gerki2013}, H-infinity \cite{Mitrishkin2011,Liu2020,Dokuka2007,Ambrosino1997,Mitrishkin2003}, and even reinforcement learning \cite{Degrave2022}), we focus on inversion-based shape control in particular because of its practical aspects of interpretability, simplicity, ease of operator tuning, and usage on existing tokamaks. The principle of IBSC is to obtain a linear mapping $T$ between the response of shaping targets to either coil currents or coil voltages. Such a mapping could be written as:

\begin{equation}\label{eq:map}
    \delta y= T_{current}\, \delta I_{c} \;\;\;\;\; \textbf{-or-} \;\;\;\;\;\; \delta y = T_{voltage} \, \delta V_c
\end{equation}

Where $y$ is the shaping parameters, $T$ is the response map, $I_c$ is the set of coil currents, and $V_c$ is the set of coil power supply voltages. If this mapping held exactly and was (right-) invertible, then we could drive any shaping error in $y$ to zero by computing the inverse mapping and applying this value to the controller. If the current equilibrium had a set of shaping errors $e$ then applying a coil current correction, 

\begin{equation}\label{eq:T_map_inverse}
    \delta I_c^{targ} = -T_{current}^{-1} e,
\end{equation}

would shift the equilibrium and drive errors to zero. A corresponding equation holds for voltage maps. Of course, in practice this map does not hold exactly and is not invertible and there are additional dynamic considerations of the system to consider, for which it may be beneficial to add additional controller features such as derivative or integral gain. However, the benefit of inversion still exists. The pseudoinverse of the response map represents linear combinations of shape errors and the corresponding actuators that strongly affect those errors. A pseudoinverse mapping helps reduce the dimensionality of the problem, so that the control designer only needs to tune a set of single-input single-output (SISO) controllers, instead of tuning a very high-dimensional MIMO controller. 

While the principle of IBSC is generally straightforward there are a number of design and implementation details to consider, and some important ones are summarized in \cref{tab:ibsc_design_choices}. These design choices center on how the response map is obtained, and the mechanism for obtaining a regularized pseudoinverse. The rest of this section discusses each of these design choices.  

\begin{table}[H] 
    \begin{center}
        \includegraphics[width=\linewidth]{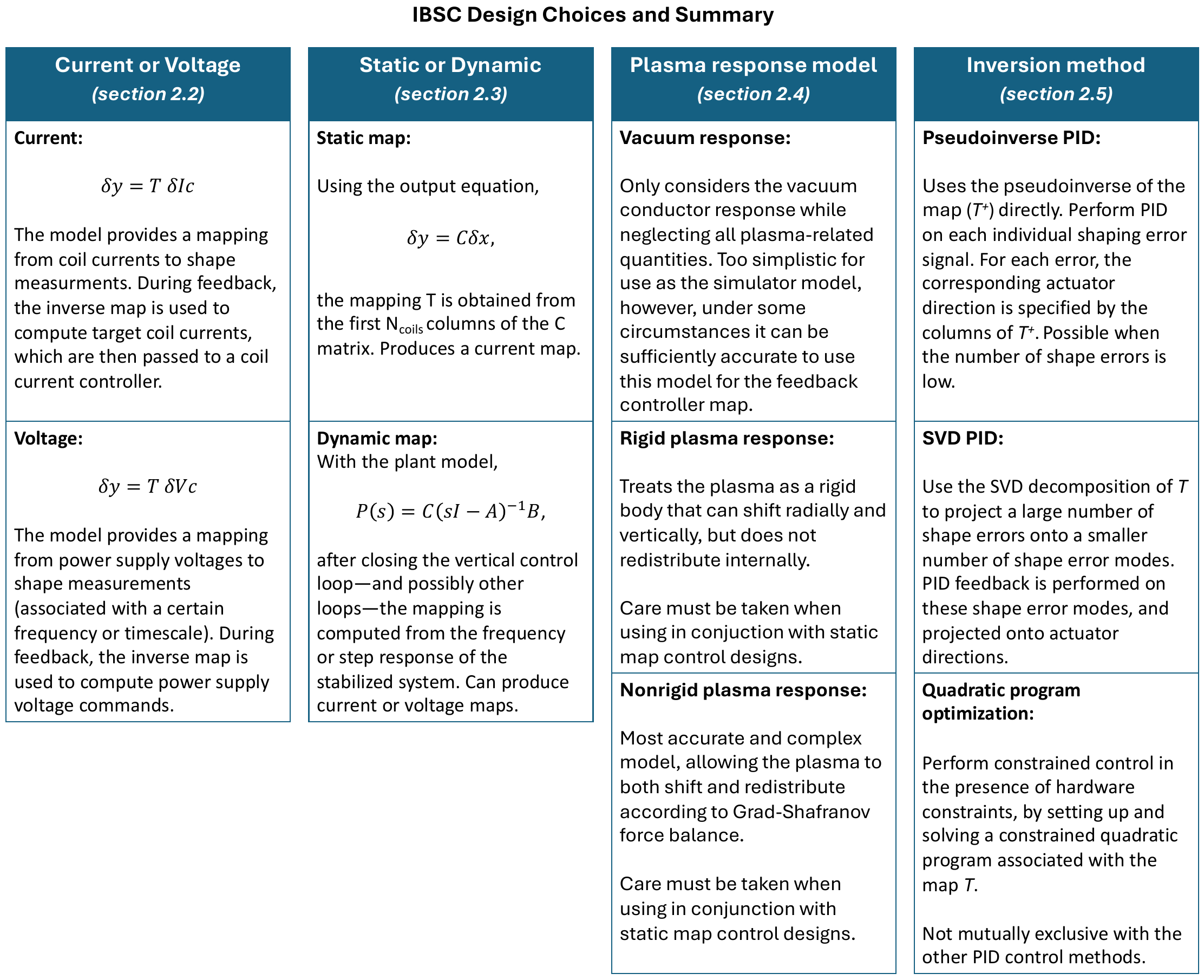}
    \end{center}
    \caption{Typical design choices for inversion-based shape control. Care must be taken that specific combinations of choices for whether static or dynamic map, and the plasma response model type, can result in worse performance. The most accurate combination is to use a dynamic map with nonrigid plasma response. See \cref{subsec:static_vs_dynamic,subsec:response_model,sec:static_plus_vertical}.}. 
    \label{tab:ibsc_design_choices}
\end{table}

\subsectionbf{Current map vs voltage map}

\begin{figure}[H]
    \centering
    \includegraphics[width=0.9\linewidth]{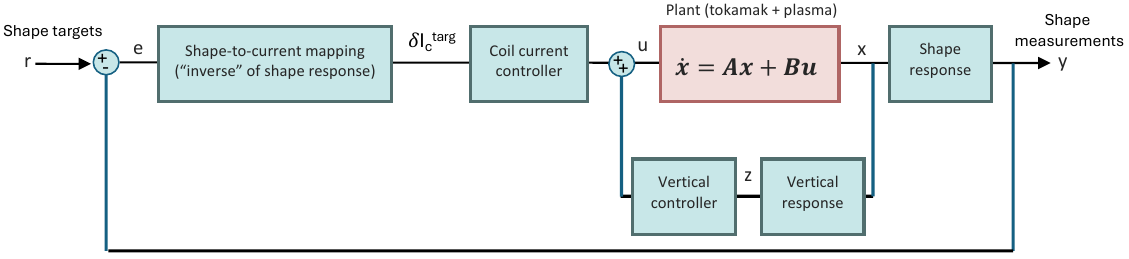}
    \caption{A typical shape control design based on response maps to coil currents. The vertical controller is often designed separately with a dedicated control loop. The shape errors are measured and mapped to coil current targets, and a coil current controller works to track these targets.}
    \label{fig:ibsc_block_diag_current}
\end{figure}

\begin{figure}[H]
    \centering
    \includegraphics[width=0.75\linewidth]{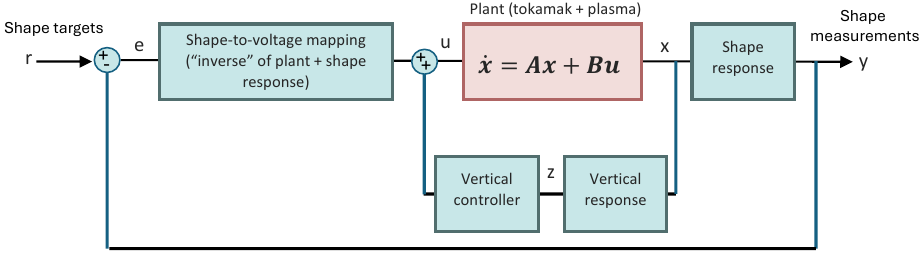}
    \caption{A typical shape control design based on response maps to coil power supply voltages. The shape errors are mapped directly to voltage commands.}
    \label{fig:ibsc_block_diag_voltage}
\end{figure}

The first design choice is whether the map corresponds to coil currents or voltages. As is typical, we assume that the input for the state-space model of the plant is the coil voltages. If a voltage map is used, the shape controller output is in units of voltage which can be passed as direct commands to the system (\cref{fig:ibsc_block_diag_voltage}) If a current map is used then these currents must first pass through a coil current controller (\cref{fig:ibsc_block_diag_current}). 

\subsectionbf{Static map vs dynamic map}\label{subsec:static_vs_dynamic}

The next distinction is whether the map is based on a static output linearization or whether it incorporates the dynamic response of the plant. This choice is also related to the previous choice of current versus voltage maps. To illustrate, we recall that the shape control state-space model takes the form:

\begin{equation}
    \begin{aligned}
        \dot x &= \vec A x +  \vec B  u \\
        \delta y &=  \vec C \delta x
    \end{aligned}
\end{equation}

\subsubsection{Static map}

The $C$ matrix represents the partial derivatives of outputs w.r.t axisymmetric conductor currents obtained from linearizing the equilibrium. In a static approach, the response map is obtained directly from this $C$ matrix. For example, if the first $n_{coils}$ columns correspond to the response of the shaping coils, the static map used by the controller is: 

\begin{equation}
    T = C \begin{bmatrix} I_{n_{coils}} \\ 0 \end{bmatrix} 
\end{equation}

\subsubsection{Dynamic map}
The dynamic approach incorporates the model dynamics into the map. The state-space model, in Laplace form, is:

\begin{equation}
    \begin{aligned}
        y(s) &= P(s) u(s) \\
        P(s) &:= C(sI-A)^{-1}B        
    \end{aligned}    
\end{equation}

For elongated plasmas, $P(s)$ contains an unstable pole associated with the vertical stability. A pragmatic approach is to first stabilize the system by closing the vertical control loop. Let the vertically-stabilized system be represented by $P_{vs}(s)$ with input-output relationship such that,

\begin{equation}\label{eq:Pvs}
    y(s) = P_{vs} u(s),
\end{equation}

where, with abuse of notation, we have re-defined $u(s)$ to correspond to only the voltages from the shape controller. The total voltage commands would be the sum of shape controller voltages and vertical control voltages. Note that \cref{eq:Pvs} is a functional map between shape control voltages and shaping parameters. For the IBSC design it is useful to approximate this functional map into a real, linear map. One approach would be to take a time-domain step response of this system evaluated at a particular time horizon, $\tau$: 

\begin{equation}\label{eq:T_step}
    T = step(P_{vs}(s)) \bigg |_{t=\tau}
\end{equation}

An alternative frequency-domain approach is to select a frequency $\omega$ and use the frame alignment algorithm ALIGN to compute a real approximation to the pseudoinverse of the complex matrix $P(j\omega)$. For details, see \cref{app:align_algorithm}.

\begin{equation}\label{eq:T_align}
    T^\dagger = ALIGN(P_{vs}(j\omega)) 
\end{equation}

\begin{figure}[H] 
    \begin{center}
        \includegraphics[width=0.75\linewidth]{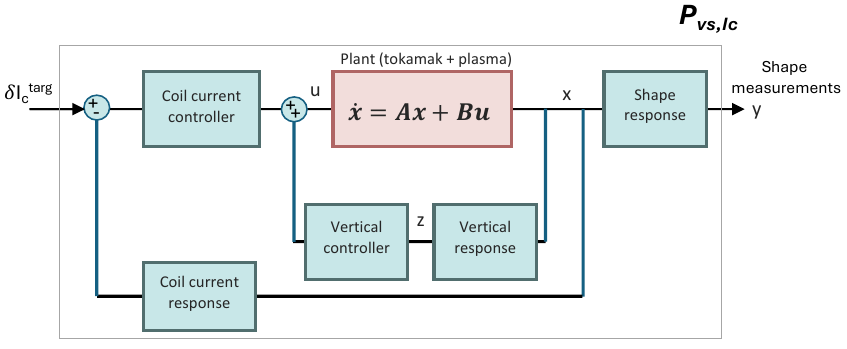}
    \end{center}
    \caption{Depiction of $P_{vs,Ic}$, the system transfer function from coil current target requests to shape parameters. This system can be ``inverted'' to determine the dynamic shape-to-current response map. This is a natural choice of system for this task, since it is equivalent to the controller design of \cref{fig:ibsc_block_diag_current} but without the shape-to-current map.}
    \label{fig:coil_current_stabilized_tf}
\end{figure}

Note that the maps from \cref{eq:T_step,eq:T_align} are in units of voltage, since the plant model input is voltage. However, dynamic maps for current could also be obtained. For example, we can close the coil current controller loop as in \cref{fig:coil_current_stabilized_tf}, and denote the transfer function for this combined system as $P_{vs,Ic}(s)$. This system describes how the shaping parameters change in response to coil current requests. Dynamic current maps are obtained from step or frequency responses of $P_{vs,Ic}$. 

\subsectionbf{Plasma response model}\label{subsec:response_model}

The choice of plasma response model refers to the underlying model that is used to linearize the equilibrium. We consider three model variations of increasing fidelity. First is the vacuum response, which does not include any plasma response but only the vacuum conductor responses. Second is the rigid plasma response, which models the plasma motion as rigid radial and vertical shifts \cite{Walker2006}. Third is the nonrigid plasma response model, which uses full Grad-Shafranov linearizations of the equilibrium and including both plasma current motion and redistribution. It should be noted that the nonrigid response model is the most accurate and the obvious choice of model for performing simulations. However, this is a distinct choice from whether it is the most appropriate model for building the controller's internal map between shape errors and coil currents. The best choice depends on whether the mapping is built using a static or dynamic response. 

If using a dynamic map design, that is, deriving the map from the plant transfer function, then the nonrigid plasma response model does indeed produce the most accurate controller maps. This can be argued from first principles, as this combination (nonrigid + dynamic) is the most complete and accurate model. 

\begin{figure}[H] 
    \begin{center}
        \includegraphics[width=8cm]{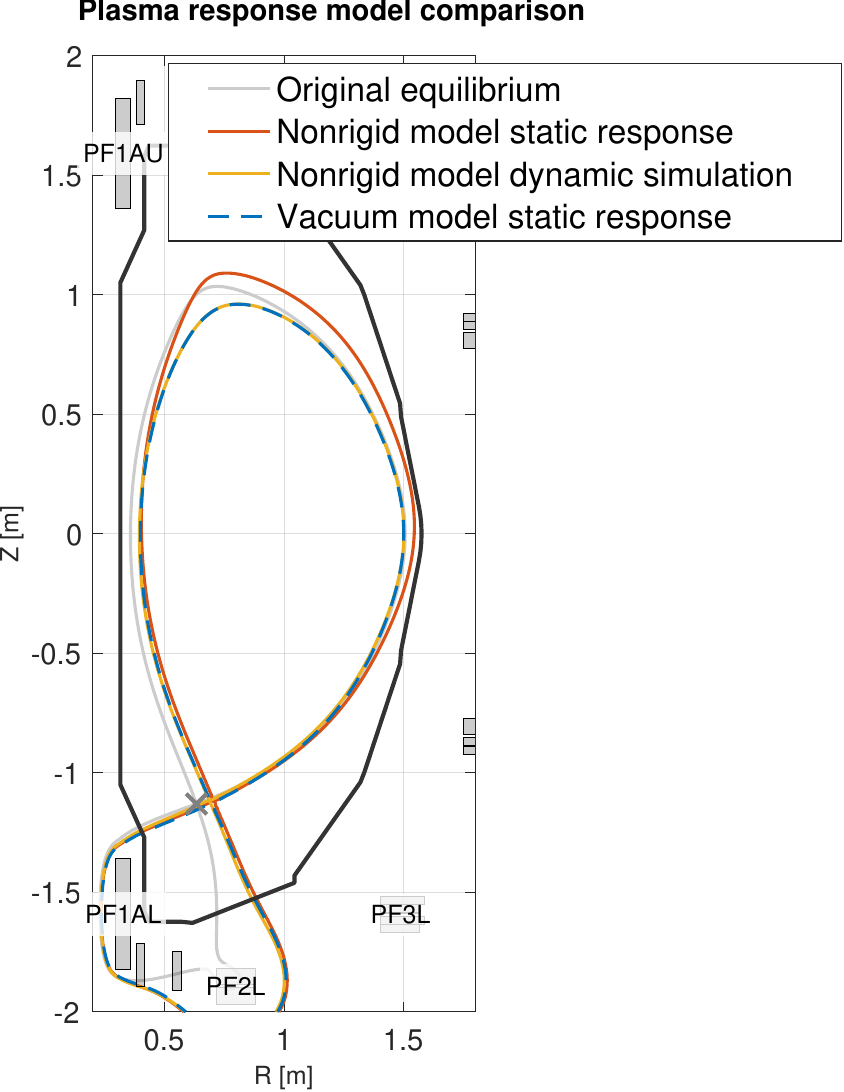}
    \end{center}
    \caption{Comparison of several flux response maps, to PF2U current perturbations. It is considered that the nonrigid dynamic response map is the highest fidelity map since this includes both nonlinearities of the equilibrium linearization and dynamics of the system. Surprisingly, the static vacuum response map is in better agreement with this than the static nonrigid response map. This behavior is generally consistent across different coils. }
    \label{fig:vacuum_vs_nonrigid_flux_response.png}
\end{figure}

However, when using a static map controller design, which is very common in the tokamak magnetic control literature, the best choice of model from which to derive the map can be an open question. In some cases, the vacuum response model can be a sufficient and even superior choice compared to the higher fidelity nonrigid response model. The NSTX-U equilibrium linearization shown in \cref{fig:vacuum_vs_nonrigid_flux_response.png} illustrates this somewhat counterintuitive result that the vacuum response model is a stronger choice. In this NSTX-U example, we obtain 3 different maps and plot the resulting flux response to a change in PF2U current. The three different maps are obtained with the following procedures: 

\begin{enumerate}
    \item Nonrigid static map: obtained directly from the nonrigid plasma response linearization, computed with the TokSys Gspert code \cite{Welander2005}. 
    \item Vacuum static map: obtained from the vacuum flux mutual inductances. 
    \item Nonrigid dynamic map: to obtain this map, we build the full dynamical system model using the nonrigid plasma response. We vertically stabilize the plant using the vertical controller designed in \cref{sec:nstxu_vert_control}. We also close the loop on coil current control, by designing a controller as in \cref{sec:nstxu_coil_control}. The resulting system is equal to that shown in the block diagram \cref{fig:coil_current_stabilized_tf}, and represents the vertically-stabilized transfer function from coil current reference target to flux on the grid. We perform a step response of this transfer function and evaluate the flux after 100ms, which is moderately beyond the settling time for coil current control. The resulting flux perturbation is plotted in yellow in \cref{fig:vacuum_vs_nonrigid_flux_response.png}. 
\end{enumerate}

Surprisingly, the static vacuum map aligns very closely with the dynamic nonrigid map, suggesting that the vacuum static map is a reasonable choice for the internal control model. By contrast, the static nonrigid map is significantly different and is even directionly wrong for some features, such as the outer/upper gap. This result is consistent with the fact that the NSTX-U shape controller gains were designed from the vacuum linearization \cite{Boyer2018} and generally achieved the desired performance. We do not have a full explanation for why the static vacuum map aligns well with the dynamic nonrigid map, and this may be coincidence, though we speculate that it could be due to the vessel conducting structures filtering and dampening the plasma motion. This would imply that, in general, the static nonrigid response model over-predicts the amount of plasma motion that actually occurs on the relevant control timescales. The strong performance of the vacuum map is not necessarily expected to be consistent across all tokamaks, and likely depends on the relative strength of vessel current screening in a tokamak. 

Again, we reiterate that the most accurate controller mappings are derived using the nonrigid response model and accounting for the dynamics. However, if wanting to use a static map controller design, as is often done in practice due to simplicity then choice of map should be tested carefully. 

\subsectionbf{Control inversion methods}

This discussion will assume we are using a coil current map although an analogous discussion holds for voltage maps. Having obtained the forward map, 

\begin{equation}
    \delta y = T \delta I_c,
\end{equation}

there are a few natural controller designs that arise from the idea of inverting this map. These include the pseudoinverse decoupling PID controller, the SVD controller, and a quadratic program constrained controller. These all play on variations of how to obtain the pseudoinverse of a matrix. For example, given a set of shaping errors $e$ each of the following give equivalent results for the corresponding coil current perturbations:

\begin{equation}
    \begin{aligned}
        \text{Moore-Penrose pseudoinverse: }& \delta I_c^{targ} = T^\dagger e \\
        \text{Singular value decomposition: }& \delta I_c^{targ} = VS^{-1}U^T e, \; \text{with SVD(}T) = USV^T\\ 
        \text{Unconstrained quadratic program: }& \delta I_c^{targ} = \argmin J(\delta I_c^{targ}), \;\; J := || T \delta I_c^{targ} - e||_2^2 
    \end{aligned}
\end{equation}

In practice, it is useful to regularize the inversions with weighting matrices to account for relative importance of minimizing any particular shaping error. Additionally, the inversion methods lend themselves to slightly different forms of interpretation and integration with control designs. These will be covered below for each inversion type. 

\subsubsection{Pseudoinverse decoupling PID control}
For example, to regularize the Moore-Penrose pseudoinverse we provide diagonal weighting matrices $W_y$ and $W_{I_c}$ that weight the importance of achieving shaping parameters and the usage of each coil actuator. The regularized pseudoinverse is:

\begin{equation}\label{eq:weighted_pseudoinverse}
    \bar T^\dagger = \begin{bmatrix} W_y \,T \\ W_{I_c} \end{bmatrix}^\dagger \begin{bmatrix} W_y \\ 0_{nc \times ny}\end{bmatrix}
\end{equation}

The pseudoinverse has the interpretation that each column of $\bar T^\dagger$ is a linear combination of coil currents that impacts a corresponding element in the shape error vector. These columns are sometimes called `control vectors' or `virtual circuits'. The principle of the pseudoinverse decoupling PID controller is to tune individual PID loops for each shaping error and sum the contributions, since the control vectors are approximately orthogonal. 

\begin{equation}
    \delta I_c^{targ} = \sum_i \bar T^\dagger_{\{i\}} \times PID_{\{i\}} \left (  e_i \right) 
\end{equation}

This approach works well if there are only a limited number of shape errors to control and is perhaps the most commonly employed magnetic control design. However, if there are a large number of shaping variables, then the errors will not decouple effectively and it makes more sense to attempt to control combinations of errors. A more effective approach for this is the SVD-controller, based on SVD inversion. 

\subsubsection{SVD Control}
The SVD controller is based on the SVD inverse, where, again, it is important to apply weights to regularize the inverse. This often becomes part of the tuning process for the controller. For the SVD inverse, we can regularize the usage of coil actuators by dropping the terms corresponding to the lowest singular values, only considering the first $r$ modes. These modes correspond to combinations of errors that are simultaneously not weighted as important and require significant actuator usage to control. The shaping variables are regularized with weights and undergo decomposition as:

\begin{equation}
    \begin{aligned}
        SVD(W_y \, T) &= USV^T \\ 
        \delta I_c^{targ} &= V_r S_r^{-1} U_r^T W_y e
    \end{aligned}
\end{equation}

Note that the singular value decomposition is performed on the weighted map, not the map itself. If we want to add PID gains on the singular modes, this gives an SVD controller design as follows: 

\begin{equation}\label{eq:svd_controller}
\begin{aligned}
    \hat e &:= U_r^T W_y e \\
    \delta I_c^{targ} &=  \sum_i^r \left( V_r S_r^{-1} \right)_{\{i\}} \times PID_{\{i\}} \left(\hat e_i \right ) 
\end{aligned}
\end{equation}

The principle of the SVD controller is that, instead of controlling individual shaping errors we reduce the number of controlled variables by projecting the errors onto a lower dimensional subspace. The $\hat e$ are linear combinations of weighted errors, where the linear combination is found from the singular vectors and describes the groups of errors that are most highly correlated. We tune individual PID controllers for each error group, and then these get mapped back to the target coil currents. This control design is effective for controlling a large number of shaping parameters and was the basis for the eXtreme Shape Controller (XSC) employed on JET. 

\subsubsection{Quadratic program constrained control}

In the presence of hardware actuator constraints a simple modification is to use a quadratic program to solve for the target coil currents. The goal is to minimize shaping error while simultaneously respecting actuator limits. We can regularize the quadratic program to add weights on the shaping errors and coil current usage. The regularized cost function becomes:

\begin{equation}\label{eq:J_regularized}
    J(\delta I_c^{targ}) = || W_y (T\delta I_c^{targ} - e )||_2^2 + ||W_{I_c} \delta I_c^{targ}||_2^2
\end{equation}

Adding the constraints, and specifying this in standard QP form gives:

\begin{equation}\label{eq:shape_inv_QP}
\begin{aligned}
\mathbf{minimize:} \;\; J &= \delta I_c^{targ,T} H \delta I_c^{targ} + 2f^T \delta I_c^{targ} \\
H &= T^T W_y^2 T + W_{I_c}^2 \\
f &= -T^T W_y^2 e \\
\textbf{subject to: }  A \delta I_c^{targ} &\leq b \\
\end{aligned}
\end{equation}

This is a small optimization problem (several dozen of variables) and can be solved by many commercial solvers. Note that this concept of achieving constrained control by reformulating inversion as a constrained quadratic program is complementary to the pseudoinverse PID controller and SVD controller designs. For example, instead of solving for the absolute currents we could reformulate the QP to solve for the singular value mode currents, and then also apply PID gains on the solution. 

Also note that the unconstrained solution to the QP is 

\begin{equation}
        \delta I_c^{targ} =  -H^{-1}f e,        
\end{equation}

which is a linear controller (and equivalent to the weighted pseudoinverse). Thus we can tune the weights and use the linear unconstrained controller to get the nominal performance that we would like, and then test the resulting controller in constrained-control applications. 

\section{Static plasma response maps and implications for vertical decoupling}\label{sec:static_plus_vertical}

One important subtlety is that a shape control model based on static plasma response maps can be misleading with respect to vertical motion. For a vertically unstable plasma, the static response suggests controller actions that are initially in the wrong direction. Consider the following step response for an elongated NSTX-U equilibrium with the vertical control loop closed with a PD-controller: 

\begin{figure}[H] 
    \begin{center}
        \includegraphics[width=8cm]{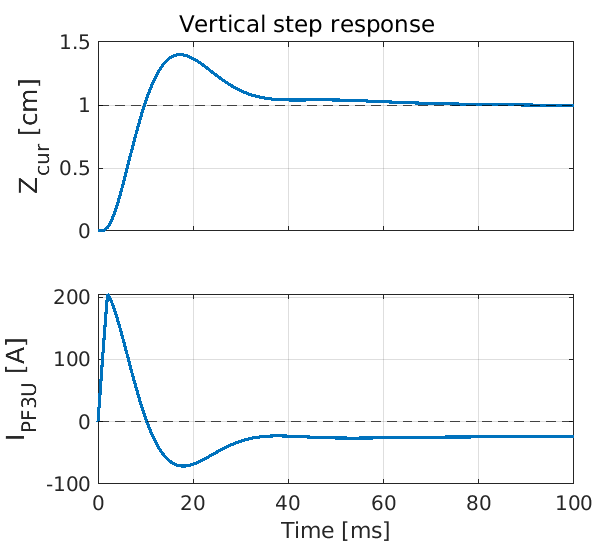}
    \end{center}
    \caption{Step response of the plasma vertical position with a proportional-derivative vertical controller applied to the (PF3U-PF3L) coil combination. The long-term, steady-state solution to move the plasma upwards requires the current in PF3U to decrease. However, to achieve the vertical motion PF3U must transiently increase (by a relatively much larger amount!) before it decreases. }
    \label{fig:zcur}
\end{figure}

Observe the behavior of the current in the PF3U coil. Consistent with the  nonrigid static plasma response model, in steady-state after all the other currents have decayed there is a negative correlation between the vertical position and coil current. However the path to reach this point is complicated. First the current had to move up, and then down in the opposite direction. Note that this type of inverse response is typical for dynamic systems containing right-half plane (RHP) zeros. In fact, it is the combination of closing the vertical control loop and truncating the plasma response map to correspond only to coil currents that introduces a RHP zero into the shape control dynamics. This is described in more detail in Appendix B. 

\begin{figure}[H] 
    \begin{center}
        \includegraphics[width=9cm]{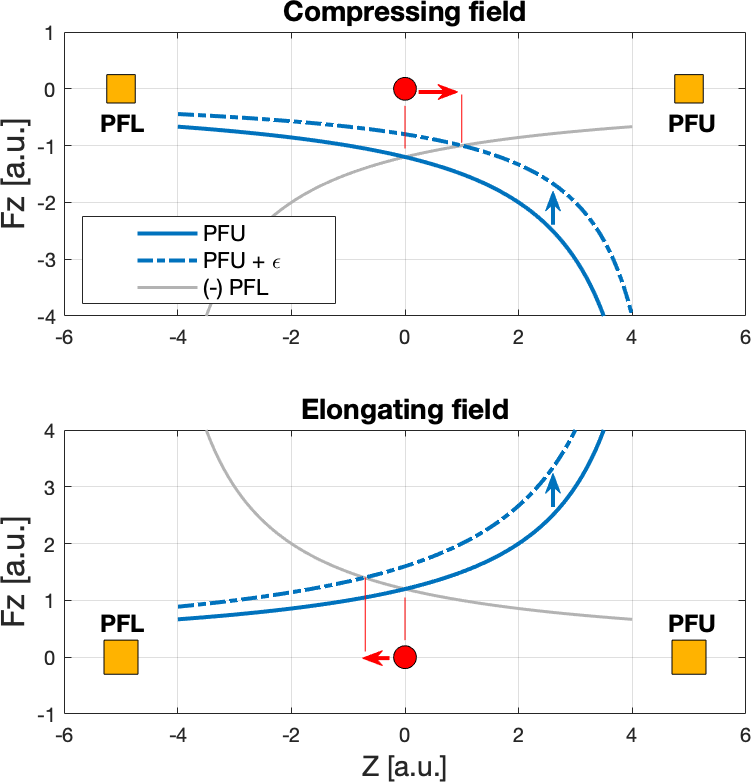}
    \end{center}
    \caption{Physical intuition of the vertical effects. \textbf{Top)} The plasma is in a compressing field and lies stably at the $z=0$ position. Both upper and lower PF coils push on the plasma with force proportional to $1/$distance-from-coil. If we perturb the current in the upper PF coil, the plasma experiences net vertical force upwards and the new equilibrium position is also vertically upwards. \textbf{Bottom)} The plasma is in an elongating field and lies at the unstable $z=0$ position, with both PF coils pulling on the plasma. We perturb the current in the same way as before (increase current, blue line shifts upward) which introduces a net upward vertical force as before. However the new equilibrium position is at a \textit{lower} vertical position. }
    \label{fig:vs_elong}
\end{figure}

The counter-intuitive relation can also be understood through equilibrium force balance in an elongating field \cref{fig:vs_elong}b. Does increasing the current (in the plasma co-current direction) in the upper shaping coil move the plasma up or down? Wires carrying current in the same direction attract, so one may think it exerts an upward force on the plasma and moves the plasma upward. On the other hand, the static plasma response model indicates $\partial z_{cur} / \partial I_{PFU}$ is negative, because with increased upward force exerted by the coil, the equilibrium position for the plasma is lower. Both intuitions are correct. The steady-state response is consistent with the static plasma response, but the initial dynamic correlation is in the opposite direction. 

The main point is that shaping corrections based on the static plasma response are initially in the wrong direction for elongated plasmas. This can result in tuning challenges as the shape controller destabilizes the vertical control. It is therefore important to employ vertical decoupling in order to improve system performance. We suggest the following guidelines for performing decoupling: 

\begin{itemize}
    \item All voltage commands for adjusting the net vertical position should be handled by the vertical controller. 

    \item The shape controller actuation should always stay orthogonal to the vertical controller actuation.  

    \item Any shape controller adjustments that are vertically anti-symmetric should be slowed down or dampened. 
\end{itemize}

This discussion was motivated by the use of static response maps. Dynamic response maps have the advantage that they are generated with the vertical controller in the loop, and therefore tend to be more accurate with respect to magnitude and direction of the suggested control actuation. However, these vertical decoupling principles are still beneficial to apply even when using dynamic maps. This improves robustness, since the vertical response correlation changes as a function of frequency, and the dynamic map was generated at a single equilibrium at a single frequency. 

\section{Proposed design procedure}\label{sec:design_procedure}

Given the number of IBSC design combinations and implications of map types, we attempt to synthesize this information into a step-by-step design procedure for vertical and shape control, including feedforward and feedback. The proposed procedure is shown in \cref{fig:control_design_procedureA}, and supplementary background material and tutorials on following this procedure for NSTX-U \censor{and SPARC designs are given in \cref{app:background_material,app:nstxu_tutorial,app:sparc_static_tutorial,app:sparc_dynamic_tutorial}}{is given in \cref{app:background_material,app:nstxu_tutorial}. \censor{The results in the next sections are obtained from applying this procedure to NSTX-U and SPARC controller designs. }{}

\begin{figure}[H] 
    \begin{center}
        \makebox[\textwidth][c]{\includegraphics[width=17cm]{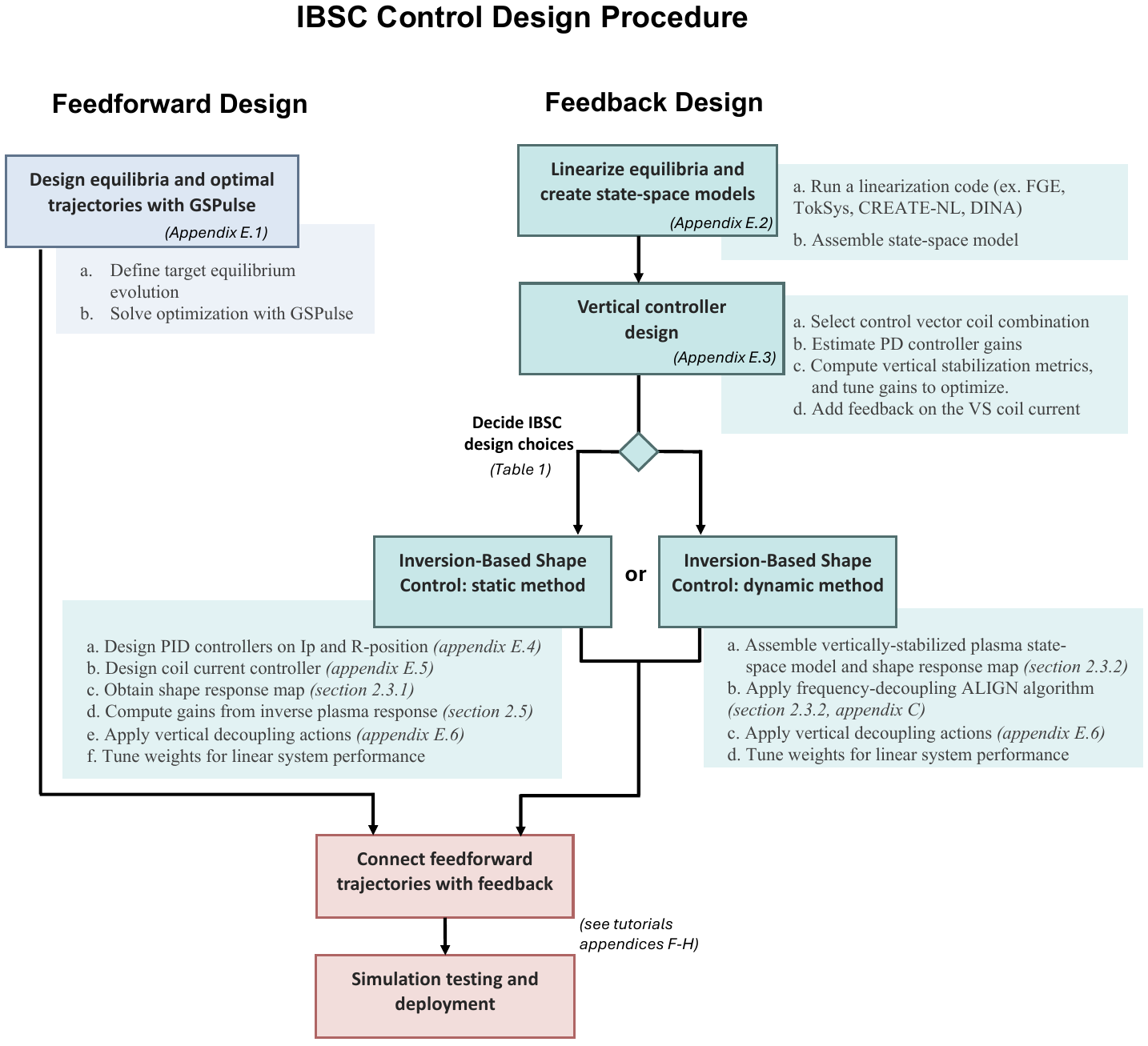}}
    \end{center}
    \caption{Proposed procedure for systematic design of integrated vertical and shape control, including both feedforward and feedback control.}
    \label{fig:control_design_procedureA}
\end{figure}

\section{NSTX-U shape control: summary of main findings}\label{sec:nstxu_results}

\subsectionbf{Improving vertical control phase margin with PF1 and PF2 usage}

The original vertical stability controller in NSTX-U used only the PF3 upper and lower coils as actuators. The combination of (PF3U - PF3L) current produces a net radial field that provides net vertical force on the plasma. However, as shown in \cref{fig:spatial_br}, employing slight usage of PF1 and PF2 is able to produce a more uniform field than just PF3 alone. This is not the only metric to evaluate, since the dynamic response is also important, but it hints that there is room for improvement by including PF1 and PF2 in the vertical control vector. It is interesting to note that the output RHP zero direction associated with the coil current observer also contains PF1 and PF2 in the coil current combination (see \cref{app:rhp_zero} and \cref{fig:rhp_zerodir}). 

\begin{figure}[H]
    \centering
    \includegraphics[width=0.8\linewidth]{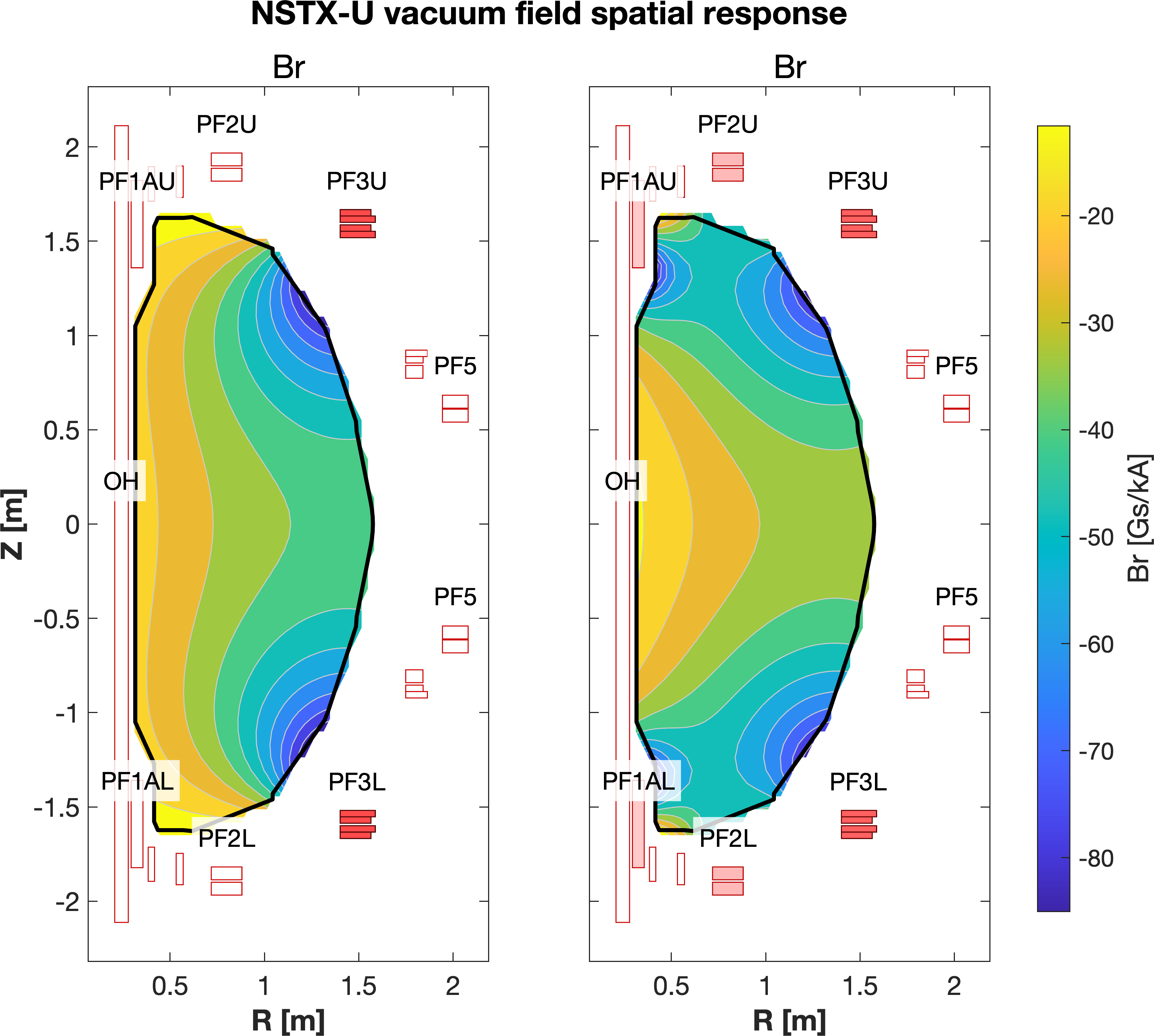}
    \caption{\textbf{Left)} Radial field distribution produced by the PF3U-PF3L coil current combination.  \textbf{Right)} The radial field distribution produced with light usage of PF1 and PF2 (20\%) is slightly more uniform.}
    \label{fig:spatial_br}
\end{figure}

\Cref{fig:vs_coil_combo} illustrates the performance gain achieved by tuning a vertical controller that includes PF1 and PF2 in the control vector. After some tuning, we find the best control vector direction is primarily in the PF3 direction, but with about 20-30\% use fraction on PF1A and PF2 as well. For a representative equilibrium, we observe that overshoot is reduced by 15\% and the phase margin increases from 32$^\degree$ to 38$^\degree$. While these are modest performance gains, they come for free since there is no additional algorithm complexity or hardware requirements. 

\begin{figure}[H] 
    \begin{center}
        \includegraphics[width=0.95\linewidth]{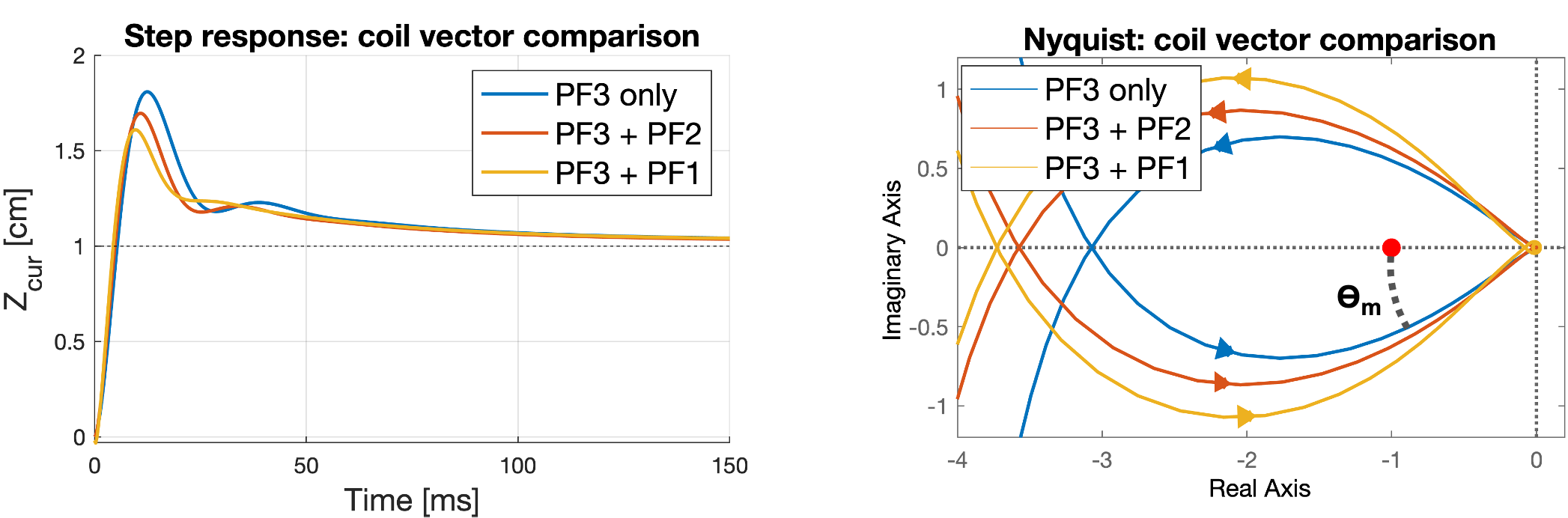}
    \end{center}
    \caption{The simple change of adding PF1 and/or PF2 to the vertical controller can reduce the overshoot and increase control phase margin.}
    \label{fig:vs_coil_combo}
\end{figure}

\subsectionbf{Applying vertical actuation penalties}

As discussed in \cref{sec:static_plus_vertical}, the shape controller can give commands that destabilize the vertical control. An example of this is shown in \cref{fig:nstxu_shape_vertical_decoupled}, in which we simulate a step response to change the plasma elongation. This particular shape perturbation was chosen because the elongation change and vertical shift for the target shape interact with the vertical control. The two controller cases are:

\begin{enumerate}
    \item A well-designed PD vertical controller, and an IBSC static vacuum pseudoinverse controller that has equal weights on all coil currents. 
    \item The same vertical and shape controller, except that the shape controller current weights have been set to penalize up-down antisymmetric coil currents with a 10X penalty. This allows the controller to retain speed in directions that don't affect vertical motion, but dramatically slows down the shape actuations for vertical changes, which improves performance and removes the disturbance oscillations.
\end{enumerate}

\begin{figure}[H] 
    \begin{center}
        \includegraphics[width=8cm]{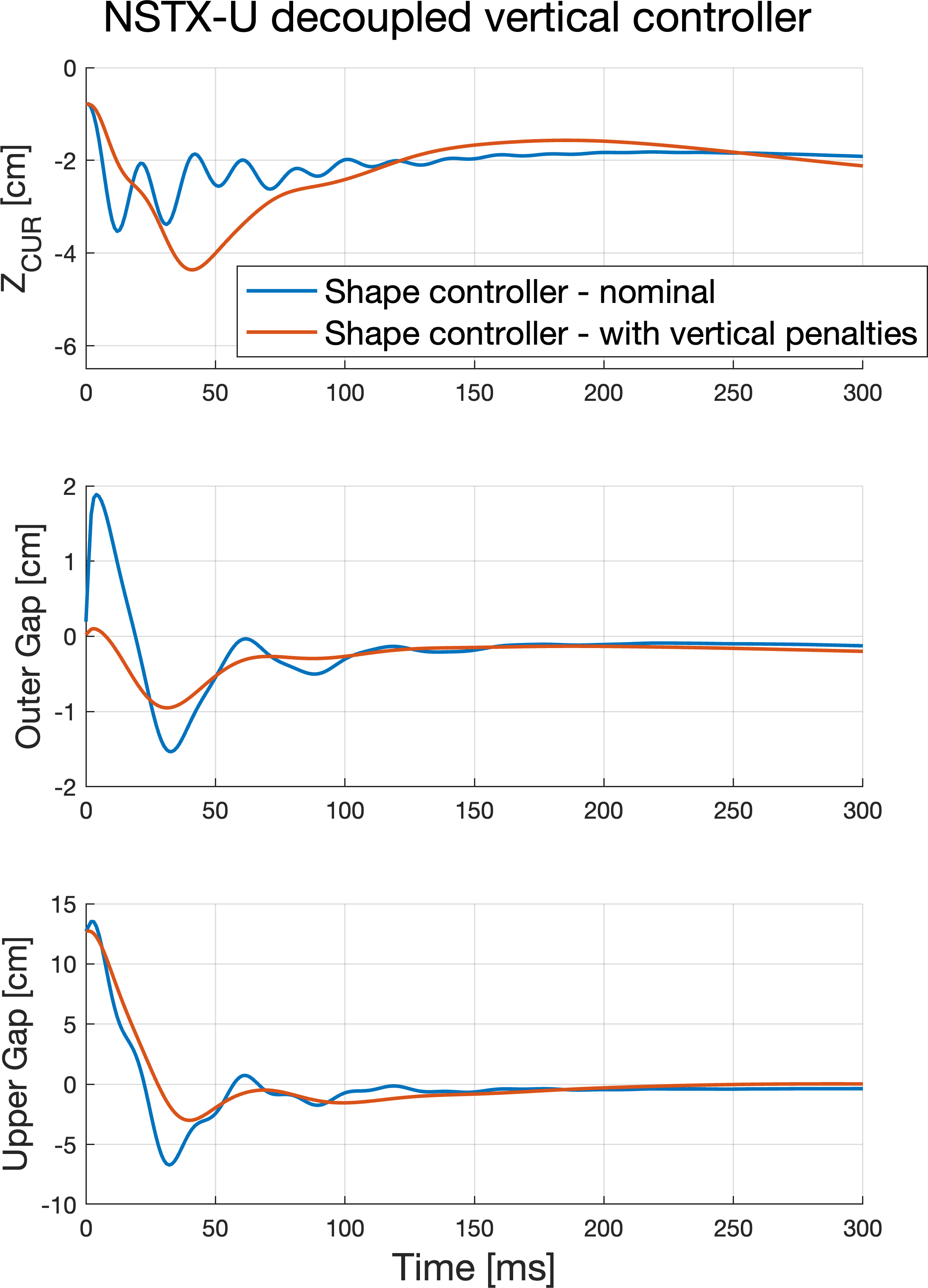}
    \end{center}
    \caption{Vertical and shape control interactions can be reduced significantly applying vertical actuation penalties to the shape controller. This simulation tracks a step response to the plasma elongation, seen for example in the large change in upper gap position. Without vertical penalties, the plasma suffers oscillations not only in the z-position but also in other shaping parameters like the outer gap.}
    \label{fig:nstxu_shape_vertical_decoupled}
\end{figure}

\subsectionbf{Nonlinear simulations via GSevolve}

The NSTX-U magnetic controller is tested in several nonlinear full-pulse simulations using the plasma flight simulator gsevolve. One particular control difficulty that was present in many NSTX-U shots was the vertical bobble, an undesired vertical oscillation behavior. A typical example is shot 204660 shown in \cref{fig:bobble}. The presence of a vertical bobble is not surprising given that the vertical control system does not have significant phase margin and that explicit shape-vertical decoupling was never employed during prior operations. This bobble behavior was loosely re-created in the gsevolve simulations using the original PCS controller. Although the oscillation frequency differs, the same undesired lower-vs-upper null switching behavior is observed along with vertical excursions of $\pm3.5cm$. 

\begin{figure}[H]
    \centering
    \includegraphics[width=0.9\linewidth]{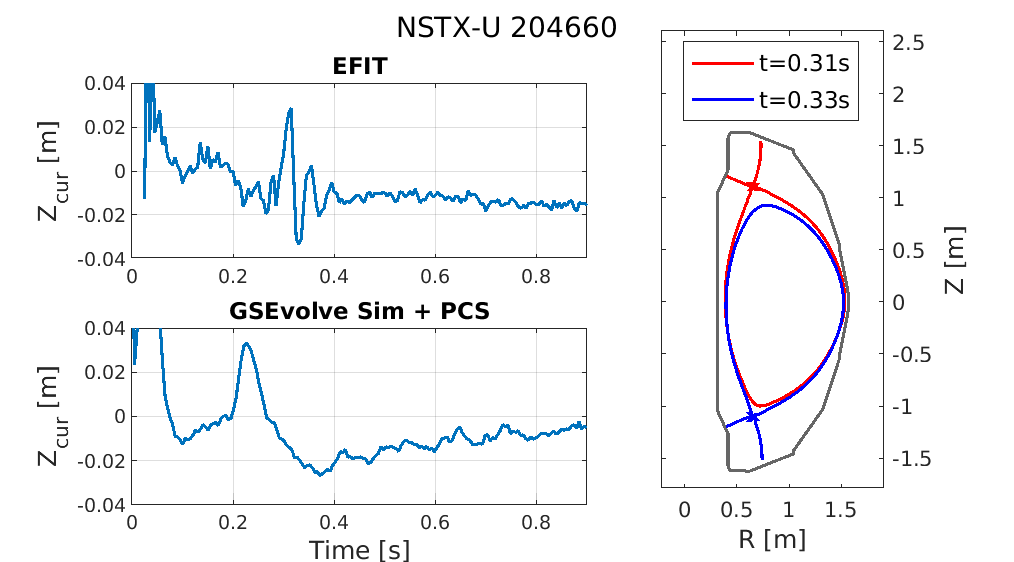}
    \caption{Example of the vertical bobble on NSTX-U from 204660, and recreation of the bobble with the original PCS controller and GSevolve simulator. The bobble is not reproduced exactly but we do capture unwanted $\pm$3.5cm vertical excursions. The corresponding flux surfaces are shown on right.}
    \label{fig:bobble}
\end{figure}

To show that our vertical decoupling approach has improved this behavior, we design a scenario with equal elongation and demonstrate intentional manipulation of the lower and upper single null switching. This is illustrated in \cref{fig:sim29c}, where we simultaneously track a shifting reference z-position and a switching target x-point. We start with an USN equilibrium, and then while moving the plasma vertically upward switch to a LSN and then back again. 

\begin{figure}[H]
    \centering
    \includegraphics[width=0.7\linewidth]{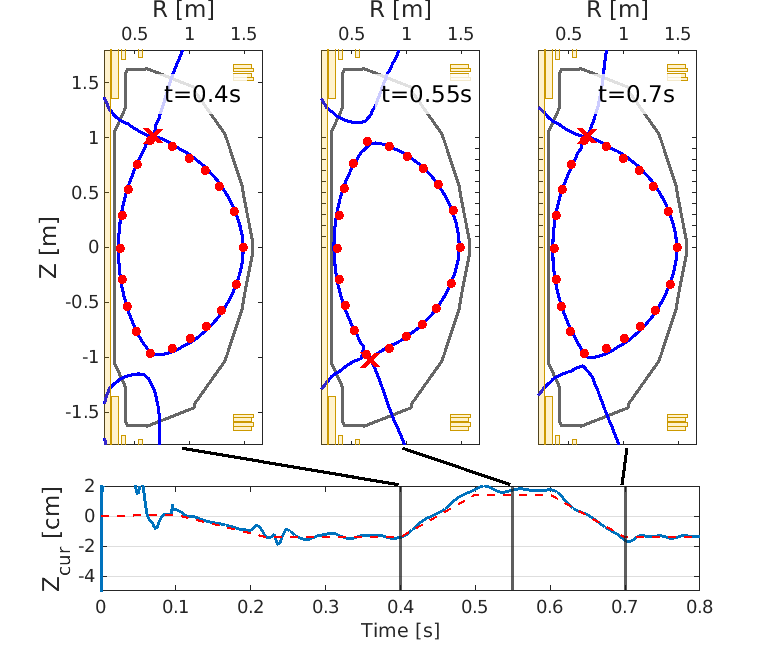}
    \caption{NSTX-U GSevolve simulation demonstrating improved controllability of the vertical position and dr-sep (lower vs upper single null) compared to the original controller design. Intentional and controlled upper-to-lower null switching behavior is executed.}
    \label{fig:sim29c}
\end{figure}

The highest elongation reached on NSTX-U was $\kappa=2.3$ \cite{Menard2017}. \Cref{fig:sim48b} is a simulation to ramp and elongate the plasma and achieve this challenging elongation. The vertical position is controlled closely throughout the pulse and shaping errors also remain below 2cm for the entire pulse. It appears that this is near the performance limits of the device. Even with slight theoretical improvements in the vertical controller performance, attempts to simulate higher elongation equilibria result in VDEs. 

\begin{figure}[H]
    \centering
    \includegraphics[width=0.7\linewidth]{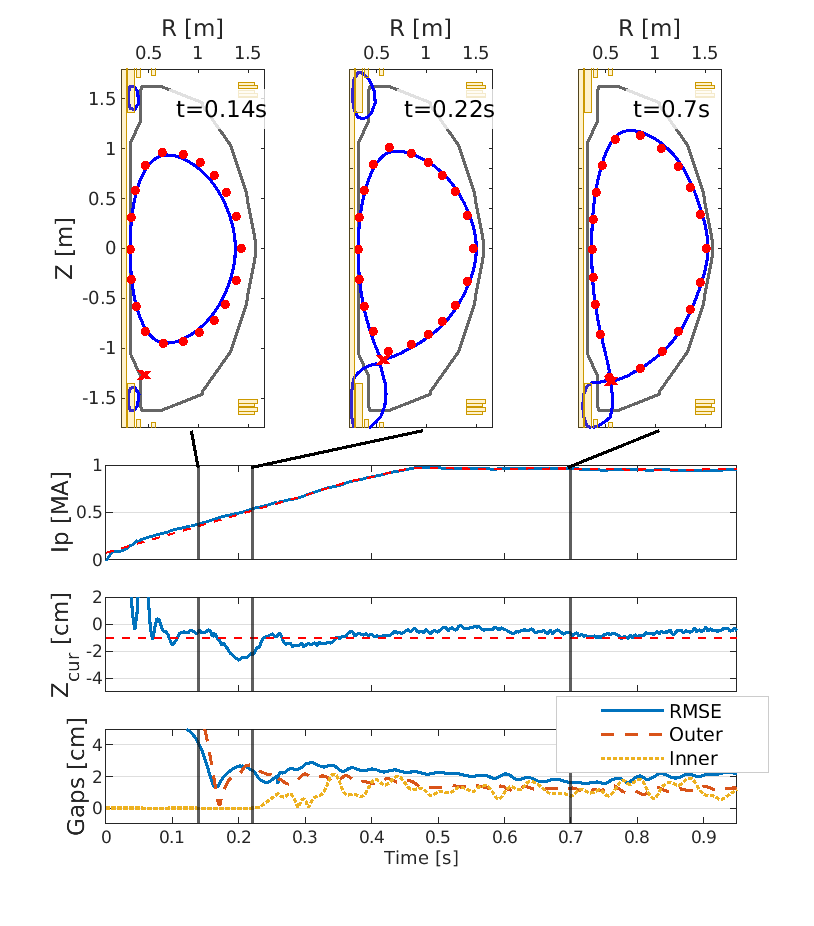}
    \caption{NSTX-U GSevolve simulation demonstrating good tracking performance and shaping for a highly vertically unstable $\kappa=2.3$ plasma.}
    \label{fig:sim48b}
\end{figure}

\Cref{fig:sim39} illustrates a nonlinear simulation of de-elongating the plasma. In this scenario, we consider a normal pulse that receives a signal that the plasma is experiencing an off-normal event and the plasma should retreat to a safe state. We attempt to ramp down $I_p$ quickly to a lower level, while simultaneously de-elongating the plasma to improve stability, but avoiding limiting on the wall. We are able to meet these objective, reducing the elongation from 1.9 to 1.2 and ramping down the plasma current by 600kA within a few 100ms. 

\begin{figure}[H]
    \centering
    \includegraphics[width=0.7\linewidth]{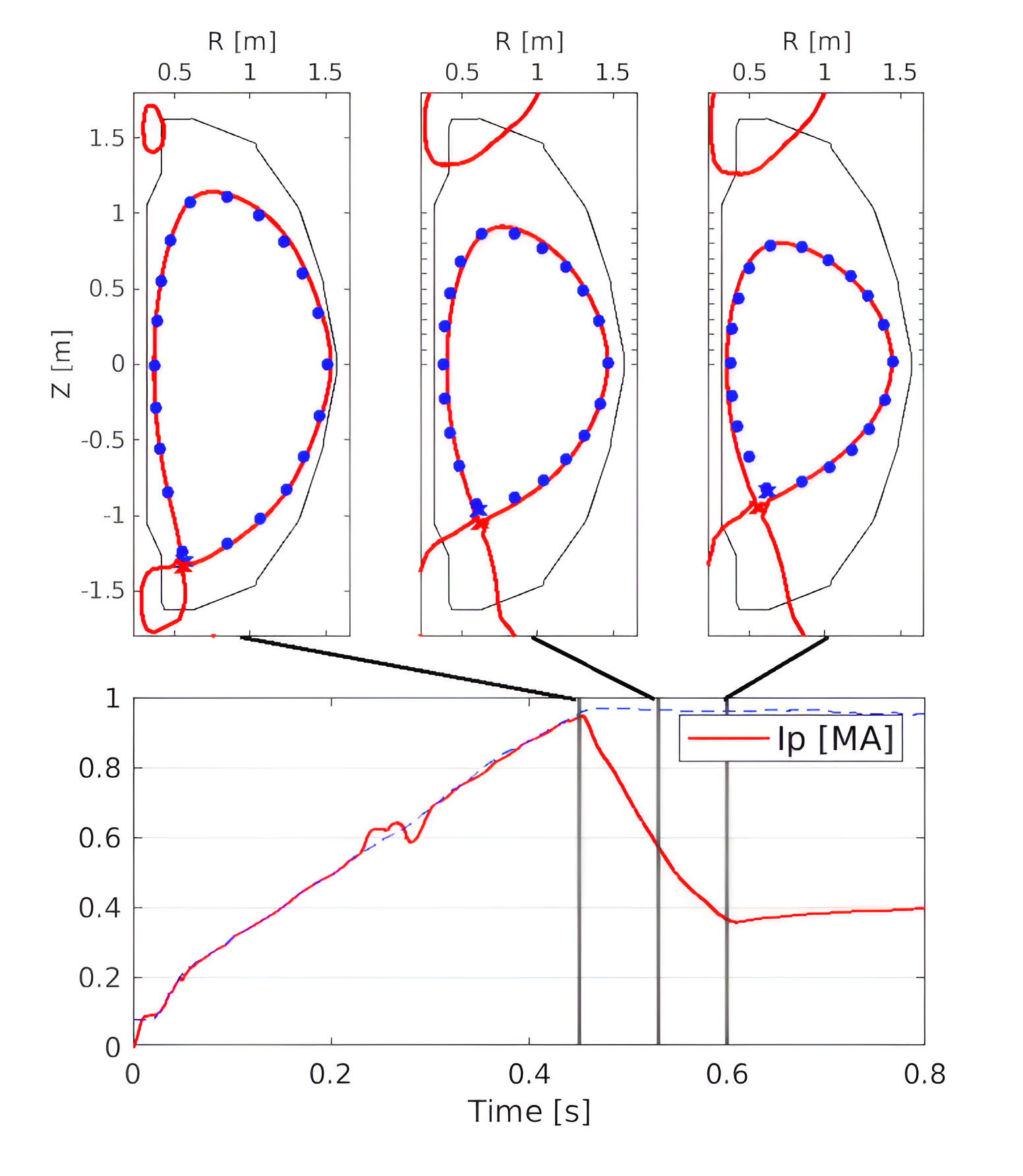}
    \caption{NSTX-U GSEvolve simulation of an emergency rampdown and compressive shrinking of the plasma, with the goal of keeping the plasma diverted.}
    \label{fig:sim39}
\end{figure}

The collection of these simulations is intended to demonstrate that the controller designed in this tutorial is versatile across a range of scenarios and can effectively track shaping targets and plasma current without disturbing vertical stability. 

\censor{
\section{SPARC shape control: design plans and progress}\label{sec:sparc_results}

\subsectionbf{Overview and plans for SPARC magnetic control}

The present vision for the SPARC magnetic controller is to use a constrained-QP controller algorithm, with a control architecture as shown in \cref{fig:sparc_ibsc_design_diag}. A description of the architecture and motivation is as follows: 

\begin{figure}[H]
    \centering
    \includegraphics[width=0.9\linewidth]{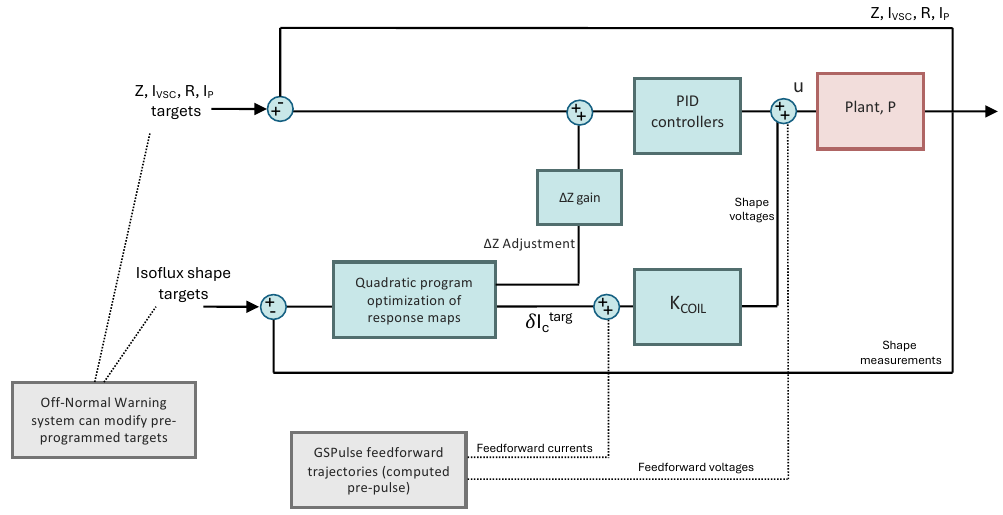}
    \caption{Baseline SPARC magnetic control design. The controller consists of dedicated loops for the vertical position and VSC current as well as radial outer gap and plasma current. All other shaping isoflux parameters are passed through a quadratic program optimization to obtain target PF coil currents. Feedforward trajectories are computed with the GSPulse code and can be injected as either currents or voltages. The off-normal warning system can modify pre-programmed shaping targets, leading to the desire that the response maps be generated in real-time.}
    \label{fig:sparc_ibsc_design_diag}
\end{figure}

\begin{itemize}
    \item With respect to the IBSC design options, we select a \textbf{QP-constrained controller}, based on \textbf{static} mappings of the \textbf{vacuum flux response}.

    \item The use of quadratic program optimization is desired because, moreso than present machines, SPARC will be hardware-constrained by both coil current and voltage limits during operation. The QP optimization is a natural extension of matrix inversion for these types of constrained problems. The QP optimizer has been implemented in the PCS and shown to be real-time capable ($<<$1ms cycle time) for typical SPARC-sized problems. 

    \item The use of static vacuum flux maps is motivated by the desire to have flexibility in responding to off-normal events, (and also deemed acceptable from the discussion of section \cref{subsec:response_model}). The SPARC PCS is being built to have a sophisticated off-normal warning system (for example, to provide warnings about imminent disruptions), and the warning system will have the capability to modify shaping targets from the nominal trajectory. 
    
    Because of the desire for flexibility, instead of building response maps along a pre-programmed trajectory, the response maps are actually computed in real-time based on the current equilibrium and targets. As shown later (\cref{sec:sparc_results}), the static map design can in general have lower performance than dynamic maps, but there is a tradeoff in that the dynamic maps are more complex to generate in real-time and sometimes require manual verification. This design choice and tradeoff will continue to be evaluated as more test cases and off-normal event simulations are developed. 

    \item Before a pulse, feedforward trajectories will be computed with the GSPulse code \cite{Wai2025}. GSPulse provides both feedforward currents and the associated feedforward voltages. These can each be optionally injected into controller as shown in \cref{fig:sparc_ibsc_design_diag} by using a weighted sum that defines the contribution from the feedback and feedforward terms. There are pros and cons to using feedforward currents vs feedforward voltages, and the controller is structured to support both options. Feedforward currents are more reliable in achieving the desired equilibrium as using voltage alone is subject to the coil currents drifting due to model mismatch. However, feedforward voltage has the advantage that there is no phase delay from the coil controller block, which can be useful for tracking dynamic trajectories like strikepoint sweeping. 

    \item This design uses dedicated PID controllers vertical position ($Z$), radial position ($R$), plasma current ($I_p$), and control of the current in the vertical stabilization coil ($I_{VSC}$). These are isolated as separate controllers in order to support tokamak commissioning and a staged implementation of the controller. In early operations, ($Z,R,Ip,I_{VSC}$ will be the most important controlled variables and a reasonable pulse could be created with just feedback control on these quantities and feedforward programming of everything else. Having these as dedicated controllers will support early operations and facilitate operator tuning. $Z$ and $I_{VSC}$ should always remain as dedicated control loops due to the vertical coupling interaction (\cref{sec:static_plus_vertical}) but there could be some performance benefit by folding $R$ and $I_p$ into the QP optimization. This will be re-evaluated if necessary.  

    \item For the coil current controller ($K_{coil}$), we will use the frame alignment technique to determine the matrix gains, which shows some performance benefit compared to a controller based on the coil-to-coil mutual inductances. 
    
\end{itemize}

SPARC will make heavy use of pre-shot simulation testing, motivated by the urgency of the SPARC mission and sub-goal of minimizing operational time spent in commissioning the control system. This is expected to be of particular use for the magnetic control due to the maturity of simulators for the free-boundary equilibrium evolution problem. The SPARC PCS is designed with a modular application style that allows for easy packaging of algorithms for deployment against various sources, including the real SPARC tokamak, hardware-in-the-loop test harnesses, and hardware-out-of-the-loop test harnesses. For magnetic control, the plan is to design a library of shape control simulation episodes and test the controller performance against each of these episodes before deployment against SPARC. This will allow us to gather metrics, test the controller algorithms and gain values, and evaluate which scenarios are likely to stress the performance. High fidelity simulations will be run against the FGE code, while lower-fidelity, real-time simulations can be run against the a purpose-built lower fidelity simulator called COMET (Control-Oriented Model of the Entire Tokamak). The real-time COMET simulations will be used to evaluate not only shape control performance but also hardware-in-the-loop and timing performance. This framework is depicted in \cref{fig:sparc_pcs_hootl_sims}. 

\begin{figure}[H]
    \centering
    \includegraphics[width=0.7\linewidth]{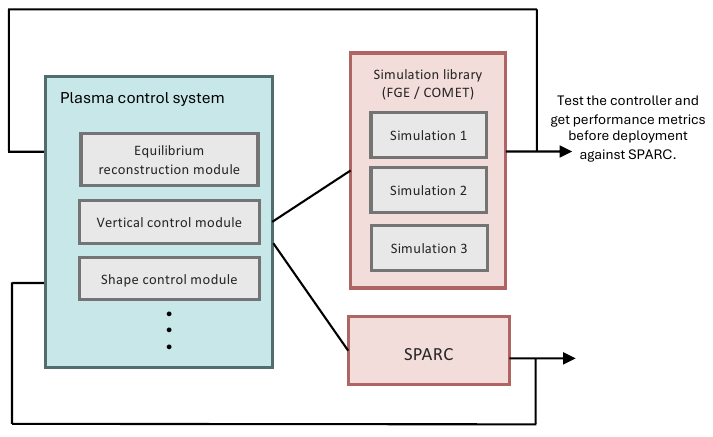}
    \caption{A modular PCS software infrastructure will support the ability to run against different targets, so that the magnetic controller can be tested against a library of simulation episodes before a pulse.}
    \label{fig:sparc_pcs_hootl_sims}
\end{figure}

\subsectionbf{Summary of SPARC magnetic control findings}

\subsubsection{Vertical control}

The vertical controller is obtained from following the tutorial in \cref{app:sparc_static_tutorial}. As can be seen in \cref{fig:sparc_vertical_nyquist}, the vertical controller has very robust phase margins of more than $100^\degree$ (for the nominal $I_p=8.7MA$, $\kappa=1.9$ PRD equilibrium). Note that phase margin is a metric used to evaluate robustness against model uncertainty, but does not consider other performance limitations like hardware constraints. Additional vertical stability performance was conducted in \cite{Nelson2024} and concluded acceptable but not highly robust operation with a $\Delta Z_{max}/a = 6-7\%$ . 

\begin{figure}[H] 
    \begin{center}
        \includegraphics[width=8cm]{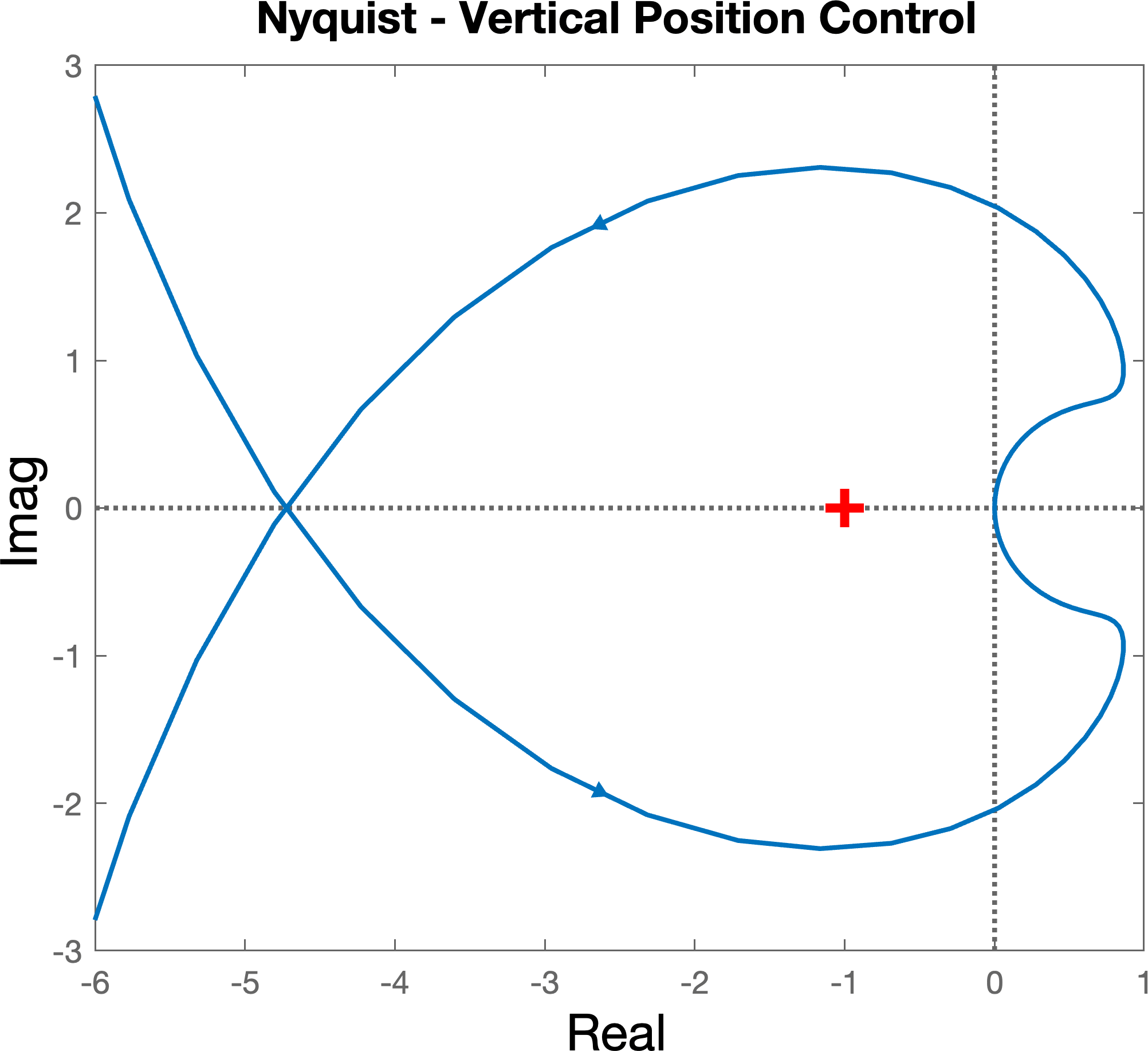}
    \end{center}
    \caption{Nyquist plot of the vertical control loop gain, indicating robust phase margins of more than $100^\degree$.}
    \label{fig:sparc_vertical_nyquist}
\end{figure}

SPARC has a dedicated in-vessel vertical stability coil (VSC) to perform fast vertical stabilization. The VSC consists of 4 windings and has a normal operating limit of 40kA per turn. In order to preserve headroom in the coil current it is necessary to perform feedback on the VSC current itself to keep the current centered around zero. The $I_{VSC}$ controller \cref{app:sparc_static_tutorial} consists of a linear combination of PF1-3 that supply radial field and apply proportional feedback based on the VSC current measurement. A step response of the combined $Z+I_{VSC}$ system is shown in \cref{fig:sparc_ivsc_control}. For a 1cm excursion in the Z position, the peak VSC current is 6.5kA (neglecting the initial spike), illustrating that SPARC should reasonably expect to keep the VSC current within limits for disturbances of multiple cm. Adding $I_{VSC}$ control does not affect the overshoot, and the settling time is slightly extended by $\sim20$ms. It takes about the 100ms to fully drive the VSC current to zero which is limited by the flux penetration time of the PF coils. 

\begin{figure}[H] 
    \begin{center}
        \includegraphics[width=0.5\linewidth]{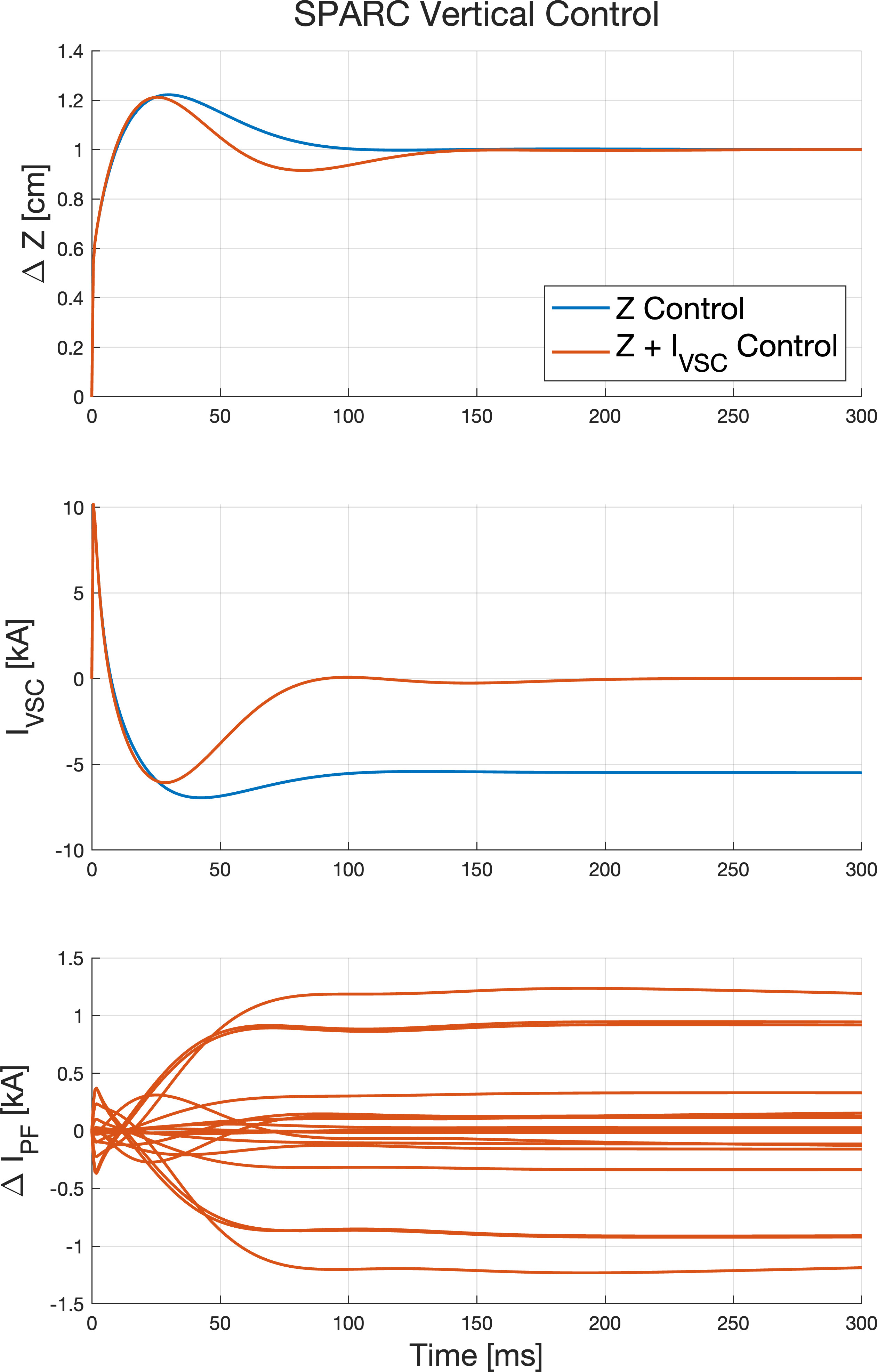}
    \end{center}
    \caption{Comparison of step responses for the $Z$-only controller and the $Z+I_{VSC}$ controller. A simple proportional controller on the VSC current shows almost zero degradation of the vertical control, and is able to drive VSC current to zero in about 100ms.}
    \label{fig:sparc_ivsc_control}
\end{figure}

\subsubsection{Radial and Ip control}

The radial and Ip controller performance is shown in \cref{fig:sparc_r_ip_step}. Here we observe that the timescale for radial outer gap control is somewhat slower with a 100ms rise time for a 1cm response. In order to maximize plasma volume and couple effectively to the ICRF heating antennas, the outer gap distance between the plasma boundary and limiter is targeted to be around 1cm. The exact tolerance for effective radial control will depend on the disturbance characteristics, but given that the response time 100ms, this could prove challenging to meet sub-cm tolerances, and may require increasing the target outer gap distance for machine protection. We are able to effectively decouple radial and Ip control, such that a 50kA change in plasma current affects the outer gap position by roughly 3mm. 

\begin{figure}[H] 
    \begin{center}
        \includegraphics[width=0.7\linewidth]{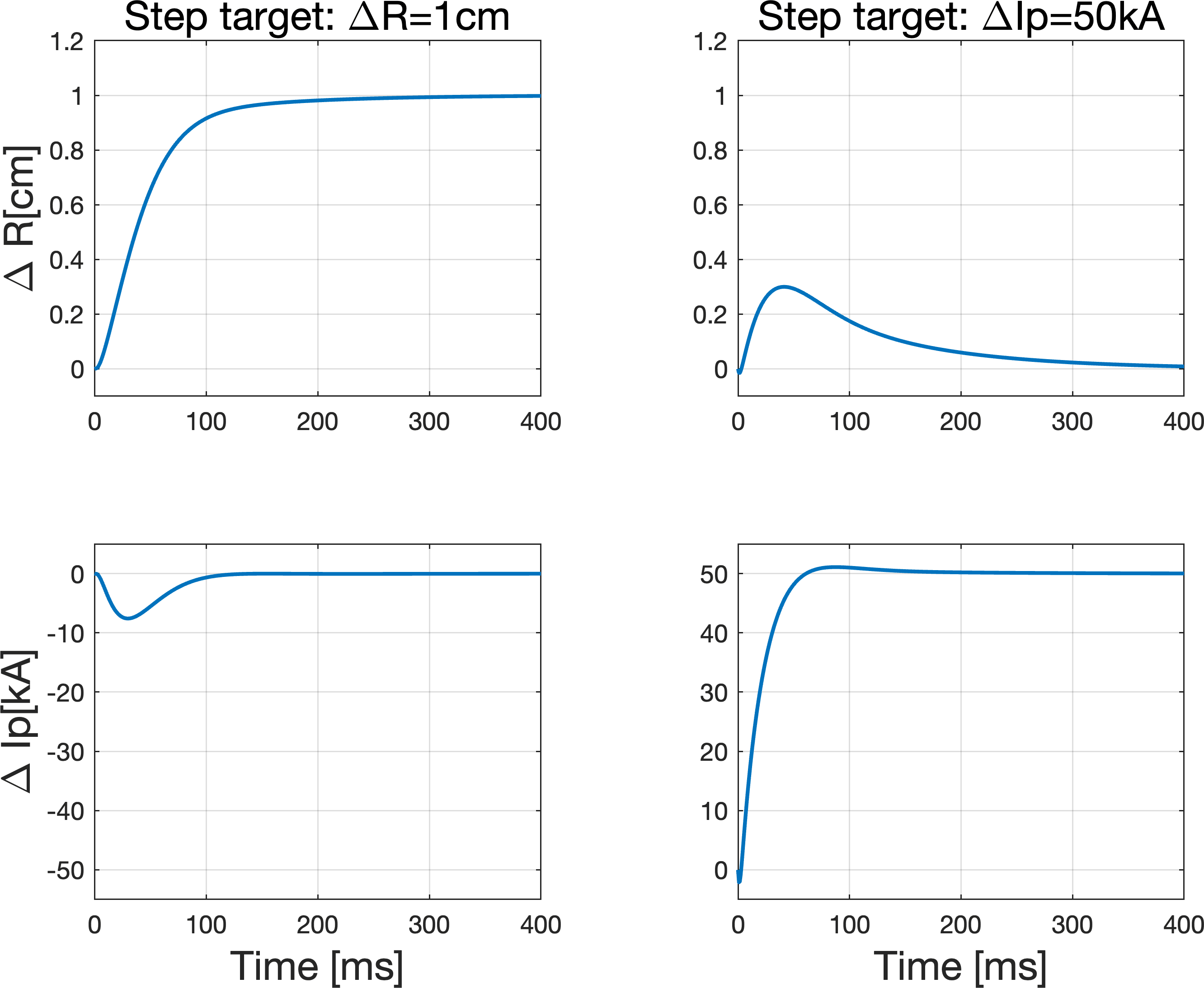}
    \end{center}
    \caption{Step responses for radial and Ip control.}
    \label{fig:sparc_r_ip_step}
\end{figure}

\subsubsection{Coil current control}

Coil current control in SPARC is performed using the frame alignment algorithm \cref{app:align_algorithm}, which results in a highly-effective decoupling controller. \Cref{fig:sparc_coil_controller} compares this controller against a common controller design based on the coil-to-coil mutual inductances. Here we perform a step response on only the PF4U coil, with a stepsize that also saturates the voltage limits. PF4 is the biggest coil on SPARC and has the highest inductive coupling to other coils. With the ALIGN decoupling controller, even under these conditions the other coils are able to reject the flux disturbance from PF4 and only show induced currents of $2-3\%$ the PF4 change. 

\begin{figure}[H] 
    \begin{center}
        \includegraphics[width=0.8\linewidth]{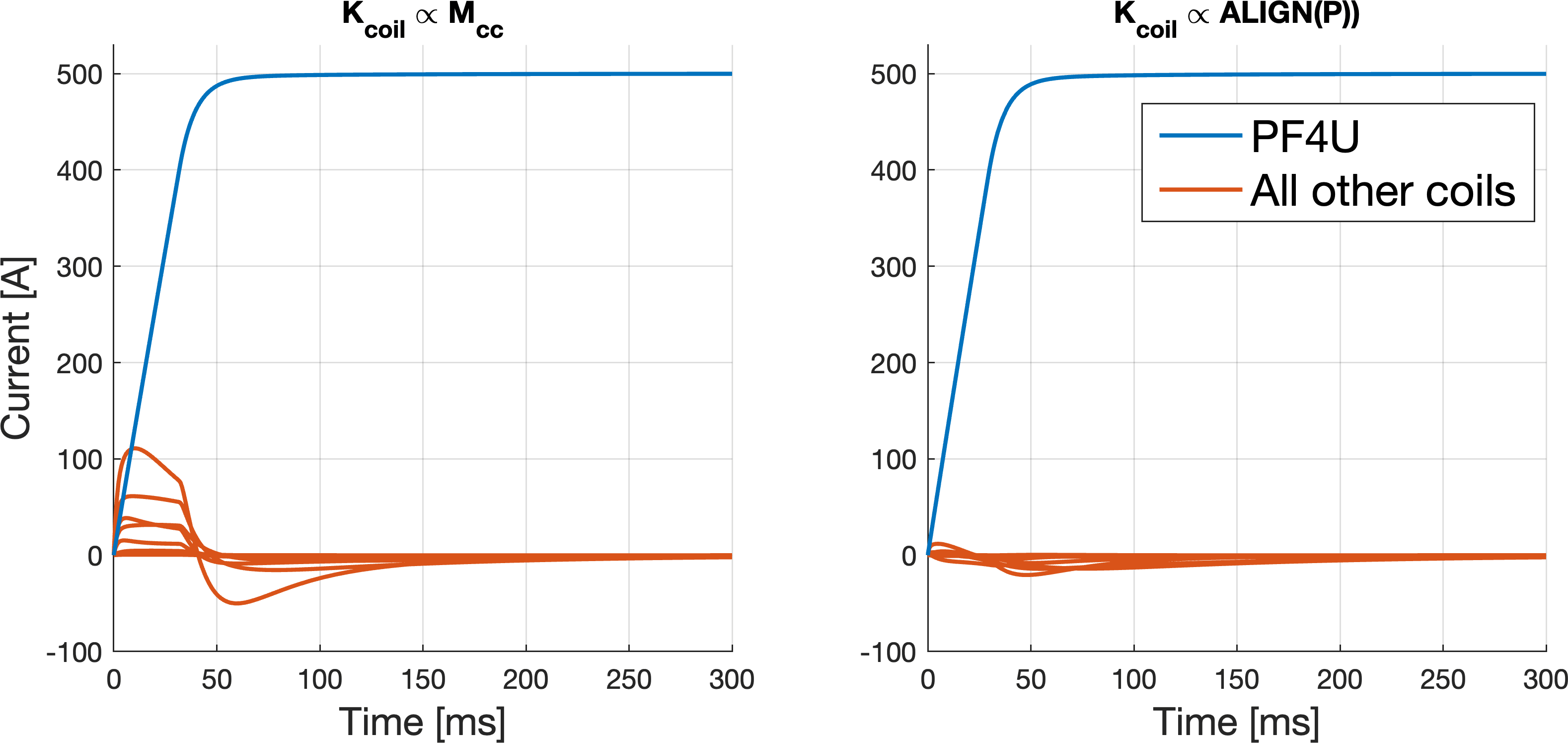}
    \end{center}
    \caption{Coil current control based on frame alignment provides effective decoupling, as demonstrated by a step response on the PF4 coil, which has the highest inductive coupling value to the other coils. The target step size of 500A is sufficient to cause voltage saturation during the response. Even under these conditions, the ALIGN controller has disturbance values of only 2-3\% of the step size.}
    \label{fig:sparc_coil_controller}
\end{figure}

\subsectionbf{IBSC with static maps: feedforward + feedback simulations}\label{sec:sparc_ibsc_results}

In \cref{app:sparc_static_tutorial}, we design an IBSC controller using static mappings of the vacuum flux response, following the proposed control design. Simulations are performeed using the nonlinear flight simulator FGE code \cite{Carpanese2021}. For references to specific coils, see the coil geometry layout of \cref{fig:geo_sparc}. 

\Cref{fig:sparc_betap} illustrates the performance of this magnetic controller to reject a disturbance in the plasma $\beta_p$ during the flattop period ($Ip=8.7$MA) of the primary reference discharge. A 30\% disturbance decreases $\beta_p$ form 0.28 to 0.2 to simulate an internal plasma disturbance such as an MHD event or ELM. At the plasma pressure drops, the radial outer midplane ($R_{OMP}$) decreases 1.5cm and then slowly recovers as the controller responds. The radial gap is mostly controlled by the PF4 upper and lower coils, and the voltage for these power supplies responds nearly instantaneously. However, it takes several 100ms for tthe PF4 current to respond, and nearly 0.5 seconds for the system to settle completely (note that PF4 is ramping throughout this process, to compensate for the ohmic current drive and shaping which occurs simultaneously). As discussed, one potential limitation for SPARC is the relatively slow response time for radial control, and any fast disturbances ($<$100ms) cannot be compensated with feedback control. On the other hand, as seen in even this large $\beta_p$ disturbance only moved the outer gap by 1.5cm, and even though this elongates the plasma, the combined magnetic control system is still able to keep the plasma vertically stable and control the plasma shape without introducing additional shaping errors. 

\begin{figure}[H]
\subfloat{\includegraphics[height=9cm]{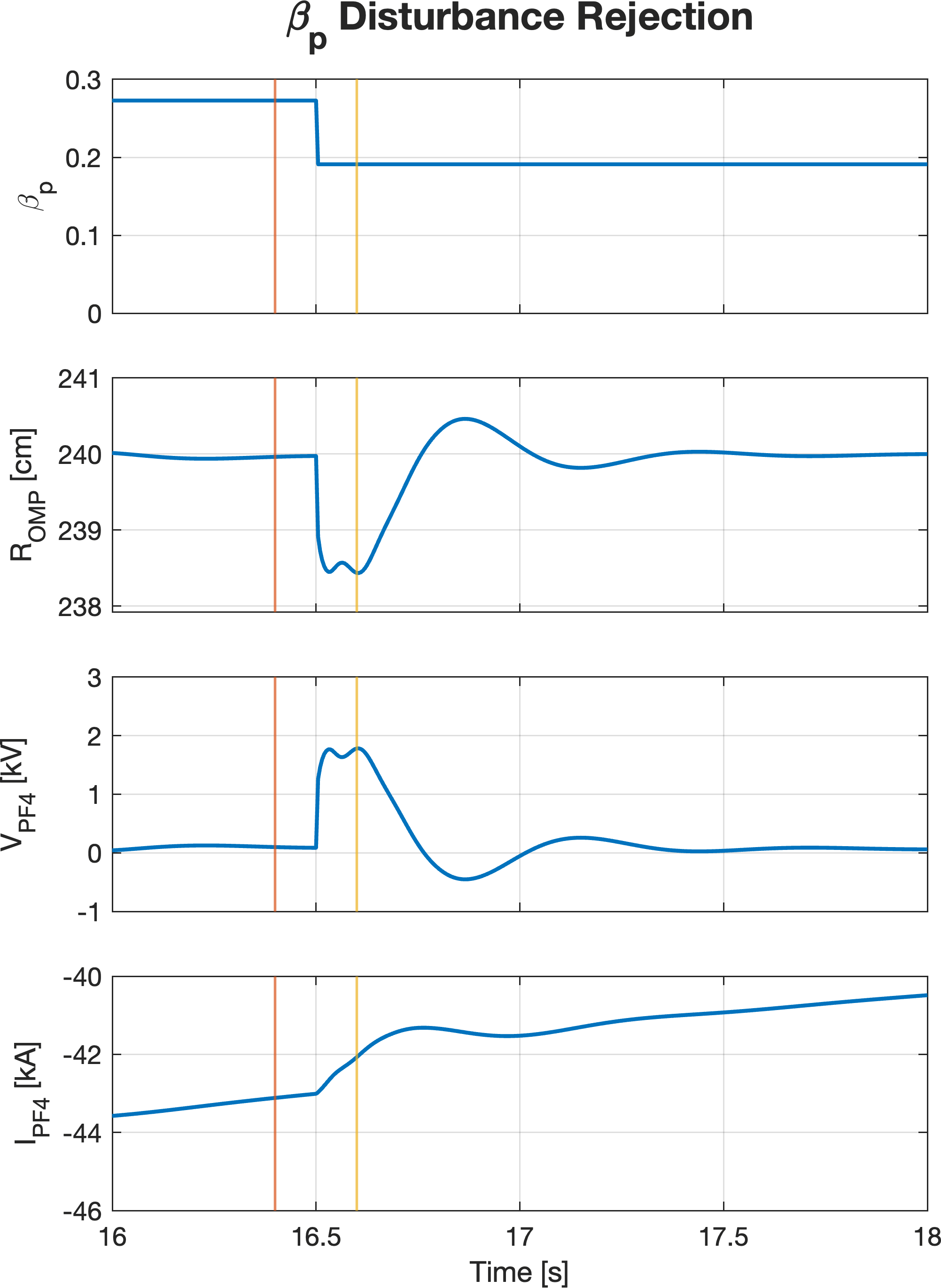}} 
\subfloat{\includegraphics[height=9cm]{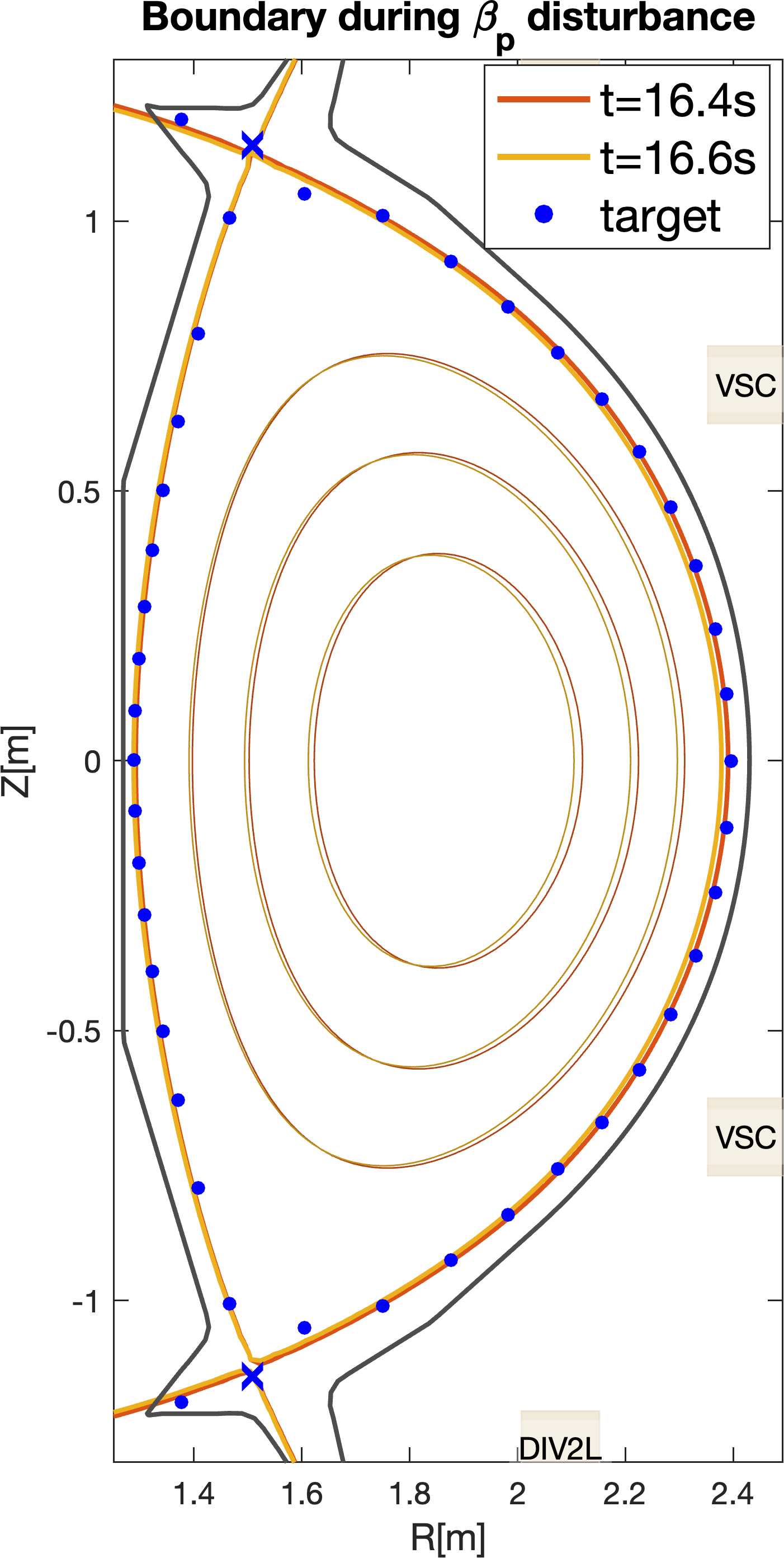}}\\
\caption{Control performance in rejecting a $\beta_p$ step disturbance representing an MHD event. The controller immediately applies a compensating voltage on PF4, but it takes several 100ms for the PF4 to fully respond and cancel the disturbance. Shape and vertical control are maintained throughout the response. }
\label{fig:sparc_betap}
\end{figure}

\Cref{fig:sparc_constrained_control} demonstrates control performance during a 24 second, 8.5MA, lower single null pulse, using the constrained QP control approach. During this pulse, the unconstrained controller violates current limits on CS3L and PF1L in the middle of the pulse. By contrast, the constrained controller is able to keep the currents within limits. As shown in \cref{fig:sparc_constrained_control}b, at $t=12$s when the current constraints were violated, the unconstrained controller is able to match the same shape to imperceptible differences. Note that the equilibrium solution found by the QP-constrained controller is different than what would be achieved by just applying limits to the unconstrained controller; for example, the CS3U current trajectory is changed even though this coil did not saturate originally. Also note that these coil current limits were down-rated slightly from the maximum hardware limits for this simulation. 

\begin{figure}[H]
\subfloat{\includegraphics[height=8cm]{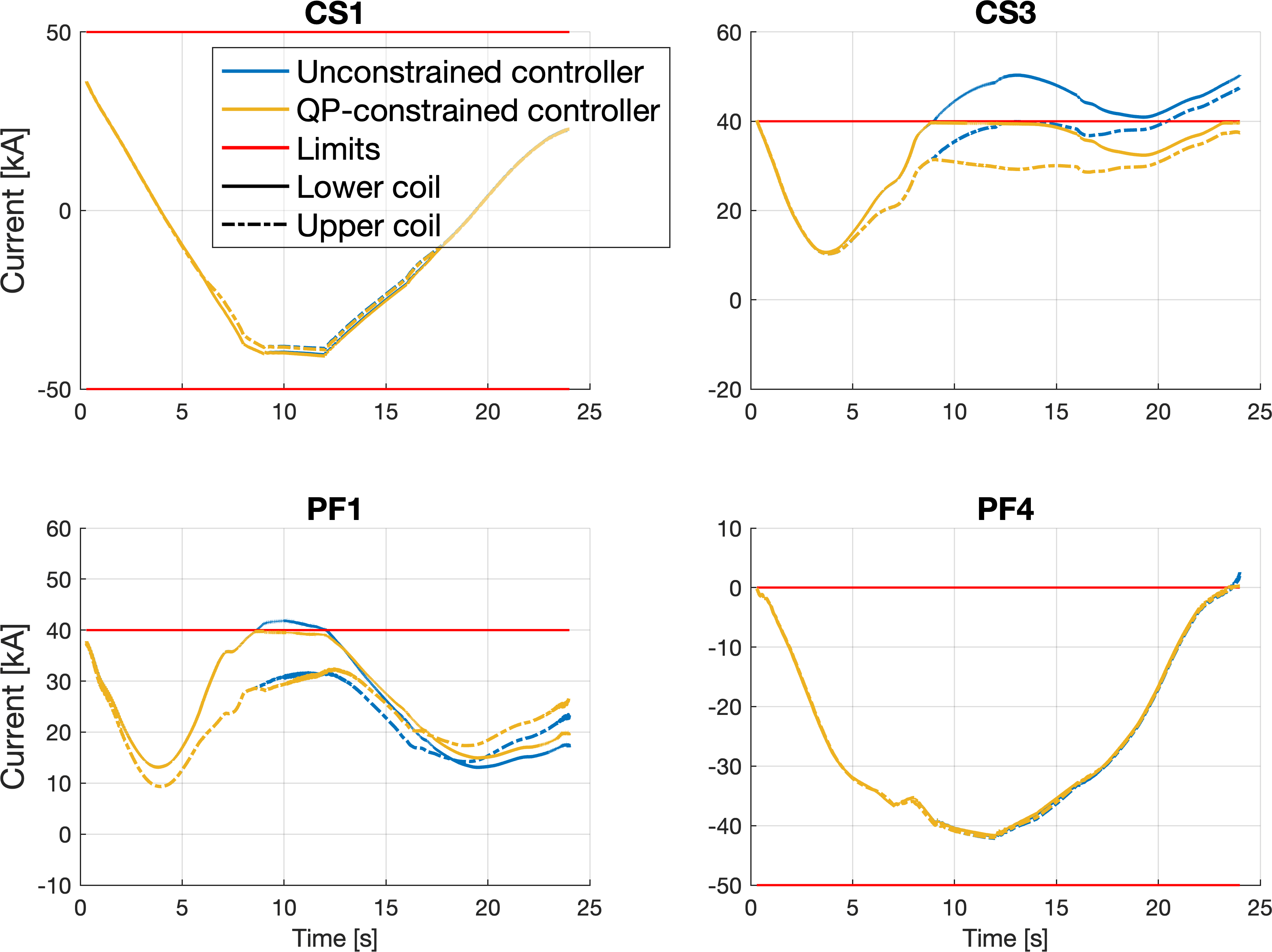}} 
\subfloat{\includegraphics[height=8cm]{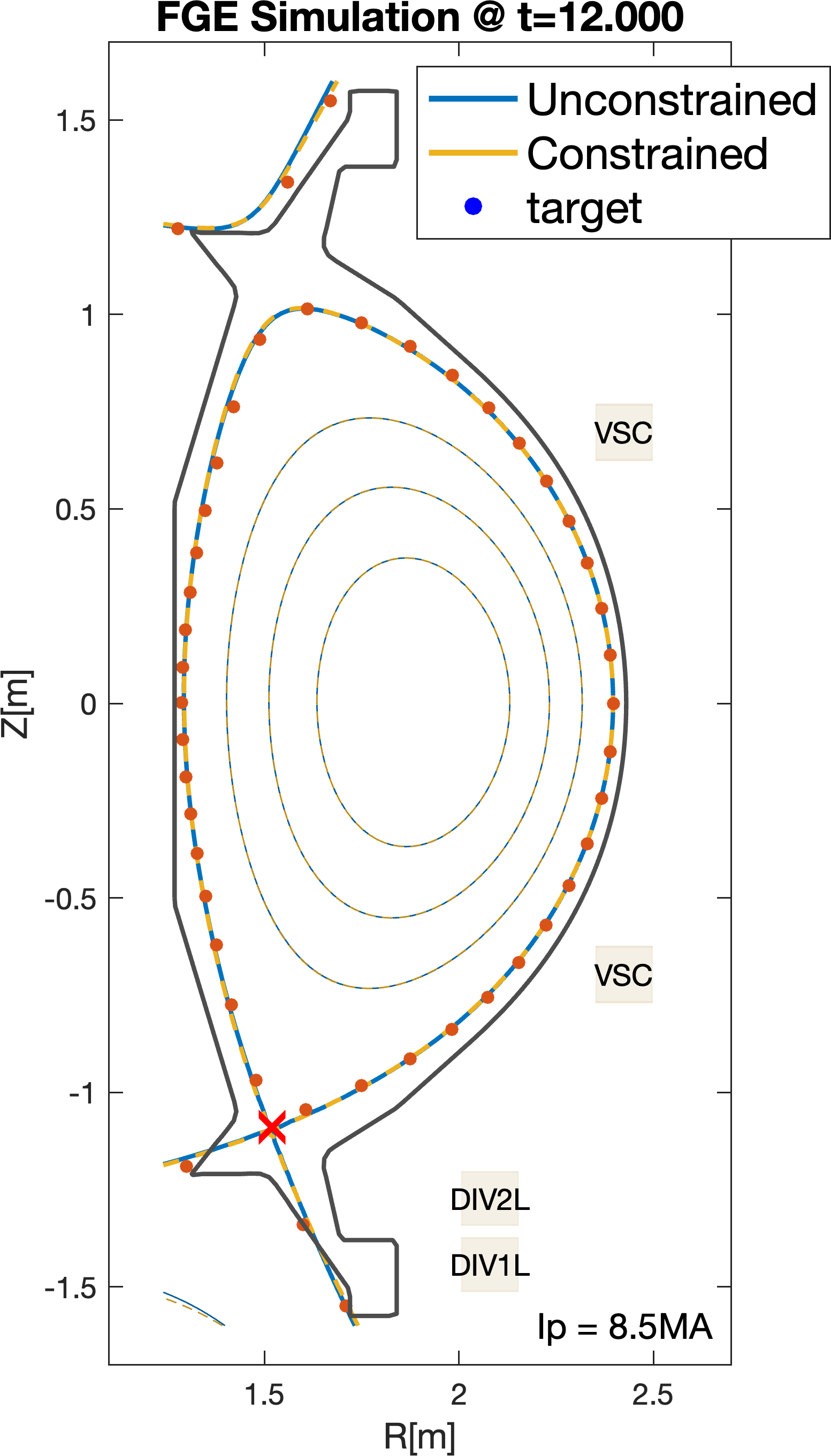}}\\
\caption{Performance of the QP-constrained constrained controller during an 8.5MA pulse. CS3 and PF1 are critical for obtaining the desired strikepoint positions and are pushed to their limits in this configuration. The QP controller keeps all currents below limits with nearly imperceptible change in the overall shaping, while the unconstrained IBSC controller violates limits on CS3 and PF1 (and PF4, at the very end of pulse).}
\label{fig:sparc_constrained_control}
\end{figure}

\Cref{fig:sparc_ff} shows controller performance during a 3MA, limited, 10-sec pulse. This is an example scenario that SPARC could run during an early campaign as it demonstrates capability towards running more aggressive pulses on the way to $Q>1$. In the feedforward case, we apply vertical control, radial control, and plasma current control. Shape control is not used, and the pulse is set up to track feedforward current targets that were pre-computed offline using the GSPulse code \cite{wai2025}. As in a real pulse, the feedforward currents are not fully accurate and use different assumptions about plasma properties than are used in simulation. As can be seen in \cref{fig:sparc_ff}, the first few seconds of simulation do reasonably achieve the target shape evolution of expanding the plasma on the inner limiter and elongating it. As the pulse progresses, the feedforward trajectories become less and less accurate and the plasma elongation is inconsistent with the target. In the feedback case, the shape controller is turned on and is able to re-balance mostly the CS2/PF2/PF3 coils in order to achieve the target elongation. This is an example of how early commissioning of SPARC controllers might work. An advantage of the IBSC approach is that it can be implemented in stages and portions of the control verified. For example, the first pulses will likely be run in feedforward only. Then radial control, Ip control, and vertical control can be added, and as confidence in the models increases, additional controller features can be added without re-tuning these systems. 

\begin{figure}[H]
\subfloat{\includegraphics[width=0.8\linewidth]{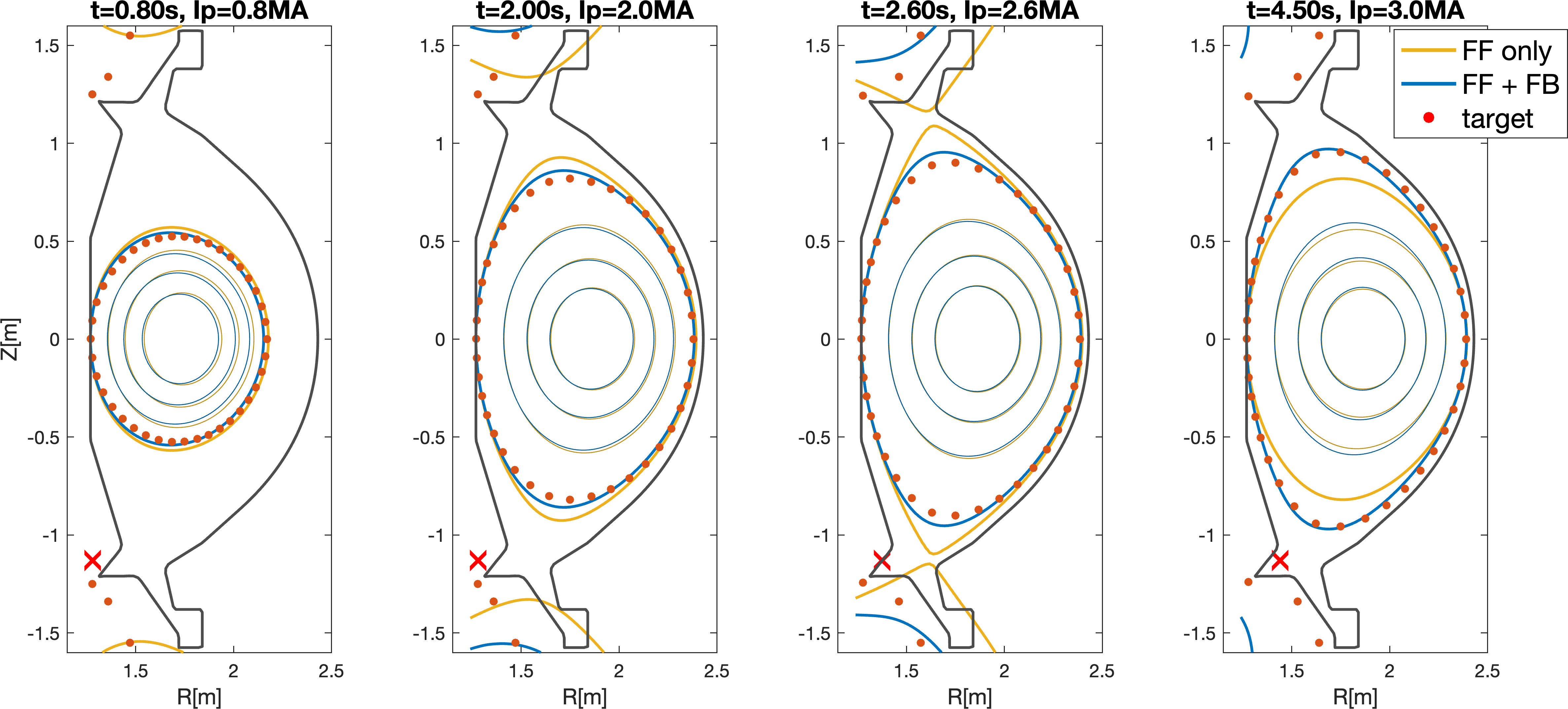}} \\
\subfloat{\includegraphics[width=0.8\linewidth]{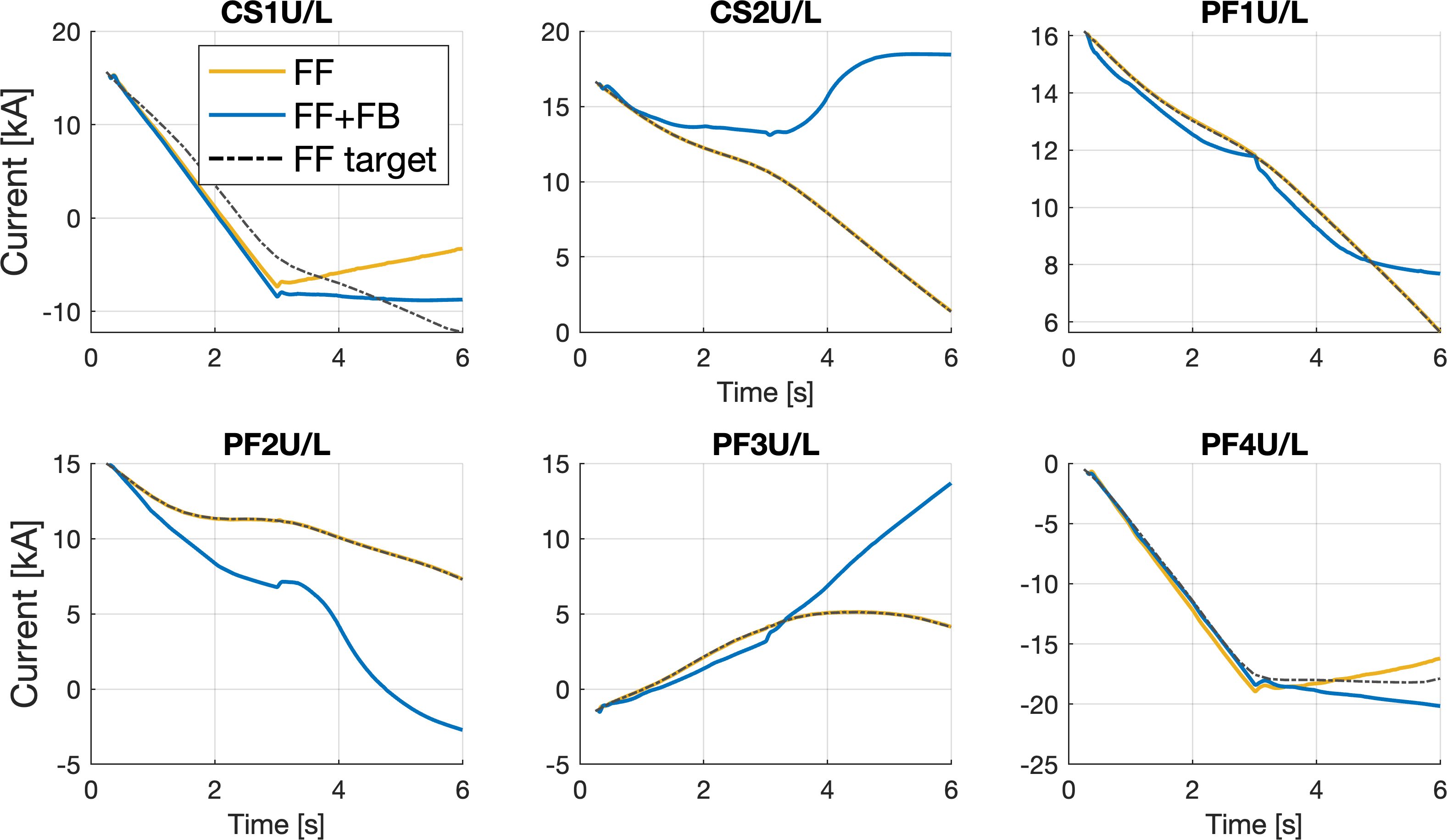}}
\caption{Control of an early campaign pulse using radial, vertical and Ip control and feedforward-only on the shape, versus using full-feedback. Since the feedforward currents use differing assumptions for plasma properties, the feedforward simulation is only able to track the shape for the first few seconds before the plasma elongation starts to differ from the target. Once shape feedback is added the controller tracks the target as desired for the full pulse. }
\label{fig:sparc_ff}
\end{figure}

\subsectionbf{IBSC with dynamic maps}\label{sec:sparc_dynamic_ibsc_results}

The baseline controller for SPARC uses static maps but may be upgraded to use dynamic maps. To explore this possibility we design \cref{app:sparc_dynamic_tutorial} two decoupling PID controllers and apply these to a target tracking problem as shown in \cref{fig:sparc_align_geo,fig:sparc_align_vs_pid_traces}. The controllers consist of 11 PID loops each, providing feedback on the plasma Z-position, plasma current, VSC current, 6 shape control points, and $(B_r,B_z)$ at the target x-point. For this case we attempt to track a ramp in the lower triangularity and x-point position while preserving the upper shape. The 9cm ramp is a very large change in the x-point position. As seen in \cref{fig:sparc_align_vs_pid_traces}, both controllers respond reasonably well to the aggressive shape change, though the dynamic ALIGN controller is superior with lower tracking errors and fewer disturbance oscillations. 

\begin{figure}[H] 
    \begin{center}
        \includegraphics[width=5cm]{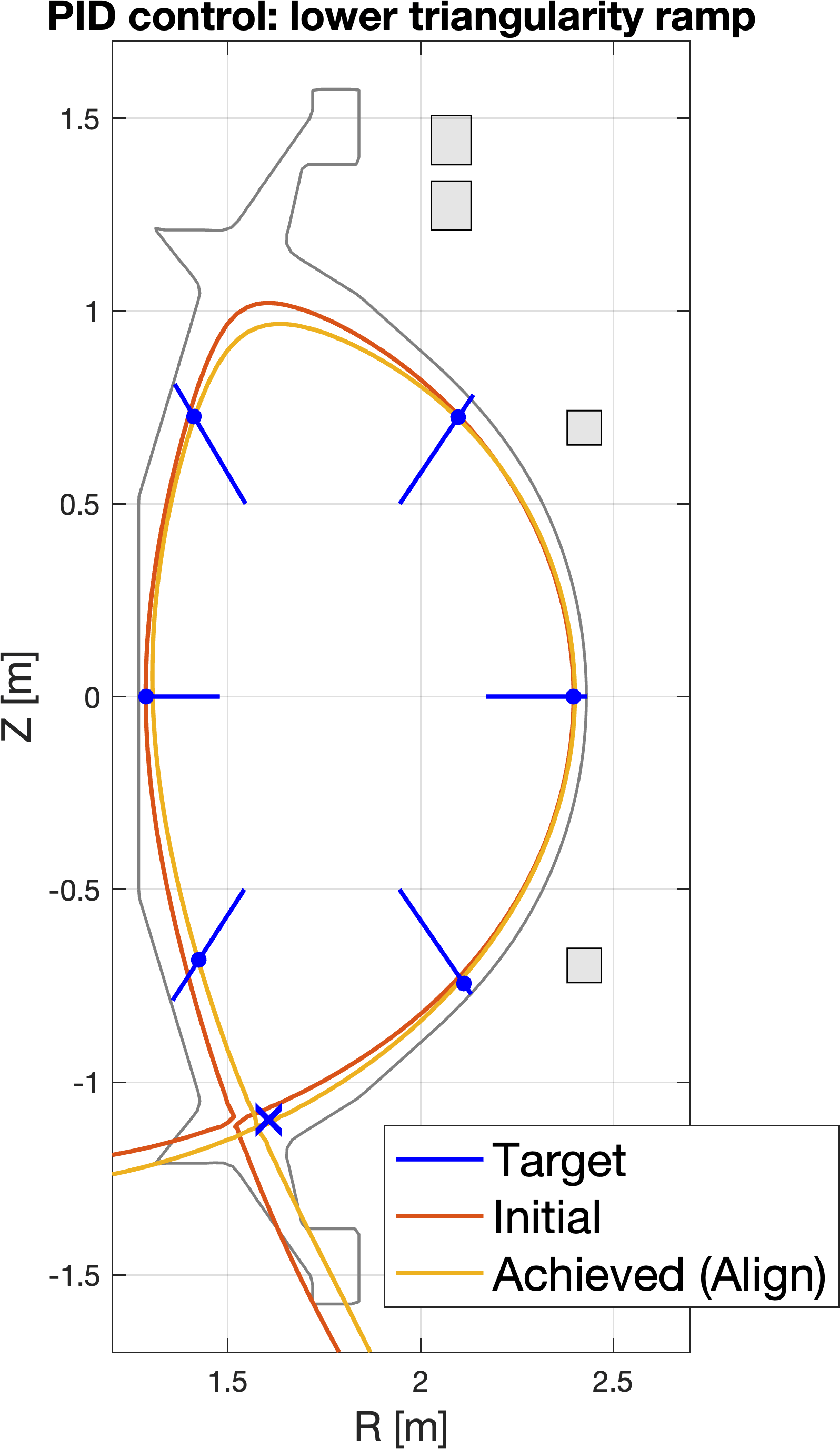}
    \end{center}
    \caption{Initial and target shapes used for the target-tracking simulation, with the objective of tracking a ramp in the lower triangularity.}
    \label{fig:sparc_align_geo}
\end{figure}

\begin{figure}[H] 
    \begin{center}
        \includegraphics[width=0.7\linewidth]{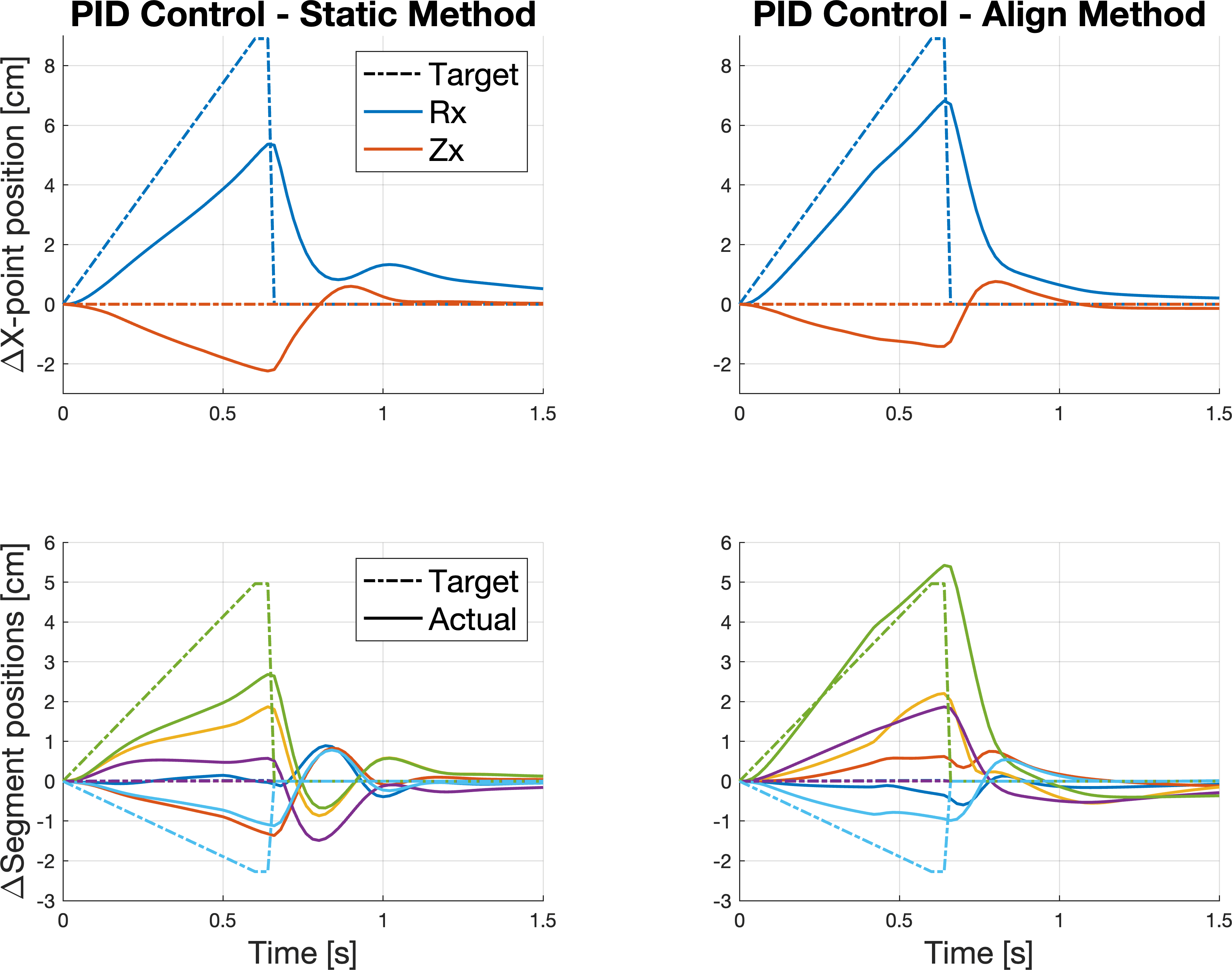}
    \end{center}
    \caption{Comparison of decoupling PID controller performance based on static and dynamic maps in tracking a lower triangularity ramp. The dynamic map controller (ALIGN) is able to move the x-point 6.5cm out of a target 9cm, compared to only 5cm for the static map controller.}
    \label{fig:sparc_align_vs_pid_traces}
\end{figure}

\Cref{fig:sparc_align_vs_pid_sensitivity1} gives further insight into the improved decoupling performance by showing the cross-sensitivity of the response of two different shape control points (inner gap and inner-upper gap). At low frequencies (5 rad/s), both methods have similar sensitivity values indicating that the near steady-state performance of both controllers is similar. However, the controller based on dynamic maps was designed to maximize decoupling at a target bandwidth of 31 rad/s, and indeed shows superior decoupling across the frequency range. As discussed previously, the performance advantages of the dynamic map controller have to be weighed against tradeoffs such as the ability to easily update and trust static maps generated in real-time. These tradeoffs will continue to be explored and the final controller may be influenced by both methods. 

\begin{figure}[H] 
    \begin{center}
        \includegraphics[width=0.8\linewidth]{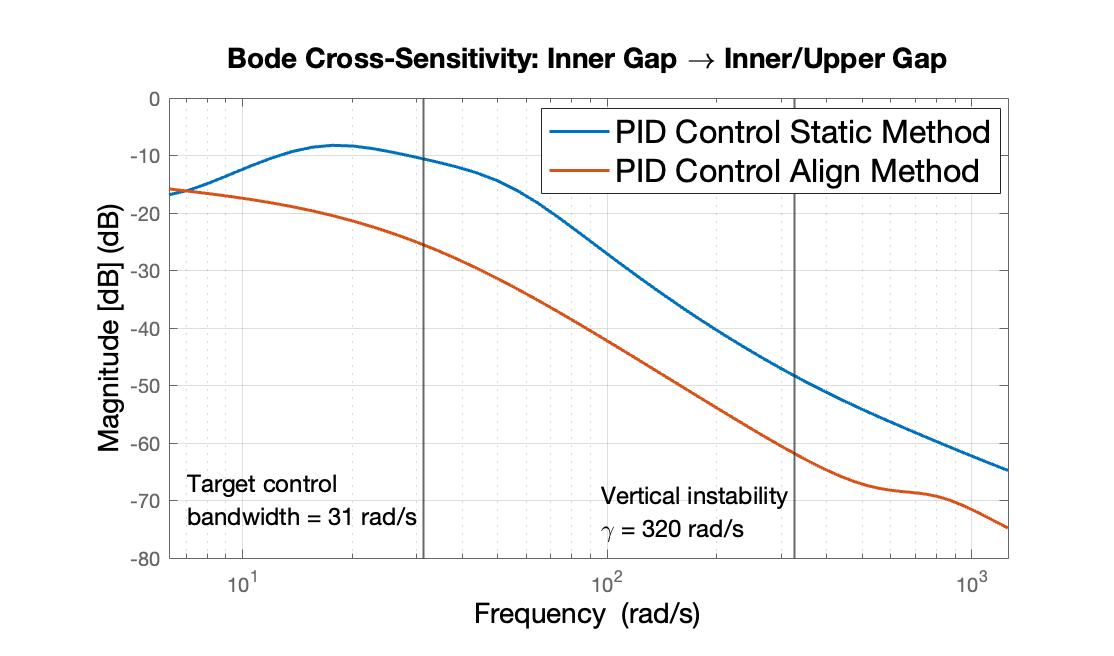}
    \end{center}
    \caption{The dynamic map controller using ALIGN has lower coupling between different shape control points, across the relevant frequency range.}
    \label{fig:sparc_align_vs_pid_sensitivity1}
\end{figure}
}{}

\newpage 
\section{Conclusion}

In this work we have described the inversion-based shape control (IBSC) framework and discussed design strategies and subtleties associated with specific design choices. Tutorials and additional discussion are included in the appendices. IBSC control designs have been created for NSTX-U\censor{ and SPARC}. For NSTX-U, the vertical bobble is removed by applying vertical/shape decoupling and simulations indicate a stronger ability to control the vertical position and whether upper or lower null. \censor{For SPARC, simulations of the controller indicate strong vertical control performance and that the hardware can support the design scenarios for SPARC including breakdown, ramp up, x-point formation and control, strikepoint sweeping, and rampdown. The SPARC controller is designed to be able to commission and test in stages, and early pulses will likely consist only of vertical, Ip, and radial control, and feedforward trajectories provided by the GSPulse code. Additionally, an integrated PCS pulse planning and simulating workflow is being created, with the vision that the feedforward is automatically computed for each target pulse, and that before each pulse the PCS controller can be tested in simulation against a library of control episodes.}

\newpage
\pagebreak
\vspace{1in}
\appendix

\newpage 
\begin{center}
    {\textbf{\huge Appendices}} 
\end{center}

\newpage
\section{Table of notation}\label{app:notation}
\begin{longtable}{ p{.20\textwidth}  p{.80\textwidth} } 
\captionsetup{font={Large}}
\toprule
$\delta$ &= perturbed value relative to a value at a reference equilibrium \\
$y$ &= vector of controlled shaping parameters (flux error at shape control points, field at target x-point, etc) \\
$T$ &= current or voltage shape response map \\
$I_c$ &= coil circuit currents \\
$V_c$ &= power supply voltages for coil circuits \\
$r$ &= vector of isoflux shaping targets \\
$e$ &= vector of isoflux shaping errors ($e := r - y$) \\
$s$ &= Laplace variable \\
$P(s) \text{ or } G(s)$ &= generic state-space system transfer function \\
$A,B,C$ &= generic state-space control matrices \\
$x$ &= generic state-space control state vector (for shape control, $x$ is the vector of coil/vessel/plasma current) \\
$u$ &= generic state-space control input vector (for shape control, $u = V_c$) \\
$I$ &= depending on context, identity matrix or vector of currents \\
$J$ &= cost function \\
$W_y$ &= weighting matrix for the shaping parameters $y$ \\
$W_{I_c}$ &= weighting matrix for the coil currents \\
$H$ &= quadratic program (QP) standard form quadratic term \\
$f$ &= QP standard form linear term \\
$A_{ineq}, b_{ineq}$ &= QP standard form inequality constraint terms \\
$Z$ &= generic vertical position (for example, plasma current centroid vertical position) \\
$R$ &= generic radial position (for example, plasma current centroid radial position, or plasma radial outer gap position) \\
$I_p$ &= total plasma current \\
$I_{VSC}$ &= current in the dedicated vertical stability coil (VSC) \\
$K_{coil}$ &= coil current controller \\
$\kappa$ &= plasma elongation \\
$\Delta Z_{max}$ &= $\Delta Z_{max}$ metric \cite{Humphreys2009} \\
$a$ &= plama minor radius \\
$M_{cc}$ &= coil to coil mutual inductances \\
$B_r, B_z$ &= radial and vertical field \\
$\Delta^*(\cdot)$ &= Grad-Shafranov operator, $=r\frac{\partial }{\partial r} \left( \frac{1}{r} \frac{ \partial (\cdot) }{\partial r } \right) + \frac{\partial^2 (\cdot) }{\partial z^2}$  \\
$\phi$ &= toroidal direction \\
$\psi$ &= poloidal flux per unit radian $(=\Phi/2\pi$) \\
$\mu_0$ &= vacuum permeability constant \\
$J_{\phi} $ &= total toroidal current density\\
$J_{\phi}^{pla} $ &= toroidal current density in plasma\\
$J_{\phi}^{vac} $ &= toroidal current density in vacuum (all non-plasma) conductors\\
$\vec{J}$ &= current density \\ 
$\vec B$ &= magnetic field \\
$P$ &= pressure \\
$R_i$ &= resistance of element $i$ \\
$M_{ij}$ &= mutual inductance between elements $i$ and $j$ \\
$L_i$ &= self-inductance of element $i$ \\
$\Phi^{pla}$ &= total poloidal flux at element $i$ due to plasma current sources\\
$\text{subscript } g$ &= referring to a plasma ``grid'' element  \\
$\text{subscripts } c, v, p$ &= referring to coil, vessel, or plasma current elements \\
$\vec M$ &= mutual inductance matrix \\
$\vec X$ &= plasma modificiation to mutual inductance matrix \\
$\vec R$ &= resistance matrix \\
$k_P, k_I, k_D$ &= PID controller gains \\
$w_z$ &= unstable eigenvector associated with vertical instability \\
$G_{Br,x}$ &= Greens function operator for computing radial field (at a specified location) due to sources $x$ \\ 
$G_{Bz,x}$ &= Greens function operator for computing vertical field (at a specified location) due to sources $x$ \\ 
$G_{\psi, x}$ &= Greens function operator for computing flux (at a specified location) due to sources $x$, note that the Greens operator for flux is equal to mutual inductance.\\ 
\bottomrule 
\label{tab:}
\end{longtable}

\pagebreak
\section{RHP vertical coupling zero}\label{app:rhp_zero}

\Cref{sec:static_plus_vertical} introduced that there is a RHP zero associated with vertical control that is responsible for the inverse response in \cref{fig:zcur}. This appendix describes the RHP zero in more detail and its connection with the shape controller. 

To begin, we make a few observations:

\begin{itemize}
    \item A zero of a MIMO system transfer function $G(s)$,

    \begin{equation}
        y(s) = G(s)u(s),
    \end{equation}

     is a complex number $s=z$ for which $G(s)$ loses rank. At $s=z$, there exists some non-zero input $u$ that lies in the null space of $G$ for which the output $y$ will always be zero and is therefore unobservable. For MIMO systems, a zero is associated with particular input and output directions (the left and right null spaces of $G$). For example, the input-output directions for the vertically-stabilized NSTX-U control system are shown in \cref{fig:rhp_zerodir}. 
    
    \begin{figure}[H]
        \centering
        \includegraphics[width=0.7\linewidth]{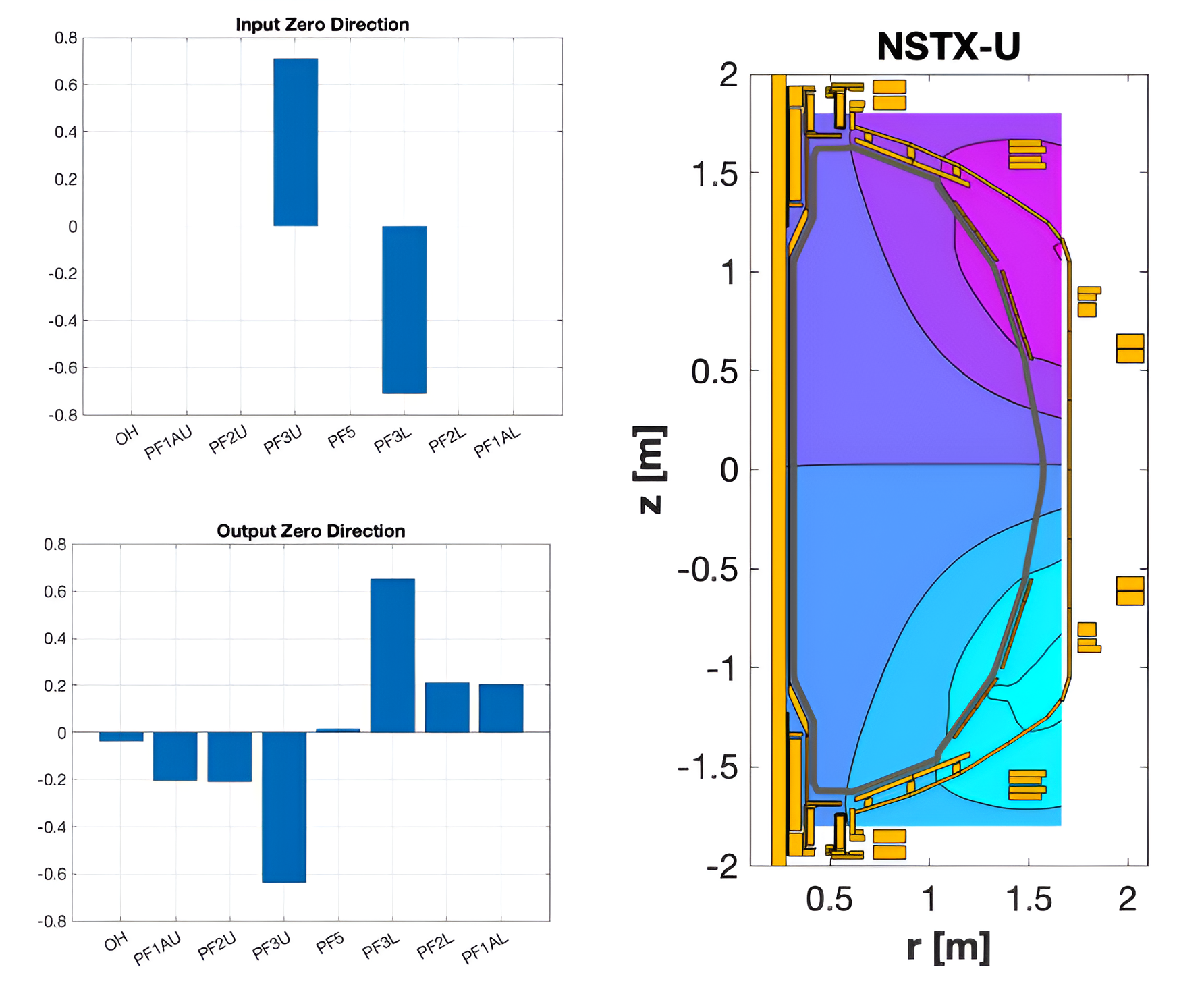}
        \caption{\textbf{Left)} Input and output directions associated with the NSTX-U RHP vertical coupling zero. The input zero is corresponds exactly to the vertical control actuation direction (PF3U-PF3L), while the output direction is associated with multiple coils. The presence of PF1AU and PF2 in the output zero direction is supportive of the conclusion in \cref{sec:nstxu_vert_control} that including these coils improves vertical control. \textbf{Right)} The output zero flux response is vertically antisymmetric.}
        \label{fig:rhp_zerodir}
    \end{figure}

    For more general information on zeros, we refer readers to \cite{Skogestad2005}. 
    
    \item Right-half plane zeros (zeros for which the real part of $z$ is greater than zero) are particularly detrimental to control. This is because they have an inverse response in the time domain \cite{Skogestad2005}, which is destabilizing if the controller is too aggressive. Consider: a controller may try to correct an error, but because of the inverse response, the error increases at first even though it would have decreased eventually. The controller sees the increasing error and responds even more strongly, increasing the error even more, and destabilizing the system. In general, RHP zeros must be accommodated by decreasing controller aggressiveness in the zero direction. 

    \item The RHP zero that is relevant for shape and vertical interactions is the zero of the transfer function from coil voltages to coil currents. This zero is described by Pesamosca \cite{pesamosca2021} (Chapter 3 and App. B.1.2), and note that this is different from a RHP vessel shielding zero originally described in \cite{Humphreys1989} and also \cite{pesamosca2021} (Chapter 3 and App B.1.3). We will frame the discussion slightly differently from Pesamosca and therefore repeat some of the analysis. 
\end{itemize}

In the static IBSC shape controller design of \cref{fig:ibsc_block_diag_current} there is a dedicated vertical control loop, and static mappings of the plasma response linearization to map from shaping errors to target coil currents. The transfer function of interest is the one from shaping actuation (the shape voltages $V_c^{shape}$ to the change in shape control current targets $\delta I_{c,targ}$.  A RHP zero in this transfer function indicates there will be control difficulties, because for some combination of voltages it will excite an inverse response in the targets that the voltages needs to track. This transfer function is shown in \cref{fig:rhp_zero_block_diag1}, and is just a re-arranged version of \cref{fig:ibsc_block_diag_current}. 

\begin{figure}[H]
    \centering
    \includegraphics[height=5cm]{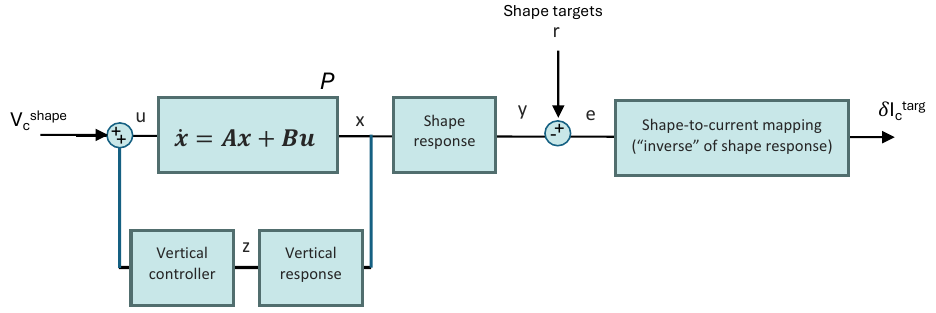}
    \caption{System transfer function from shaping voltages to coil current target changes, re-arranged from the IBSC static control diagram \cref{fig:ibsc_block_diag_current}.}
    \label{fig:rhp_zero_block_diag1}
\end{figure}

\begin{figure}[H]
    \centering
    \includegraphics[height=4cm]{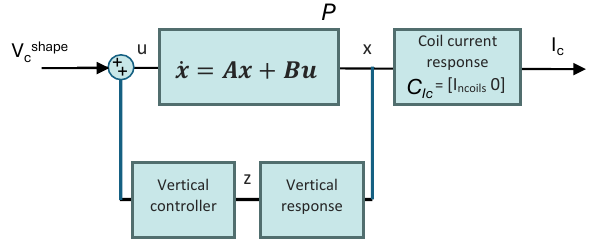}
    \caption{Simplified system transfer function, from shaping voltages to coil currents.}
    \label{fig:rhp_zero_block_diag2}
\end{figure}

We can simplify the analysis by looking at the transfer function in \cref{fig:rhp_zero_block_diag2}. This is valid because the span of the observable space for the shape response plus shape-to-current mapping is equivalent to the span of direct current observation, that is:

\begin{equation}
\begin{aligned}
    \text{span} \left( \begin{bmatrix} I_{ncoils}  & 0 \end{bmatrix} \right) &\geq \text{span} \left( C T^\dagger \right) \\ &= \text{span} \left( C \begin{bmatrix} I_{ncoils} & 0 \end{bmatrix} C^\dagger \right )
\end{aligned}
\end{equation}

The closed-loop transfer function for the system in \cref{fig:rhp_zero_block_diag2} is: 

\begin{equation}
    G = C_{I_c}P \left ( I + K_z C_z P \right )^{-1}
\end{equation}

This indicates that the zeros of G also include the zeros of $C_{I_c}P$, i.e. the zeros of the state-space system $ss(A,B,C_{I_c},0)$. The zeros of any state-space system can be obtained from identifying non-trivial solutions to the Rosenbrock system equations,

\begin{equation}
    \begin{bmatrix}
        zI-A & B \\
        C & D
    \end{bmatrix} 
    \begin{bmatrix}
        x \\ u
    \end{bmatrix} = 0,
\end{equation}

which for this case becomes

\begin{equation}
    \begin{bmatrix}
        zI-A & B \\
        [I_{n_c} \;\;\; 0] & 0
    \end{bmatrix} 
    \begin{bmatrix}
        x \\ u
    \end{bmatrix} = 0.
\end{equation}

In general this would have to be solved numerically. However, there are identifiable nontrivial solutions for the case where the coil resistances are infinite, that is $R_c \rightarrow \infty$. The solution is obtained with $u=0$, $x$ an eigenvector of $A$, and $z$ the associated eigenvalue. 

We have, independently, that:

\begin{equation}
\begin{aligned}
    0 &= Bu, && \text{  because $u=0$}. \\ 
    0 &= (zI-A)x,  && \text{  because $(z,x)$ is an (eigenvalue, eigenvector) pair of $A$}.\\
    0 &= \begin{bmatrix} I_{n_c} & 0 \end{bmatrix} x, && \text{  because with $R_c=\infty$, the unstable eigenvector $x$ is} \\
    & && \text{    zero at all the coil elements.}
\end{aligned}
\end{equation}

For an elongated plasma the dynamics matrix $A$ contains a RHP eigenvalue and eigenvector, and a corresponding RHP zero solution to the Rosenbrock system equations. This shows that the RHP shape control zero is fundamentally related to the control interaction with the unstable pole. In the less general case where $R_c \neq \infty$, the zero value and zero vector shift away from the eigenvalue and eigenvector of the plant, and in \cite{pesamosca2021} Pesamosca showed that for a simplified case that the zero is always slightly faster than the unstable pole. 

We reiterate that the general strategy for accommodating RHP zeros is to make the control less aggressive in the zero direction. For IBSC static shape control, this means either making the shape control completely orthogonal to the vertical control direction, or at least tune the gains so that the control in that direction is meaningfully slower than the RHP zero.

\pagebreak
\section{ALIGN algorithm and discussion}\label{app:align_algorithm}

\subsectionbf{Intuition and discussion}
The frame alignment algorithm \cite{EDMUNDS1979} (ALIGN) is a useful technique for transforming a complex frequency response into a real, linear mapping. Consider the plant transfer function relationship: 

\begin{equation}
    y(s) = P(s) u(s)
\end{equation}

The inverse plant $P^{-1}(s)$ is a function of frequency and in general may not exist or could be difficult to realize. In the frame alignment method, we specify a particular frequency such as the bandwidth frequency $w_{b} = 1/(\text{desired response time})$ and evaluate the plant at this frequency, $P(jw_b)$, which will be a complex matrix. For control purposes it is useful to find a real approximation $P^{-1}(jw_b)$ because these represent constant decoupling directions and are much easier to implement. 

Obtaining a ``real approximation'' to $P^{-1}(jw_b)$ is not a trivial task. A naive approach would be to project this matrix directly onto the real-plane by dropping the imaginary components. This is not an effective method, as demonstrated by the following: suppose one of the complex elements of the matrix is $r\,exp(j\theta)$. The magnitude of the response is determined by the magnitude $r$ while $\theta$ represents the amount of phase delay between input and output. If $\theta=90^\degree$ the element is purely imaginary and the important magnitude information is lost even if $r$ is large and there is strong coupling. This example illustrates that it is important to capture the strength of the coupling which can be hidden due to the input-output phase relationship $\theta$. The ALIGN algorithm is a method for computing a real approximation to a complex matrix without losing information that is encoded in the phase delays. 

Recall that a matrix inverse $A = \{a_i\} = P^{-1}$ is defined by the relationship,

\begin{equation}
    P a_i = e_i
\end{equation}

where $e_i$ is the standard basis unit vector for the $ith$ dimension. With the restriction that $a_i$ is a real vector, this relationship cannot be satisfied in general. The ALIGN algorithm uses the modification:

\begin{equation}
    P a_i = exp(j\theta) e_i + \epsilon_i,
\end{equation}

where $\epsilon_i$ is an error term to be minimized, and $\theta$ is a free parameter that encodes phase information. Allowing the phase $\theta$ to be a free parameter ensures that we are not missing input-output correlations that are obscured just because the phase is unaligned. The solution to the ALIGN algorithm is the set of vectors $a_i$ that perform a least squares minimization of the error term $||\epsilon_i||_2^2$. The point is that, in the least squares sense, the vectors $a_i$ are the best \textit{real} vector decoupling directions associated with the given plant and frequency. 

Once this real mapping is obtained it can be used directly within the IBSC framework to design decoupling PID, SVD, or QP-based controllers on the map. Note that $ALIGN(P(jw))$ gives the real approximation to the pseudoinverse of $P(jw)$. As an example, the weighted pseudoinverse (\cref{eq:weighted_pseudoinverse}, repeated here) is: 

\begin{equation}
    \bar T^\dagger = \begin{bmatrix} W_y \,T \\ W_{I_c} \end{bmatrix}^\dagger \begin{bmatrix} W_y \\ 0_{nc \times ny}\end{bmatrix}
\end{equation}

The analagous weighted inverse using ALIGN would be: 

\begin{equation}\label{eq:weighted_align}
    \bar T^\dagger := ALIGN \left( \begin{bmatrix} W_y \, P(jw) \\ W_{c} \end{bmatrix} \right) \times \begin{bmatrix} W_y \\ 0_{nc \times ny}\end{bmatrix}
\end{equation}

\subsectionbf{ALIGN solution}

The solution to ALIGN was discovered in \cite{EDMUNDS1979} and is repeated here for convenience. For a complex matrix $V$, let $A$ be the real approximation to the inverse of $V$. The solution is: 

\begin{equation}
    \begin{aligned}
        D &= V^* V + (V^*V)^T \\
        \psi &= \angle \left [ \text{diag (} V D^{-1} V^T \text{)} \right ] \\
        A &= D^{-1} \left [ V^T \text{exp}(-j\psi/2) + V^*\text{exp}(j\psi/2) \right ]
    \end{aligned}
\end{equation}

In Matlab, this can be implemented as (equation references are from \cite{EDMUNDS1979}): 

\begin{lstlisting}[language=Matlab]
function A = ALIGN(V)
D = V'*V + (V'*V).';          % Eqn 4.6
Di = pinv(D);  
psi = angle(diag(V*Di*V.'));  % In the text near equation 4.10
A = Di * (V.' * diag(exp(-0.5j*psi)) + V' * diag(exp(0.5j*psi))); % Eqn 4.10
\end{lstlisting}

\newpage
\section{Shape control background material}\label{app:background_material}

\subsectionbf{Outline}
In order to support the instructive nature of this paper, this appendix provides background material on the shape and magnetic control problem for tokamaks. The appendix describes the axisymmetric conducting model and the free-boundary equilibrium evolution (FBEE) governing equations, and illustrates the relation to the state-space control model. We also discuss a number of intuition-building key concepts such as the relevant control timescales, Ip flux swing dynamics, and shape control isoflux approach that are commonly used within the field. 

\subsectionbf{Intro to shape control}

Shape control refers to controlling the geometry and features of the magnetic flux surfaces within the tokamak. The shape of the plasma is determined by the balance of forces between the internal pressure of the plasma and magnetic forces from currents flowing in the tokamak conductors and plasma itself. This force balance relation is described by the Grad-Shafranov equation which will be discussed in more detail later. 

Control is performed by using the poloidal field coils in a tokamak as actuators. The currents in these coils push and pull on the plasma to achieve the desired shape, as illustrated in \cref{fig:geo_nstxu}. The amount of current flowing in the PF coils, as well as current in the vessel conducting structures and plasma, is a dynamic problem with time-evolving behavior. This behavior is described by an axisymmetric circuit conductor model, as shown in \cref{fig:axisymmetric_conductor}. 

Additionally, the presence of a vertical instability in elongated plasmas requires a vertical position controller. This is not a separate problem from shape control, but often requires a dedicated treatment since the vertical stabilization problem has peculiar dynamics compared to the rest of the shaping features. The resulting controller usually needs to run with uniquely fast cycle times and perhaps dedicated hardware. 

Shape control is also tightly coupled to the plasma current control problem, since the plasma current relies on the same actuators to create a toroidal loop voltage and drive current. Also, specific features of the magnetic field -- features like strike point positions, x-point positions, flux expansion, and advanced divertor geometries -- also arise due to the interaction of currents and fields in the plasma and coils. Thus in a narrow sense shape control refers to to the formation of a particular shape for the plasma boundary, but in a broader sense refers to the formation of any 2-D magnetic geometry feature, as well as interactions with plasma current control and vertical control. All of these dynamic interactions are described by the free-boundary equilibrium evolution (FBEE) governing equations. 

\begin{figure}[H] 
    \begin{center}
        \makebox[\textwidth][c]{\includegraphics[width=5cm]{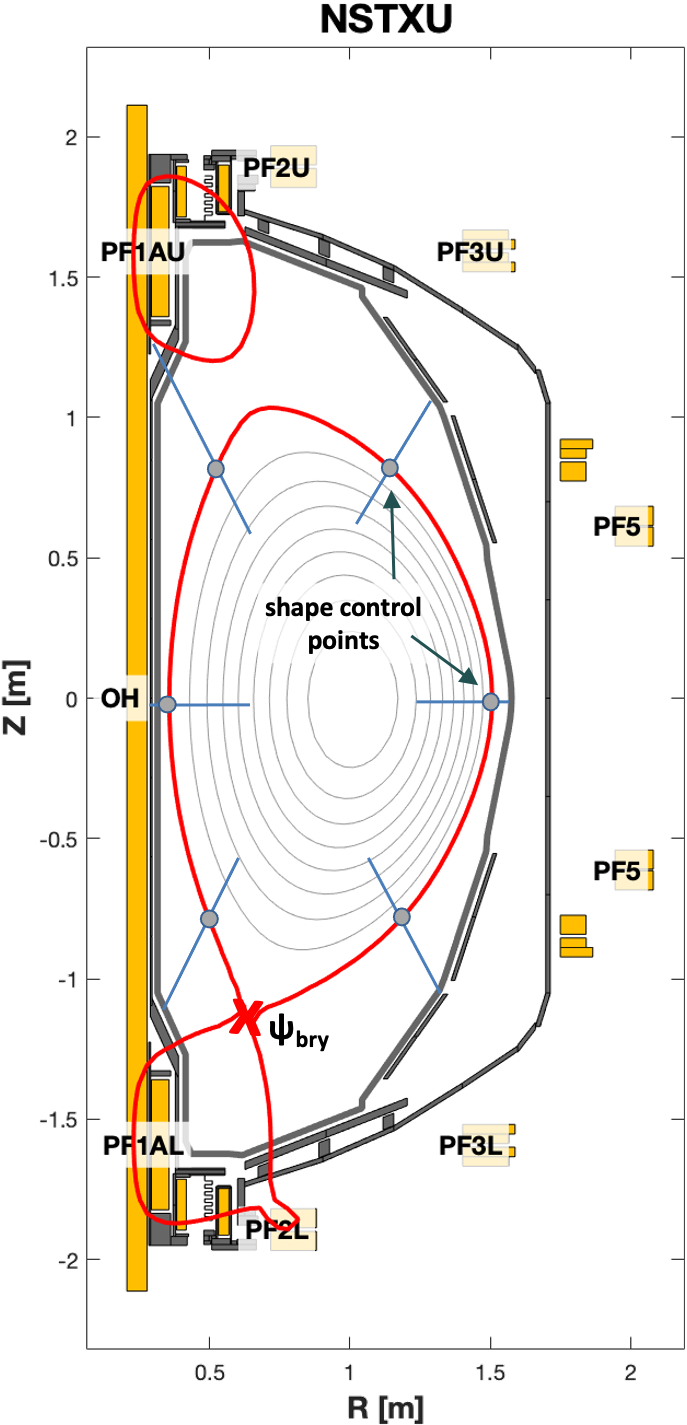}}
    \end{center}
    \caption{Cross-section of NSTX-U. In shape control, the plasma boundary (red) is feedback-controlled to pass through the desired shape control points (blue). The shape and position of the plasma flux surfaces are described by the Grad-Shafranov relation \cref{eq:GS}, one of the FBEE governing equations.}
    \label{fig:geo_nstxu}
\end{figure}

\begin{figure}[H] 
    \begin{center}
        \includegraphics[width=7cm]{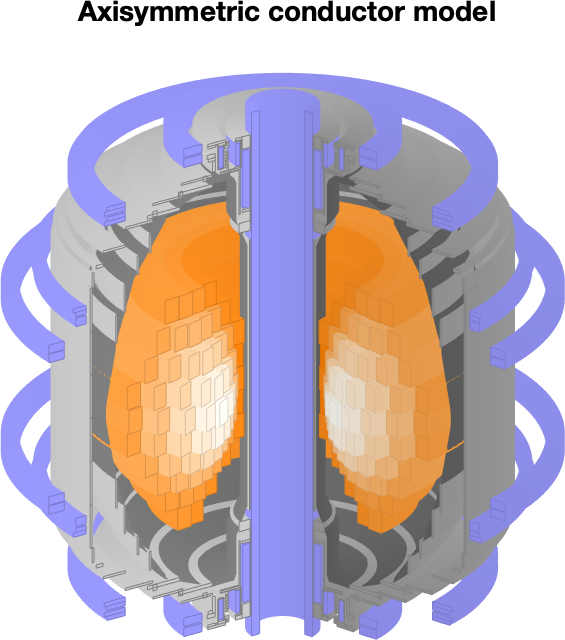}
    \end{center}
    \caption{Axisymmetric conductors of the NSTX-U tokamak. The conductors are modelled as PF coils (blue), discretized vacuum vessel elements (gray), and plasma (orange). The evolution of currents in each of these conductors is described by axisymmetric circuit conductor dynamics \cref{eq:circuit}, one of the FBEE governing equations.}
    \label{fig:axisymmetric_conductor}
\end{figure}

\subsectionbf{Free-boundary equilibrium evolution (FBEE) Governing equations}\label{sec:fbee_governing_eqns}

As discussed, the FBEE governing equations describe both the force balance and flux surface geometry, as well as the dynamic evolution of current flowing in the tokamak conducting structures. These equations consists of two parts, the first of which is the well-known Grad-Shafranov equation: 

\begin{equation}\label{eq:GS}
\begin{split}
    \Delta^* \psi &= -\mu_0 R J_\phi \\
    J_\phi &:= J_{\phi}^{pla} + J_{\phi}^{vac} \\
    J_{\phi}^{pla} &= RP'(\psi) + \frac{FF'(\psi)}{\mu_0 R}
\end{split}
\end{equation}

A table of notation for all equation variables is contained in \cref{app:notation}. The Grad-Shafranov equation describes the balance of pressure and magnetic forces within the tokamak, and is derived from the force balance equation,

\begin{equation}
    \vec J \times \vec B = \nabla P. 
\end{equation}

The Grad-Shafranov equation transforms this vector force balance equation into the above partial differential equation which is written in terms of the flux variable $\psi$. It maps the relationship between the magnetic flux $\psi$, and the current distribution within the plasma that depends on internal profiles of the plasma pressure and toroidal field source function ($F:=RB_T$).

Physically, the value of $\psi$ at any location $(R,Z)$ corresponds to the amount of magnetic flux that passes through a circular surface defined by that point, as illustrated in \cref{fig:gs_flux_diag}. Solving the Grad-Shafranov equation amounts to solving for the 2D density distribution of the flux $\psi$, which defines the magnetic surfaces of the plasma. This is illustrated in \cref{fig:magnetic_surfaces}, where it is seen that the 3D surface traced out by the magnetic field lines corresponds to a 2D contour of constant flux value. 

\begin{figure}[H] 
    \begin{center}
        \makebox[\textwidth][c]{\includegraphics[width=9cm]{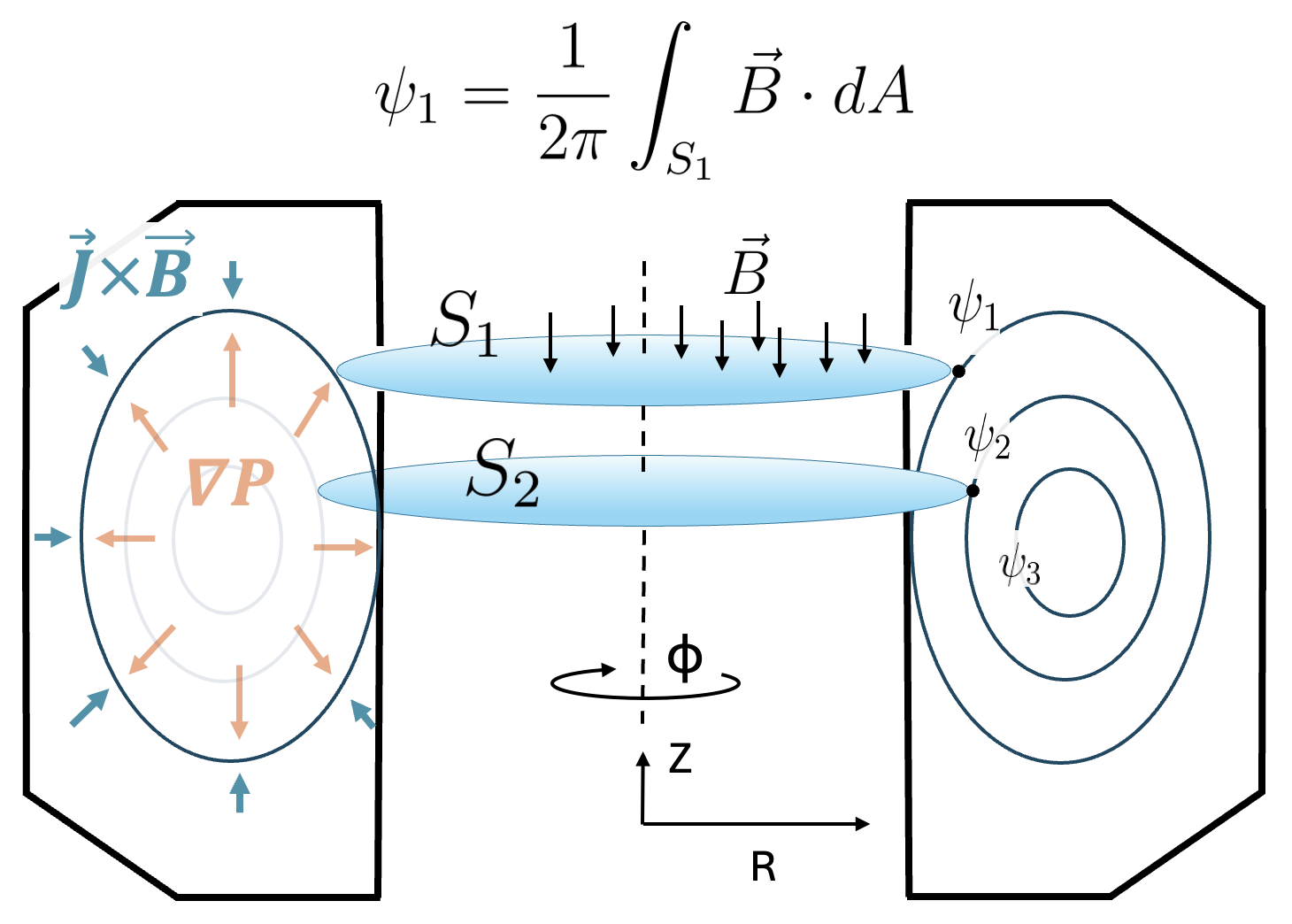}}
    \end{center}
    \caption{Depiction of the Grad-Shafranov equation. (Left side) The pressure within the plasma is balanced by opposing $\vec J \times \vec B$ forces of currents and fields that are present within the tokamak. (Center and right side) The value of flux $\psi$ at any point corresponds to the amount of magnetic field that passes through the circular ``poloidal surface'' $S$ defined by that point location. The distribution of flux contours defines the plasma equilibrium.}
    \label{fig:gs_flux_diag}
\end{figure}

\begin{figure}[H] 
    \begin{center}
        \makebox[\textwidth][c]{\includegraphics[width=5cm]{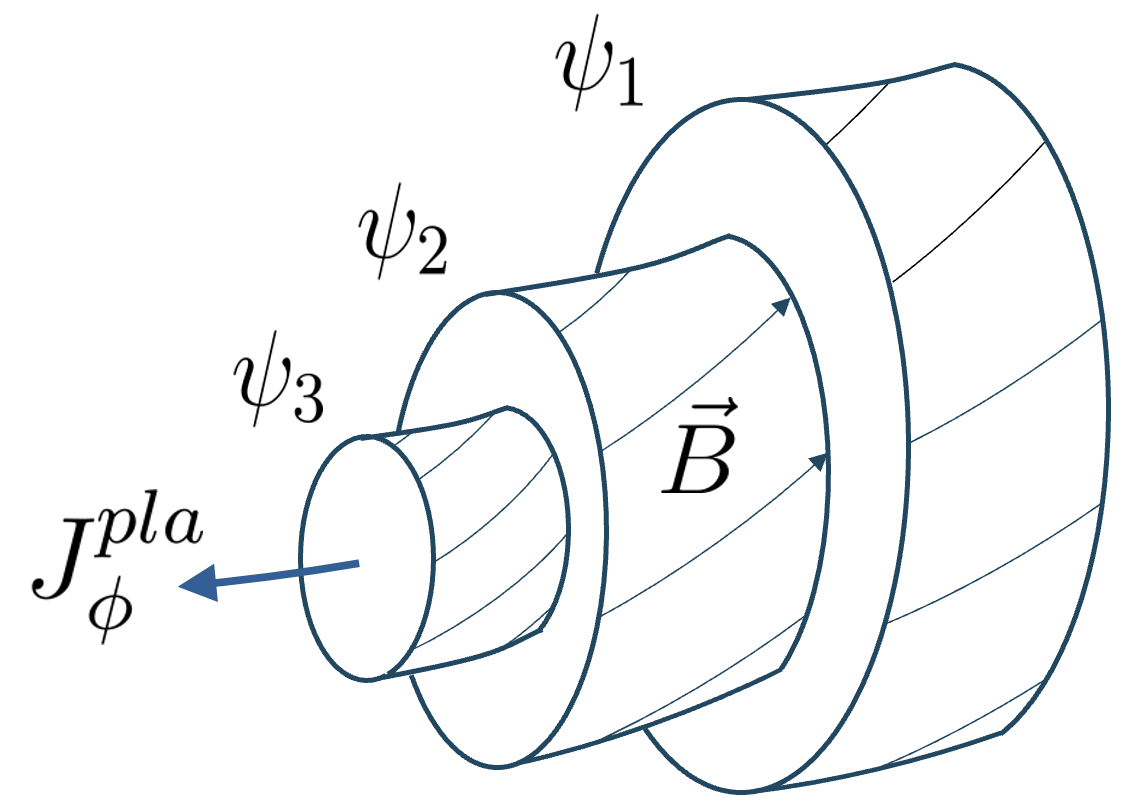}}
    \end{center}
    \caption{The 2D contours of flux $\psi$ form the 3D magnetic surfaces on which the magnetic field lines lie.}
    \label{fig:magnetic_surfaces}
\end{figure}

Any currents flowing within the tokamak's shaping coils or conducting structures enter the force balance through the $J_\phi^{vac}$ term. These currents affect the plasma shape and flux distribution by pushing and pulling on the plasma. The second half of the FBEE governing equations concerns the dynamics of these currents, and is given by the circuit equation:

\begin{equation}\label{eq:circuit}
    V_i = R_i I_i + \sum_j M_{ij} \dot I_j + \dot \Phi^{pla}_i
\end{equation}

This equation is Ampere's law applied to the tokamak conductors, and assumes the form of a conventional circuit with resistive (R) and inductive (L,M) terms, plus a plasma flux contribution. Except for the last term which concerns interaction with the plasma this is a standard description of coupled LR-circuits like that shown in \cref{fig:LR_circuit}. However, instead of only 2 coupled circuits, this is just expanded to account for all of the axisymmetric conductors within the tokamak, which are discretized and modelled as illustrated in \cref{fig:axisymmetric_conductor}. 

\begin{figure}[H] 
    \begin{center}
        \makebox[\textwidth][c]{\includegraphics[width=10cm]{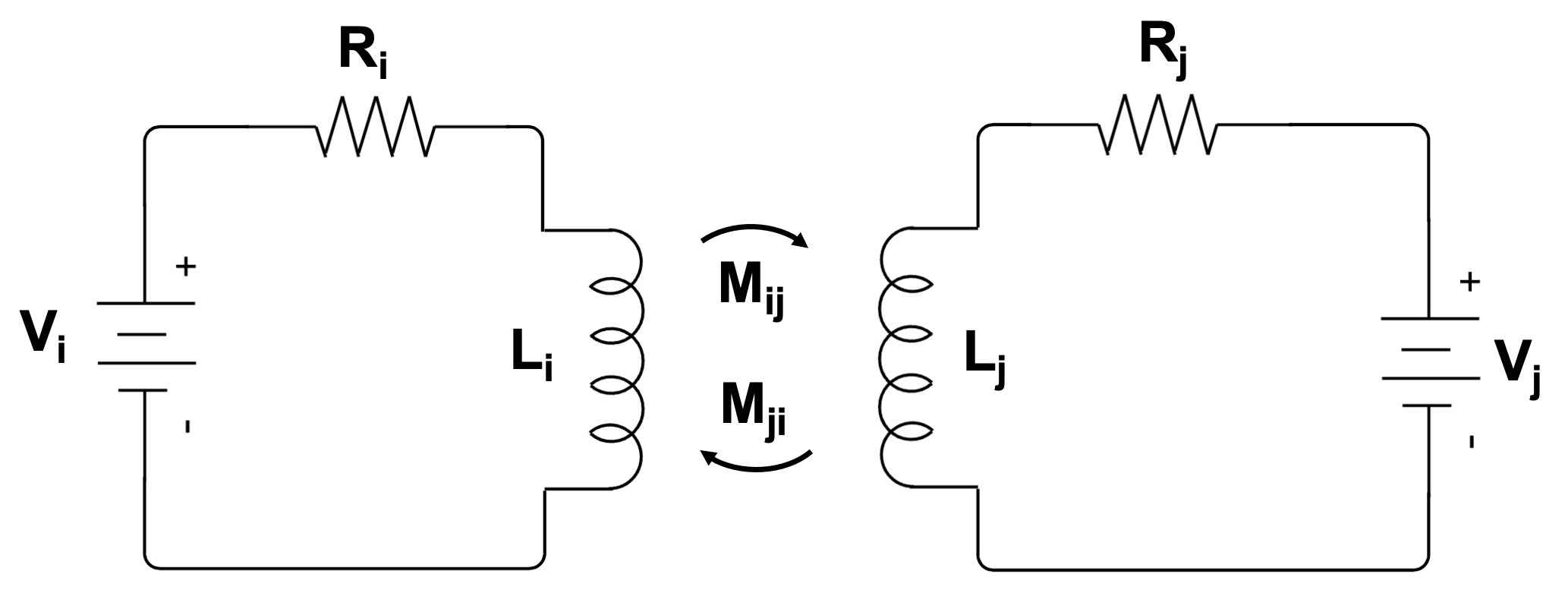}}
    \end{center}
    \caption{Circuit diagram for a pair of inductive-resistive (LR) circuits that have a mutual inductance coupling.}
    \label{fig:LR_circuit}
\end{figure}

The FBEE equations are tightly coupled and interact with each other. As discussed, the currents flowing in the conducting structure push and pull on the plasma equilibrium. Simultaneously, any motion or redistribution of the plasma current induces flux that is felt at the conductors (the $\dot \Phi^{pla}_i$ term) and interacts with the current dynamics. We will treat the plasma flux term in more detail later in \cref{sec:state_space_derivation} when these equations are transformed into a state-space model form that is useful for control simulations. Solving the FBEE equations amounts to finding evolution of currents and equilibria that satisfy both \cref{eq:GS,eq:circuit}, representing how the plasma evolves throughout a pulse. 

\subsubsection{Derivation of state-space form}\label{sec:state_space_derivation}

In the context of feedback shape control the FBEE equations are usually `linearized` around an equilibrium in order to write the equations in state-space form which is commonly used in control problems. To derive the state-space form, we make the substitution:

\begin{equation}
\dot \Phi_i^{pla} = M_{ig} \pdv{I_g}{I_j} \dot I_j
\end{equation}

This is an approximation for the flux induced at conductor $i$ as a result of the redistribution of plasma current on the grid. In this equation, $M_{ig}$ is the mutual inductance coupling between conductor $i$ and the plasma grid locations $g$, and $\pdv{I_g}{I_j}$ is the `linearization' of the equilibrium. That is, it describes how the plasma current distribution $I_g$ shifts in response to a perturbation from conducting element $j$. The perturbed plasma current redistribution must satisfy the Grad-Shafranov equation. One way to obtain the linearization is to use an equilibrium code and perturb the current in each conducting element, and apply a finite difference evaluation with the resulting equilibria. (For example, this is one of the options available in the FGE code.) Alternatively, it is possible to linearize the Grad-Shafranov equation directly, with varying degrees of assumptions involved in the procedure. In general, obtaining this term must be done with a linearization code such as with the tools available in TokSys (gspert, gsupdate, build\_rz\_rig\_model), FGE \cite{Carpanese2021}, or CREATE-NL \cite{Albanese_2015}.  

If we now make the substitution $X_{ij} := M_{ig} \pdv{I_g}{I_j}$, we can write \cref{eq:circuit} in matrix notation to obtain:

\begin{equation}
    \vec V = \vec R \vec I + (\vec M + \vec X) \vec{\dot I}
\end{equation}

Rearranging this equation gives us the state-space dynamics equation:

\begin{equation}\label{eq:dynamics}
 \dot x = \vec A x +  \vec B  u
\end{equation}

\begin{equation*}
\begin{split}
x &:= \begin{bmatrix} I_c^T & I_v^T  & I_p^T \end{bmatrix}^T\\
u &:= V_c \\ 
\vec A &:= -(\vec M + \vec X) ^{-1}  \vec R \\
\vec B &:= (\vec M + \vec X)^{-1} \begin{bmatrix} \vec I_{n_c} \\ \vec 0_{n_v} \\ 0_{n_p} \end{bmatrix}
\end{split}
\end{equation*}

In this formulation we have ordered the elements such that the state vector $x$ corresponds to (c)oil currents, (v)essel currents, and total (p)lasma current. Since the vessel currents and plasma current are not directly connected to external power supplies, the applied voltage for these elements is $0$ and the control actuators just correspond to the applied coil voltages $V_c$.  

To complete the state-space model, we also obtain an output model equation which takes the form:

\begin{equation}\label{eq:output}
\delta y =  \vec C \delta x
\end{equation}

\begin{equation*}
\begin{split}
    \delta y &:=  y-y_{eq} \\
    \delta x &:=  x-x_{eq} \\
    \vec C &:= \pdv{y}{x},
    \end{split}
\end{equation*}

where $y$ corresponds to any shaping-relevant parameter such as flux at the desired shaping control point locations. The output model also depends on the equilibrium linearization through the $\vec C$ matrix which in general is a function of the linearization. Together, \cref{eq:dynamics,eq:output} form the state-space model, and these directly correspond to the underlying FBEE governing equations \cref{eq:circuit} and \cref{eq:GS} respectively. The state-space model is a linearized approximation to the `true' governing equations.

An exact definition for some commonly-controlled $y$ shaping parameters is discussed later in \cref{sec:output_linearization}, where we will also derive precise definitions for the $\vec C$ matrix of parameter responses.

\subsectionbf{Intuition-building and key concepts}\label{sec:intuition}

In this section we discuss a variety of concepts related to equilibrium evolution and shape control, in order to define a few vocabulary terms, highlight some key features of the governing equations, and to build intuition for later sections on controller design. 

\subsubsection{Vertical instability}

A dominant feature predicted by the shape control model is the presence of the vertical instability, which shows up as an unstable (positive real part) eigenvalue of the dynamics matrix $\vec A$. Stabilizing this mode is often the most challenging demand of the shape control system, requiring the fastest control timescales and aggressive hardware performance. For example, on NSTX-U a time delay of $0.1$ms is sufficient to perceptibly impact the vertical stabilization. The growth rate of the vertical instability varies substantially with the plasma equilibrium, but some individual samples are 5-20Hz for ITER, \cite{Kolesnikov2013,Humphreys2009}, 30-120Hz on NSTX-U \cite{Wai2022}, 20Hz on KSTAR \cite{Hahn2018}, and 100-400Hz on DIII-D \cite{Sammuli2021}. The growth rate tends to be slower for larger or thick-walled machines since this creates larger induced currents that dampen the plasma motion. 

It is called the vertical instability because the plasma motion associated with the unstable mode tends to be dominantly up-down antisymmetric. However, depending on the particular equilibrium and tokamak layout the mode may also include radial motion and plasma deformation. Vertical stabilization requires a dedicated treatment, in part because of the aggressive stabilization that may be required, and also because closing the vertical control loop introduces unique dynamics to the rest of the shape control system. Vertical feedback is described in \cref{sec:notes_vert_control}.

\subsubsection{Control timescales}

There are several different relevant timescales associated with shape control. We define the vertical growth rate, vessel shielding, and coil inductive timescales as:

\begin{equation}\label{eq:control_timescales}
\begin{aligned}
    \text{vertical growth rate time}  &= 1 / \; \text{max(real(eig(}\vec A \text{)))} \\
    \text{vessel shielding} &= 1 / \; \text{max(real(eig(} -\vec M_{vv}^{-1} \vec R_v \text{)))} \\
    \text{coil inductive}  &=  (L \Delta I_{typical}) / V_{max} \\
\end{aligned}
\end{equation}
\captionof*{figure}{$^*$all timescales given in seconds, when using MKS units}
\parindent15pt

These calculations are simple to perform before diving into the detailed controller design and give a reasonable view of the possible control performance. We make a few remarks:

\begin{itemize}

\item The faster the growth rate the more difficult control will be. This is often the fastest magnetic control timescale and in general, hardware and software delays need to be 1.5-2 orders of magnitude faster than this timescale in order to not have detrimental effects. 

\item The vessel shielding timescale is obtained from the slowest eigenvalue of the matrix $-\vec M_{vv}^{-1} \vec R_v$. (This is the dynamics matrix for a state-space model of the vacuum vessel elements only, note the similarity with \cref{eq:dynamics}.) Physically, this represents how currents in the vessel conducting structure will decay exponentially over time due to resistive dissipation.

\begin{figure}[H] 
    \begin{center}
        \makebox[\textwidth][c]{\includegraphics[width=14cm]{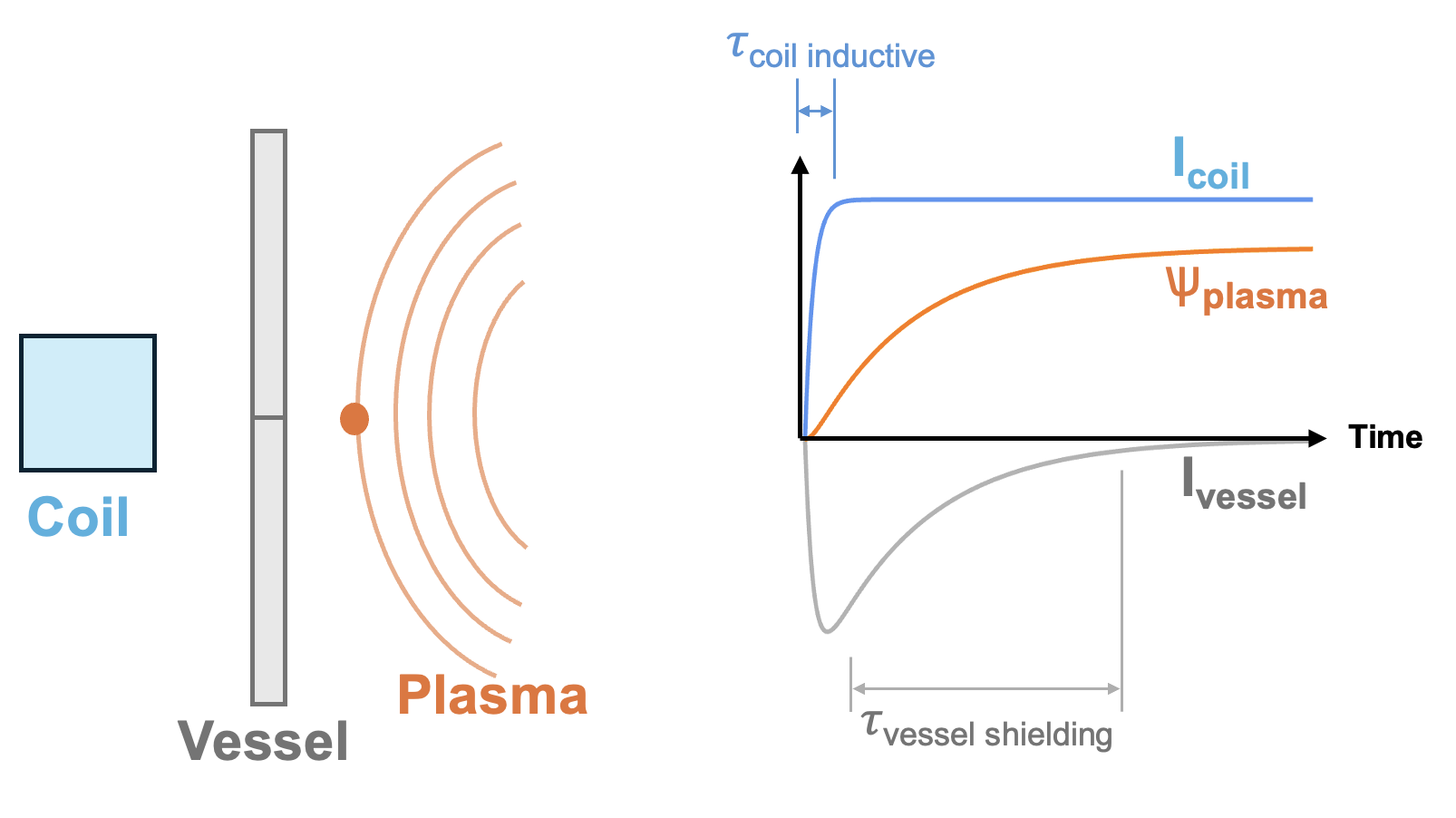}}
    \end{center}
    \caption{Illustration of the vessel shielding effect. The PF coil is used in feedback to affect the flux distribution at the plasma inside the vessel. As the coil is driven to a certain value of current (on the coil inductive timescale), image currents form in the vessel that decay resistively on the vessel shielding timescale. The flux felt at the plasma is a sum of contributions from both coil and vessel and therefore reaches a steady-state value on the vessel-shielding timescale.}
    \label{fig:timescales}
\end{figure}

The importance of this timescale stems from the vessel shielding effect, illustrated in \cref{fig:timescales}. In general, the goal is to use the PF coils to actively manipulate the flux and field inside the vacuum vessel. When a PF coil has a change in current due to inductive coupling, the vacuum vessel will form image currents in response that serve to cancel the flux and field changes from the PF coil. As these image currents decay from resistivity they stop blocking flux change from the coil, allowing flux to penetrate to inside the vessel. The significant point is that the control coils cannot effect changes to the plasma faster than the vessel shielding time, because the effect of coils will be screened by the vessel image currents.

There can be local exceptions to this rule. For example, if the vacuum vessel has an uneven distribution of inductance and resistivity, there may be specific places where flux can penetrate faster. In this case it may be necessary to perform additional analysis to determine localized shielding times. Another exception is when coils (such as the vertical control coil) are placed inside the vacuum vessel specifically for the purpose of avoiding vessel shielding.

\item The coil inductive timescales is derived from the single-variable vacuum circuit equation:

\begin{equation}
    \dot I = -\frac{R}{L} I + \frac{1}{L}V 
\end{equation}

The first term in this equation represents the resistive decay dynamics, that is, how the current will resistively decay in the absence of actively applied voltage. The second term represents the actively applied voltage dynamics. 

Generally speaking, the first resistive term is less important because it tends to be much slower and less dominant than the active-control coil inductive timescale. Although note that a high coil resistivity can influence the dynamics and may motivate the use of integrators in a feedback controller. 

The coil inductive timescale is usually much more important. The above definition represents how fast a coil could effect a change in current $\Delta I_{typical}$ (for example, we could take $\Delta I_{typical}$ as 20\% of the coil's range) when applying the maximum power supply voltage. This timescale also introduces a performance limitation, and for any particular coil inductance and voltage combination may be even more limiting than the vessel shielding timescale. 

\end{itemize}

\subsubsection{Ip dynamics and flux swing}

The circuit equation written for just the total plasma current is

\begin{equation}\label{eq:Ip_circuit}
    0 = R_p I_p + \begin{bmatrix} \vec M_{pc} & \vec M_{pv} & L_p \end{bmatrix} \begin{bmatrix} \dot I_c \\ \dot I_v \\ \dot I_p \end{bmatrix}, 
\end{equation}

or equivalently, 

\begin{equation}
    \dot I_p = -L_p^{-1} R_p I_p + L_p^{-1} \vec M_{ps} \dot I_s
\end{equation}

(This equation neglects non-inductive current drive effects such as bootstrap current). The first term represents resistive decay of the plasma current. However, the second term represents a transformer effect for driving current in the plasma. As current ramps in the other conductors, if they have strong mutual inductance coupling to the plasma, it will drive plasma current as well. This is just like the inductively coupled circuit of \cref{fig:LR_circuit} where the plasma forms one of the circuits (but the plasma does not have a voltage source of its own). Over the course of the pulse, the coils must continually ramp their current in order to counteract the resistive decay of plasma current. This means that equilibrium shaping capabilities change throughout the course of pulse since the proximity to hardware limits and balance of currents is always changing. 

\subsubsection{Shaping parametrization}

Since there are only a limited number of shaping coils the shape of the plasma can usually be well described by just a few shaping parameters. The plasma shape is often parametrized in terms of perturbations based on poloidal mode number, the first few of which are the plasma size and position, elongation, triangularity, and squareness, as shown in \cref{fig:shape_perturbations}. Achievable shapes can usually be specified almost exactly by the values of these parameters, plus a few magnetic geometry features such as the x-point and strike point locations. For exact definitions of these parameters we refer readers to the analytical forms given in \cite{Luce2013}.

\begin{figure}[H] 
    \begin{center}
        \includegraphics[width=12cm]{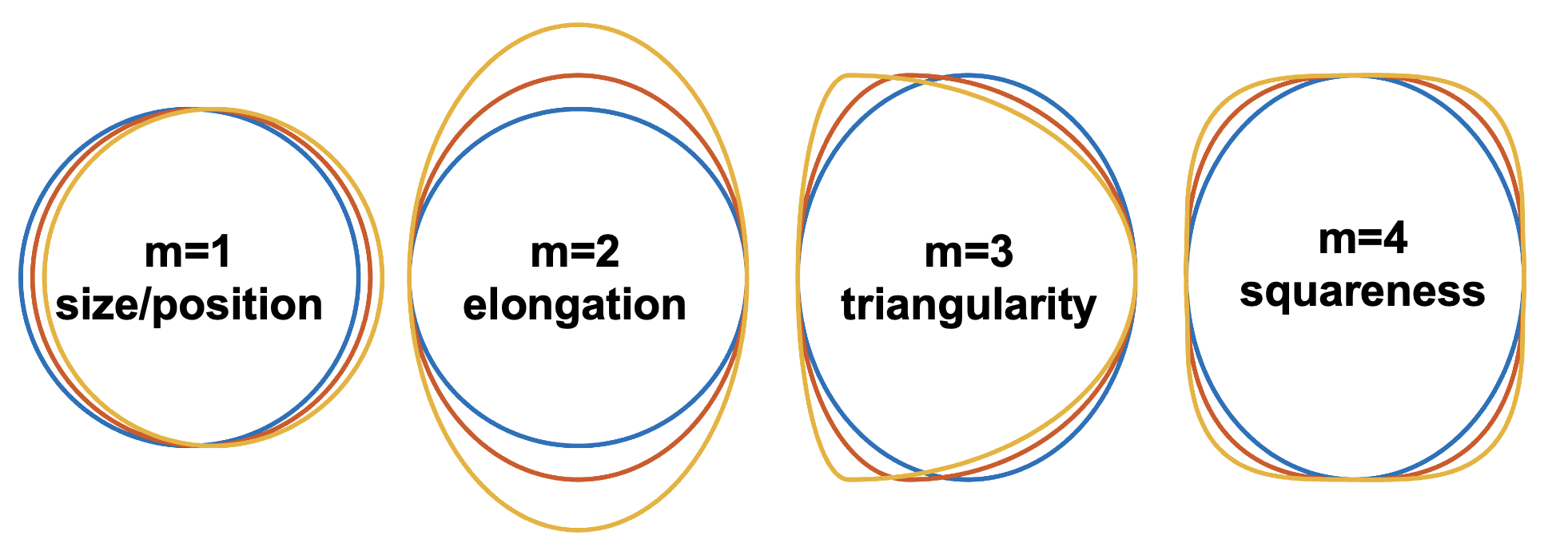}
    \end{center}
    \caption{Illustration of the position, elongation, triangularity, and squareness shaping parameters, which are associated with the $m=1-4$ poloidal mode perturbations.}
    \label{fig:shape_perturbations}
\end{figure}

\subsubsection{Shaping intuition via current-carrying wires}

What is the purpose of each PF coil? Certainly the exact effect of each is best solved computationally by an equilibrium solver or linearization code. However it is useful to gain some intuition via the analogy of current-carrying wires. Recall that if two wires carry current in the same direction there is an attractive force between them, and likewise if in opposite directions there is a repulsive force. Many tokamaks have coil layouts that are similar to that shown in the cartoon \cref{fig:cocurrent_wires}. 

\begin{figure}[H] 
    \begin{center}
        \includegraphics[width=14cm]{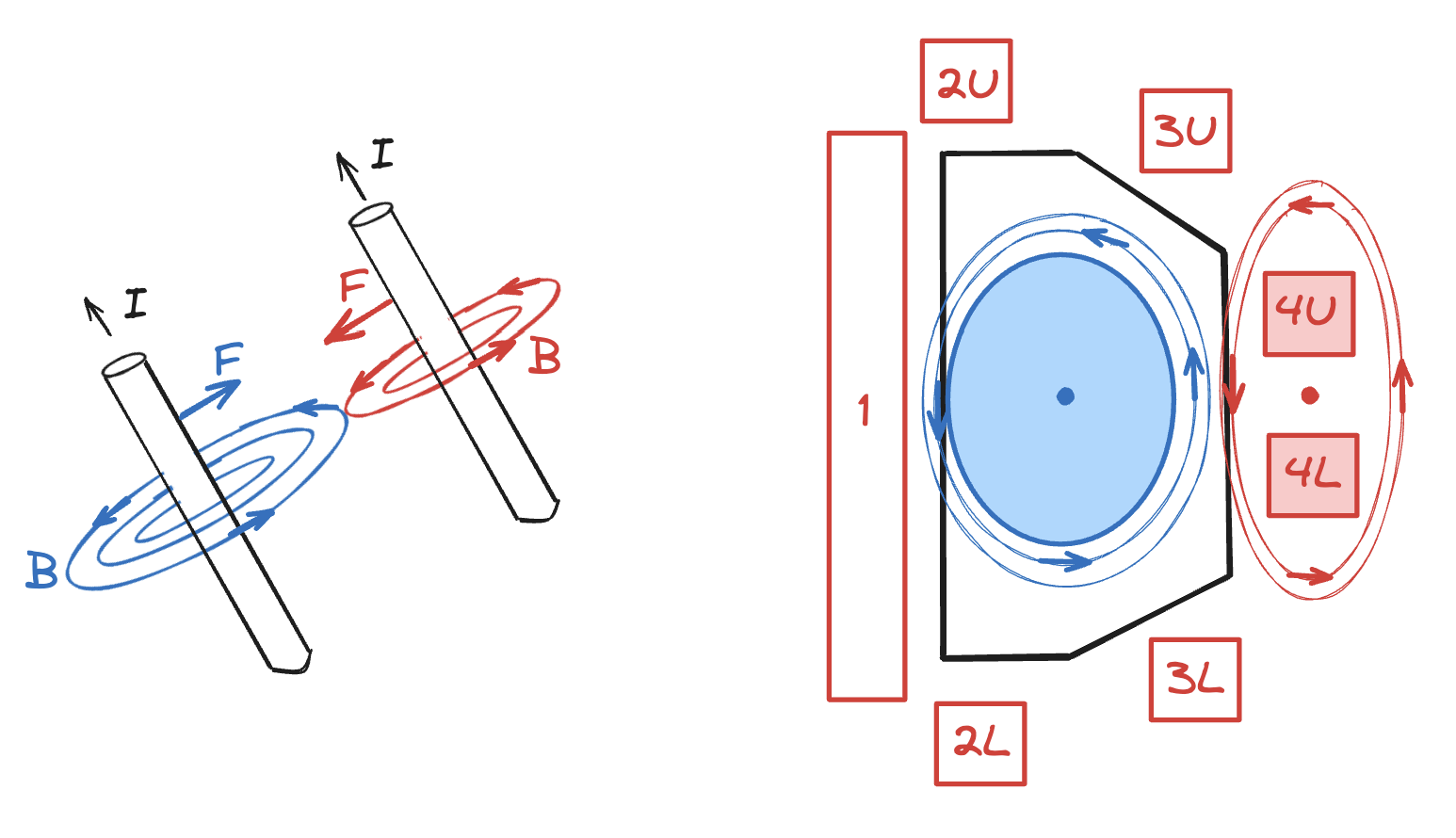}
    \end{center}
    \caption{Wires carrying current in the same direction form magnetic fields that cause an attractive force between them. This analogy provides reasonable intuition for understanding the roles of specific tokamak shaping coils, in which one of the wires is the plasma itself.}
    \label{fig:cocurrent_wires}
\end{figure}

With this analogy in mind we can think of each coil pushing and pulling on the plasma current ``wire'' and make the following generalizations: 

\begin{itemize}
    \item Coil 1 is large and stores a lot of flux. It does influence the radial position of the plasma, but usually the main purpose is to drive plasma current through the flux swing. 
    \item Coils 2U/2L are located above and below the plasma, and influence the vertical position of the plasma and its elongation, and are also often used to form x-points. 
    \item Coils 3U/3L are located above and below the plasma. They are in a good location to influence the vertical position and plasma elongation. 
    \item Coils 4U/4L are located radially outboard of the plasma, and are mainly used to adjust the radial position. 
\end{itemize}

\subsubsection{Linearization models}

Methods for linearizing the equilibrium can be classified as ``rigid plasma response models'' and ``nonrigid plasma response models''. The rigid plasma response model was introduced in \cite{Walker2006} (see also section 2.3.1 of \cite{Wai2023} for a slightly shorter derivation). In this model, the plasma current is assumed free to move vertically and radially, but the current density distribution does not change. This model is relatively easy to calculate and is even fast and reliable enough for real-time calculations \cite{Bao2020}. The rigid plasma response model is generally considered accurate enough for shape control development. One strong caveat however is that it tends to underpredict the vertical growth rate by a factor of 1.5-2. Nonrigid plasma response models are more accurate in describing perturbations to Grad-Shafranov equilibria. These models account for how the plasma current density distribution changes in response to perturbations. See \cite{Welander2005} for an example nonrigid modelling approach. 

\subsubsection{Gap vs isoflux control}\label{sec:gap_vs_isoflux}

There are two main techniques for controlling the plasma to obtain a particular shape, namely isoflux control and gap control. These are shown in  \cref{fig:isoflux_vs_gap2}. We recommend using isoflux methods instead of gap methods because of the additional nonlinearity added in the gap control method. Of course many times it is advantageous to write objective or constraints in spatial units. In this case, it is possible to keep the best of both worlds by adding pre- or post-processing steps such that outputs are human readable, but the lowest level feedback control can still operate in flux units to take advantage of linearity.

\begin{figure}[H] 
    \begin{center}
        \includegraphics[width=0.7\linewidth]{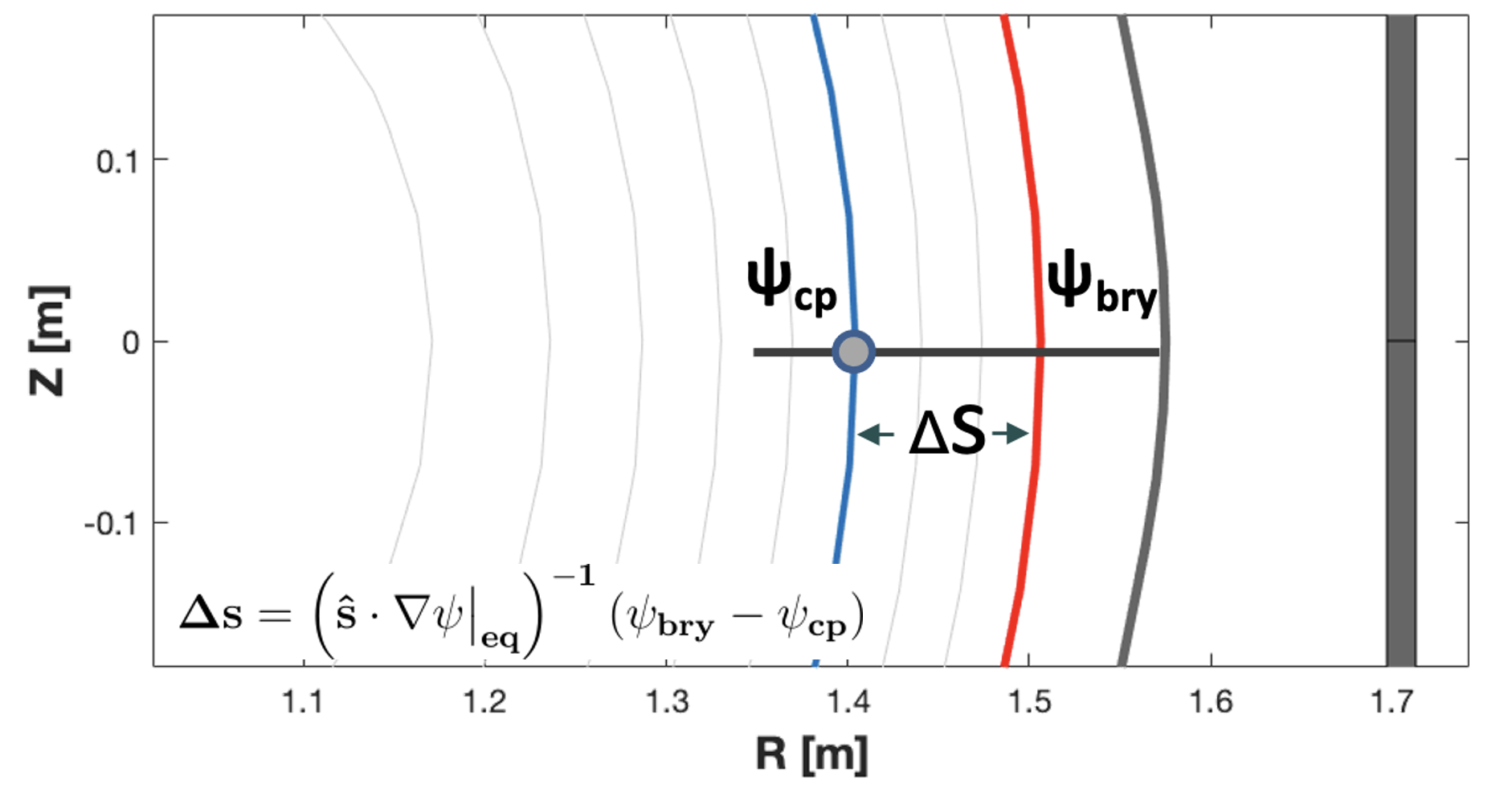}
    \end{center}
    \caption{Comparison of isoflux and gap control methods. Both methods attempt to control the plasma boundary (red) to pass through the desired control point (grey). Isoflux performs this by driving the flux error to zero, while gap control drives the spatial error to zero.}
    \label{fig:isoflux_vs_gap2}
\end{figure}

\begin{figure}[H] 
    \begin{center}
        \includegraphics[width=10cm]{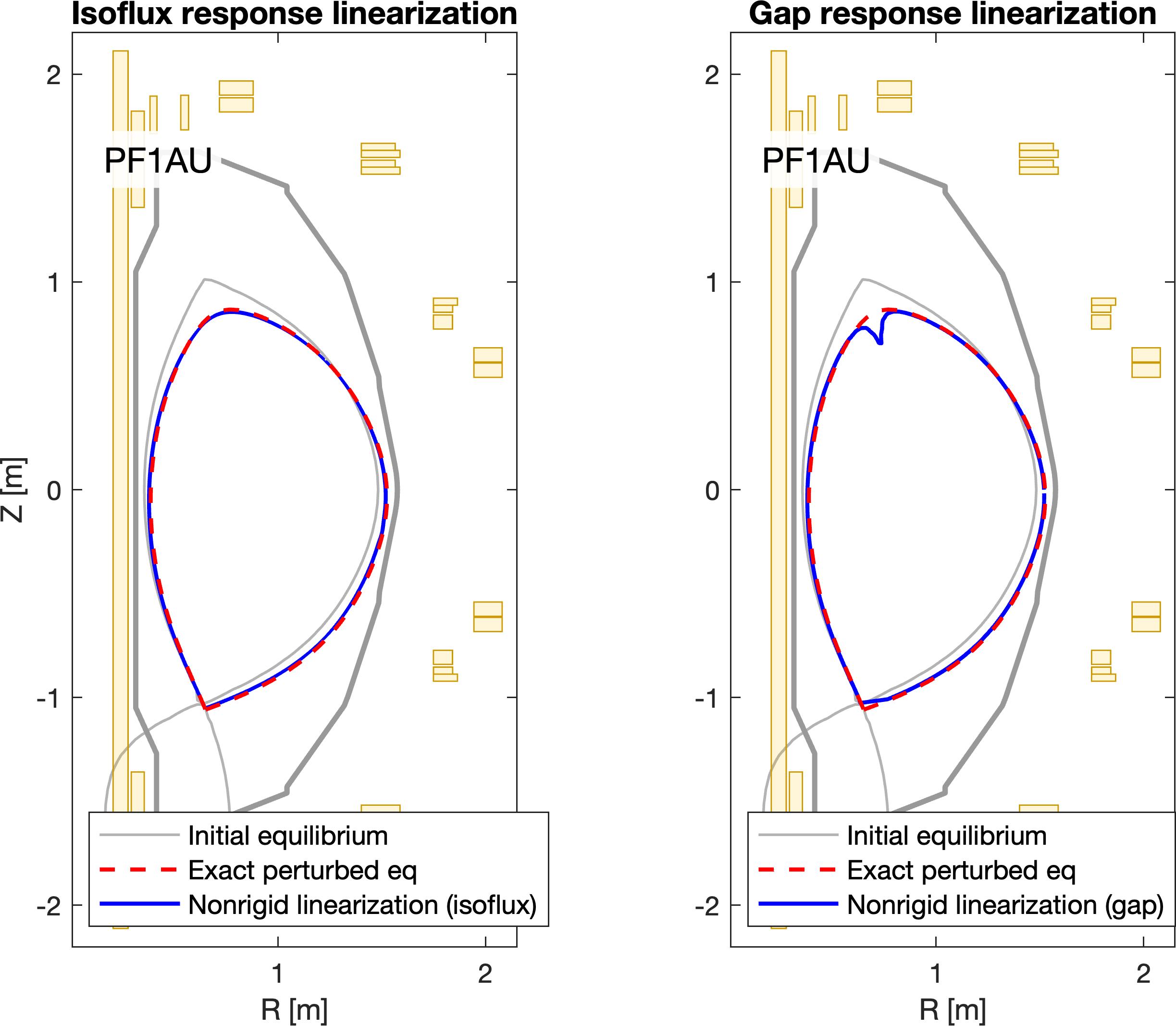}
    \end{center}
    \caption{Response of an NSTX-U plasma equilibrium to perturbations of the PF1AU current. The original equilibrium is shown in gray, and the prediction of the nonrigid plasma response linearization (blue, calculated via the GSPert code) is a good match to the exact perturbed equilibrium solution (dashed red, computed with the free-boundary equilibrium code GSDesign). In the left panel, the linearization is specified in terms of flux. In the right panel, we use the exact same model but write the linearization in terms of gap responses. This adds an additional nonlinearity that can distort the prediction especially near the x-points.}
    \label{fig:gap_linearization}
\end{figure}

In both cases, we would like the plasma boundary to pass through a particular point, designated the control point. Multiple control points are usually defined, and if the plasma passes through all of them it has achieved the desired shape. The difference between these two methods is that gap control uses physical-space units to define the errors whereas isoflux control uses flux-space units. Gap control measures the distance along the control segment to where the plasma boundary intersects it, and compares this to the distance along the segment to the control point. The controller tries to achieve that this gap error is zero:

\begin{equation}
    \Delta s := s_{bry} - s_{cp} \rightarrow 0
\end{equation}

In contrast, the isoflux control method measures the flux at the boundary and compares this to the flux at the control point and attempts to drive this error to zero. 

\begin{equation}
    \Delta \psi := \psi_{bry} - \psi_{cp} \rightarrow 0
\end{equation}

These two methods are related through the local gradient of the equilibrium, since

\begin{equation}\label{eq:gap_vs_isoflux}
    \Delta \psi = \left ( \hat s \cdot \nabla \psi \big|_{eq} \right ) \Delta s 
\end{equation}

where $\hat s$ is the unit vector for the control segment and $\nabla \psi$ is the gradient of the flux in the equilibrium. The advantage of the gap method is that the units are physically intuitive. Also, limits such as minimum wall gaps are usually expressed in terms of physical-space units. However, while less intuitive, the isoflux method has the advantage that flux is linearly related to current. ($\psi = \vec M I$). The gap response introduces an additional nonlinearity through \cref{eq:gap_vs_isoflux}. This means that the flux response has a larger region of validity compared to the gap response. One needs to be especially careful when the flux gradient is small, such as near an x-point, since then the gap response is sensitive as $\left ( \hat s \cdot \nabla \psi\right )^{-1}$ is very large. In general flux is a better indicator of how close equilibria are, that is, a trajectory that is smooth in flux evolution is also smooth in current and actuator evolution, although not necessarily smooth in gap evolution. For this reason we recommend using isoflux methods for the controller, at the cost of mapping results back-and-forth to physical space units for interpretability. 

\subsubsection{X-point position vs field control}

An x-point is a saddle of the flux distribution, satisfying the condition $B_p = \nabla \psi = 0$. When controlling the x-point we are faced with a similar design choice as above, namely whether to use a spatial-based approach or a field-based approach. In the spatial approach, the controller applies feedback based on the spatial error between a target x-point position and the actual x-point position. In the field-based approach, the controller feeds back directly on the field value at the target x-point position, since driving the field to zero is equivalent to creating an x-point at that location. 

For x-point control, we recommend that the field-based approach is superior to the spatial approach and recommend applying feedback to the value of the magnetic field (or flux gradient) at the target x-point position. The reasoning is two-fold: 

\begin{itemize}
    \item Just as is in isoflux vs gap control, the response of the magnetic field is linear with respect to the shaping coil currents and other axisymmetric conductors. However, the x-point position response is not linear since it additionally depends on local gradients of the particular base equilibrium. In other words, the x-point position response uses an additional nonlinear term not present in the field response, thus introducing more numerical sensitivity. 

    \item The field value and response are well-defined throughout the whole pulse, even before an x-point has formed. By contrast, it is not clear what it would mean to compute derivatives of the x-point position if the x-point does not exist. When using spatial methods, one has to rely on tricks such as using a known trajectory to drive the plasma to an equilibrium where an x-point exists and then transferring control over to the x-point position controller. 
\end{itemize}

\newpage
\section{Notes on design procedure steps}\label{app:design_procedure_notes}

In \cref{sec:design_procedure} and \cref{fig:control_design_procedureA}, we proposed a design procedure for IBSC. In this section, we expand on this proposed procedure and provide some additional notes on accomplishing some of these steps. This appendix is intended to be tutorial in nature. 

\subsectionbf{Notes on: feedforward design with GSPulse}

In general, feedforward actuation has beneficial effects on control performance provided that the feedforward inputs are sufficiently accurate. For example, feedforward can reduce phase delays, as well as help steer the pulse closer to a desired global trajectory that has been validated a-priori. For shape control, feedforward trajectories can take the form of currents or voltage. Voltage feedforward is a ``true'' feedforward in the sense that the power supply voltages are applied directly to the system without respect to any measurements of the system. With coil current feedforward, the coil currents are controlled in feedback along the desired current trajectory, but without respect to any shaping parameter measurements. 

GSPulse is a new tool \cite{Wai2025} designed specifically to address feedforward equilibrium trajectory design in tokamaks, and is available open source at \blue{github.com/jwai-cfs/GSPulse\_public}.

\subsectionbf{Notes on: linearize equilibria and build state-space models}

There are many tools available to linearize equilibria and build the state-space models, such as TokSys RZ-rigid \cite{Walker2006}, GS-pert \cite{Welander2005}, FGE \cite{Carpanese2021}, and CREATE-NL \cite{Albanese_2015}. Tools like these will generally provide an equilibrium and the linearized state space dynamics model:

\begin{equation}\label{eq:ss2_dynamics}
     \dot x = A x +  B u 
\end{equation}

The state-space output model is:

\begin{equation}\label{eq:ss2_output}
    \delta y = C \delta x 
\end{equation}

However, since the matrix $C := \partial y / \partial x$ depends on exactly which shaping parameters are controlled, it may or may not be directly calculated by the linearization code. If not provided directly,  \cref{sec:output_linearization} describes how to compute it for some common isoflux shaping parameters, assuming that the linearization code provides the full-grid flux response and full-grid current response.

It is important to note that the dynamics and output (\cref{eq:ss2_dynamics,eq:ss2_output}) are written in different reference frames. The dynamics is written in an absolute reference frame (relative to zero) while the output model is in a reference frame relative to some equilibrium (relative to $x_0, y_0$). Detailed analysis of this topic was given in \cite{Walker2006}. The main implication is that the user must be careful to track and account for which variables are absolute and which are perturbed. For example, employing the state-space model in a Simulink-style environment should use a block diagram as depicted in \cref{fig:plant_ref_frame}a.

\begin{figure}[H] 
    \begin{center}
        \includegraphics[width=10cm]{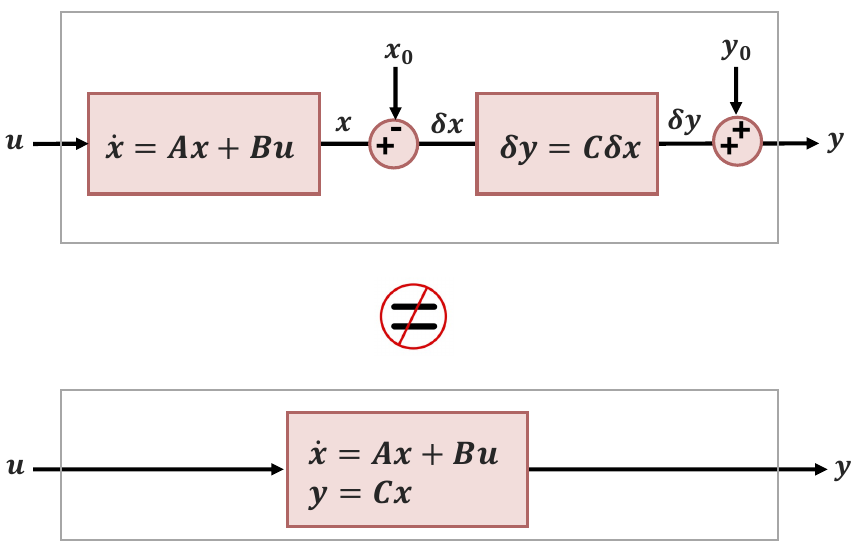}
    \end{center}
    \caption{Block diagrams of state-space models. \textbf{a)} Exact implementation of the shape control state-space model that correctly captures the different reference frames of \cref{eq:ss2_dynamics,eq:ss2_output}. \textbf{b)} A typical state-space model.}
    \label{fig:plant_ref_frame}
\end{figure}

\subsubsection{Derivation of output model for several isoflux shaping parameters}\label{sec:output_linearization}

An equilibrium linearization code may not directly provide the Jacobian of shaping parameter responses. Let's consider how to construct the responses for some typical parameters that are controlled in an isoflux approach (see also \cref{sec:gap_vs_isoflux} for how to map from isoflux to gap responses). Some typical controlled outputs in this case are: 

\begin{equation}\label{eq:y_def}
\renewcommand*{\arraystretch}{1.2}
    y = \begin{bmatrix*}[c]
        \psi_{cp}  - \psi_{xp} \\
        \psi_{cp} - \psi_{tch} \\[1.5ex]
        \dfrac{\partial \psi_{xp}}{\partial r} \\[2ex]
        \dfrac{\partial \psi_{xp}}{\partial z} \\[2ex]
        I_c \\ 
        I_p \\        
        r_{cur} \\
        z_{cur} \\                
    \end{bmatrix*}, \;\;\;\;\;\;\;\;\;\;\;\;
    \left \{
    \begin{aligned}
        \psi_{cp} &:= \text{flux at the shaping control points} \\
        \psi_{xp} &:= \text{flux at a target x-point} \\
        \psi_{tch} &:= \text{flux at a target touch-point} \\                
        I_c &:= \text{coil currents} \\
        I_p &:= \text{total plasma current} \\
        r_{cur} &:= \text{current centroid radial position} \\
        z_{cur} &:= \text{current centroid vertical position}                
    \end{aligned}
    \right .
\end{equation}

In other words, we are interested in controlling the difference in flux between some shaping control points and the x-point (if controlling a diverted plasma), or the difference versus flux at the touch point (if controlling a limited plasma). We may also be controlling the flux gradient at a target x-point location, in order to create an x-point there. Additionally, we may also directly control coil currents, plasma current, and plasma centroid position. 

The goal is to compute the output linearizations to assemble the $\vec C$ matrix. In general a linearization code would provide outputs such as:

\begin{equation}
    \pdv{\psi_g}{x}, \; \text{and} \;\; \pdv{I_g}{x}
\end{equation}

These are the total flux response and the grid in response to a coil current perturbation, and the current response on the grid. For the flux responses, we can compute these via interpolating the 2D-grid of flux response. For example,

\begin{equation}
    \pdv{(\psi_{cp}  - \psi_{xp})}{x} = \mathrm{interp2D} \left( \pdv{\psi_g}{x}, r_{cp}, z_{cp} \right) - \mathrm{interp2D} \left( \pdv{\psi_g}{x}, r_{xp}, z_{xp} \right)
\end{equation}

Similarly, to obtain the flux gradient response at an x-point,

\begin{equation}
    \pdv{}{x} \pdv{\psi_{xp}}{r} = \pdv{}{r} \left( \mathrm{interp2D} \left( \pdv{\psi_g}{x}, r_{xp}, z_{xp} \right) \right). 
\end{equation}

Because $x$ contains the coil currents and plasma current as part of its entries, the response of these items amounts to selecting the correct rows from an identity matrix. For example,

\begin{equation}
    \pdv{I_c}{x} = \begin{bmatrix} \vec I_{n_c} & \vec 0_{n_v} & 0_{n_p} \end{bmatrix}. 
\end{equation}

Lastly, the response of the current centroid position, if not provided directly, can be obtained from the total current response. For example, the centroid vertical response is: 

\begin{equation}
    \pdv{z_{cur}}{x} = \frac{1}{I_p} \int_{\Omega_p} z_g \pdv{I_g}{x}  dV 
\end{equation}

In general, modelling the response of specific parameters requires individualized care, and part of the challenge is deriving appropriate models for the output linearization.

\subsectionbf{Notes on: vertical controller design}\label{sec:notes_vert_control}
The goal of vertical control design is to stabilize the plasma vertical instability with the most margin against disturbances and variation in the equilibrium. The most widely used and successful technique is to employ a proportional-derivative (PD) controller that feeds back on the plasma vertical position, or proxy for the vertical position. As shown in \cite{Humphreys1989} there are cases where proportional control is sufficient, but in general PD control for the range of equilibria created by most tokamaks. 

Since linearization models can generally be trusted with some certainty, a reasonable approach is to perform direct numerical optimization against a variety of equilibria and scenarios. The control law is:

\begin{equation}  
    K_z = \hat u_z (k_p + k_d s) 
\end{equation}

The free parameters for this controller are: the control vector direction (virtual circuit combination) $\hat u_z$ (control vector), and the controller PD gains. Additionally the choice of feedback variable is also a free parameter. This could be, for example, the plasma vertical position, or specific combinations of sensors.  

In many cases, the choice of the virtual circuit for vertical control is obvious. For example, a tokamak may have purpose-built in-vessel coils to provide the fast field response needed. In other cases, the virtual circuit combination actually has some flexibility with important implications for vertical control. As a general rule, coils that can supply dominantly radial field (since radial field produces a vertical force) that penetrates into the vessel on a fast timescale are advantageous for vertical control. 

Lazarus \cite{Lazarus1990} discusses this issue in detail, where they optimize the virtual circuit combinations with differing control gains on DIII-D. Similarly, in the \cref{sec:nstxu_vert_control} we optimize the NSTX-U vertical controller and show that a different control design than was used in the previous campaign predicts moderate improvements. \censor{By contrast, SPARC has a thick metal wall and a pair of in-vessel coils that do not allow much flexibility in the vertical controller design. }{}

One important thing to note for numerical simulation is that it is critical to include a time delay (measurement delay + actuation delay). Of course, higher detailed models are desirable if available, but we recommend that at a minimum the power supply delay must be included when testing for stability. In the first author's experience, not including a delay will often to lead to unphysical assumptions about stability due to sensitivity of the numerics. 

\subsubsection{Common metrics for vertical instability}

There are several important metrics for evaluating the performance of the vertical controller. The most obvious is stability (all poles of the closed loop system are negative). Additionally, the phase margin and $\Delta z_{max}$ metrics must meet criteria in order to have reliable performance for realistic implementations. A summary is given in  \cref{tab:vs_metrics}.  

\begin{table}[H]
    \caption{Vertical stability performance metrics. }
    \label{tab:vs_metrics}
    \includegraphics[width=7cm]{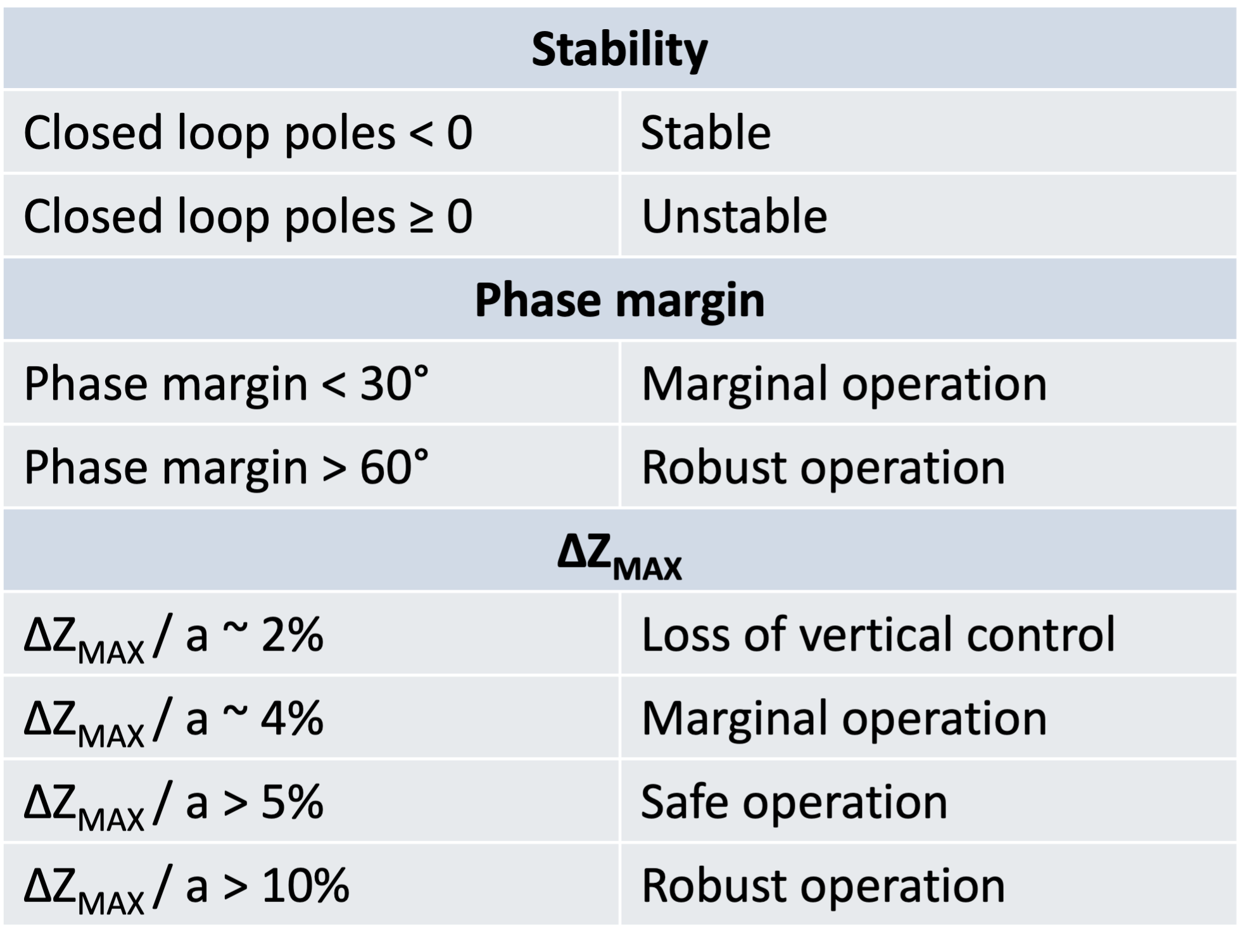}
\end{table}

Stability and phase margin are general control metrics, while $\Delta Z_{max}$ was introduced in \cite{Humphreys2009} with the following thought experiment. Suppose the plasma was shifted vertically instantaneously by a specified amount. For example, an ELM discharging current into the scrape-off-layer can have alter this plasma current distribution and have this effect. $\Delta Z_{max}$ is the maximum instantaneous vertical shift for which the controller can respond and restore the position. For a large perturbation, exponential growth of the instability is the dominant effect, while for smaller perturbations the controller can ``catch'' the plasma before it drifts too far. 

To compute $\Delta Z_{max}$ the state-space model is simulated forward in time with an initial condition given by,

\begin{equation}
    x_0 = \frac{w_z Z_0}{w_z \cdot \dv{Z}{x}},
\end{equation}

where $w_z$ is the unstable eigenvector direction, $Z_0$ is the test perturbation size, and $dZ/dx$ is the plasma vertical position response. When doing this type of analysis, it is also important to ensure that the voltage/current/force limits also stay within hardware limits, and to downgrade the metrics appropriately if they are not. 

\subsubsection{Feedback on current in the vertical stabilization (VS) coil}\label{sec:Ivsc}

On some devices it is important to also include feedback control on the current in the vertical stabilization (VS) coil. \censor{SPARC and ITER both have hardware designs that require this because of the in-vessel VS coil approach.}{} The idea is that small and fast VS coils are placed inside the vacuum vessel where they can provide control quickly without being delayed by the vessel shielding time. However, the placement inside the vacuum vessel usually limits the size of these coils so that they cannot carry significant current. In order to prevent the VS current ($I_{VS}$) from drifting over the course of the pulse, and jeopardizing VS control because of reaching a current limit, it is important to add an additional control layer to drive the low-frequency amplitude of $I_{VS}$ to be centered around zero. 

One approach for this problem is to select a linear combination of the other PF coils and apply proportional-integral feedback on the VS coil current itself. The current in these PF coils compensates for current in the VS coil, on a timescale roughly limited by the vessel shielding time and coil inductive time. One method for selecting this virtual circuit combination is to find a group of coils that supply uniform radial field (e.g. purely vertical shifting). The radial field is determined by from the coil greens functions:

\begin{equation}
    B_r = G_{Br,x} \, x
\end{equation}

This mapping between radial field at a group of points and coil currents can be inverted to find a control vector that supplies approximately uniform field over a target area. Then we use a control law of the form:

\begin{equation}
    u_{I_{VS}} = \hat w_{I_{VS}} \left(k_p + k_I / s \right) I_{VS}, 
\end{equation}

where $\hat w_{I_{VS}}$ is the control vector direction. 

\subsectionbf{Notes on: Ip and radial control}\label{sec:notes_r_ip_control}
It is common practice for plasma current control and radial control to have their own dedicated control loops. Functionally, it is possible to just include plasma current and radial position as members of the shaping parameter responses, and control these through the response map inversion. This would be a natural approach for controlling these parameters when using the IBSC framework, especially with dynamic response maps which are generally more accurate maps.  

However, it is common practice and there are some advantages to keeping Ip and radial control with their own dedicated controllers. First, vertical position, radial position, and plasma current control are generally the 3 most important global parameters. Thus it may be beneficial to spend additional effort tuning controllers for just these parameters to achieve a desired level of robustness.

Indeed relatively good performance for a pulse can be achieved by only controlling these 3 in feedback with everything else in feedforward, allowing for a staged commissioning process. \censor{(e.g. for SPARC commissioning, early pulses could be feedforward only, then feedforward + Ip control, then feedforward + R,Z,Ip control, then feedforward + R,Z,Ip,shape control). }{}

For radial control, it is worthwhile to consider what exactly the controlled variable will be. For example, some choices might be:

\begin{itemize}
    \item Radial position error of the current centroid. 
    \item Radial position error of the outer gap. 
    \item Radial flux error of the outer gap. 
\end{itemize}

PID control laws for radial and Ip control are of the form: 

\begin{equation}
    \begin{split}
    u_{r} &= \hat u_{r} ( k_P + k_I/s + k_D s) e_r \\
    u_{Ip} &= \hat u_{Ip} ( k_P + k_I/s + k_D s) e_{Ip}
\end{split}
\end{equation}

To select the virtual circuits directions $\hat u_r$ and $\hat u_Ip$, one approach is to use the Greens functions for the vertical field and for the flux (note the Greens function for flux is mutual inductance). 

\begin{equation}\label{eq:bz_flux_greens}
\begin{split}
    B_z &= G_{Bz,x} x \\
    \psi &= G_{\psi,x} x
\end{split}
\end{equation}
 
The radial position and radial force balance of the plasma is affected by the vertical field. Therefore, an ideal $R$ virtual circuit applies uniform vertical field without supplying any radial field or flux. Similarly, the plasma current is driven by loop voltage from supplied flux, and an ideal $I_p$ virtual circuit applies uniform flux without supplying any radial or vertical field. Inverting the Greens functions in \cref{eq:bz_flux_greens} can identify appropriate control vectors. 

There are additional factors to consider besides just supplying uniform flux or field. For example, if there are different timescales on which the coils respond. Another consideration for $I_p$ control is the amount of flux available in each coil. It is common for most of the flux consumption to be supplied by just a few center stack coils, in which case it may be necessary to restrict the virtual circuit to have these coils dominant so as to not hit current limits on the remaining coils over the course of the pulse. For these more complex cases, tools like the feedforward designer GSPulse can be useful in scoping out the flux consumption, shaping limits, current limits, and how these evolve over the course of the pulse. 

After vertical-stabilization, the plant dynamics for radial position and plasma current tend to act like first-order systems. That is, the R and Ip controllers can be tuned with classical PID tuning methods (e.g. Ziegler-Nichols). One simple method for tuning this system is: 

\begin{enumerate}
    \item Start with all gains zero. Increase the $k_p$ gain until the system responds at the desired speed, potentially with some overshoot. 

    \item Increase the $k_i$ gain until the steady-state error is removed. 

    \item Adjust the $k_d$ gain to dampen the system and mitigate some of the dynamic effects like overshoot. 

    \item The gains for radial and Ip control should be tuned and tested simultaneously, since there is normally coupling between these channels. 
\end{enumerate}

\subsectionbf{Notes on: coil current control}

An important part of the full shape controller is designing a decoupling coil current controller. The role of the coil current controller is to map from target coil currents to power supply voltages. 

A common approach for the coil current controller is based on the vacuum circuit equation, written for the shaping coils only:

\begin{equation}
    M_{cc} \dot I_c + R_c I_c = V_c 
\end{equation}

In the inductive limit ($R_c \rightarrow 0$), we observe the relationship between voltage and current is scaled by the mutual inductances. This suggests that the coil current controller be proportional to the mutual inductances:

\begin{equation}\label{eq:kcoil}
\begin{split}
    V_c = K_{coil} e_{Ic} \\
    K_{coil} = k_P M_{cc}
\end{split}
\end{equation}

Where the scalar gain $k_P$ can be adjusted to obtain the desired controller speed. (It may also be advantageous to add integral and derivative terms to the controller, depending on the tokamak.) This controller design generally provides good decoupling performance and is easy to compute. 

We also note an alternative method for decoupling based on the frame alignment technique that generally provides superior decoupling. In this case, we compute the vacuum transfer function for the coil current evolution in response to power supply voltages. 

\begin{equation}
    G_{Vc\rightarrow I_c} = \text{ss(} A_{vac}, B_{vac}, \begin{bmatrix} I_{ncoils} & 0 \end{bmatrix}, 0 \text{)}
\end{equation}

In the frame alignment technique, we select a frequency $\omega_b$ and use the ALIGN algorithm \cref{app:align_algorithm} to get a real approximation to the complex inverse. \censor{For example, on SPARC, with coil inductive timescales around 200ms we choose $w_b = 1/200 = 5$Hz. We take the coil current controller as a scalar gain of this result. }{}

\begin{equation}
\begin{split}
    G (i \omega_b)^{-1} &\approx ALIGN(G(i \omega_b)) \\
    K_{coil} &= k_P \times ALIGN(G (i \omega_b))
\end{split}    
\end{equation} 

Integral and derivative terms could be added as well, if deemed necessary. We note that this controller is functionally equivalent to \cref{eq:kcoil}, but just has slightly different matrix gains that usually result in superior decoupling. 

\subsubsection{Decoupling in the presence of voltage saturation}

The decoupling coil current controller is a matrix gain controller. However, matrix gains can cause problems when used in situations where voltage saturation occurs. As an example, consider two coils with strong mutual coupling, in which case there would be an off-diagonal gain between these coils. Now consider if the controller attempts to track a large change in current for the first coil. Since the request is large, the first coil voltage would be clipped, but the voltage on the second coil would still respond as if anticipating a larger change in the first coil. In other words, the second coil ``over-corrects'' based on an anticipated change that is never realized.

In order to reduce the cross-coupling while voltage-limited, we do not want to just clip the controller output voltages but instead (while satisfying limits) downscale the voltages while preserving the coil-to-coil coupling ratio. A simple algorithm for accomplishing this task is to pre-scale the errors using the following algorithm:

\begin{itemize}
    \item First, take the diagonal values of $K_p$ and calculate what the nominal voltage $\hat v$ would be for the given set of current requests:

    \begin{equation}
        \hat v = \text{diag}\left(K_p\right) \otimes e_I 
    \end{equation}
    
    \item Next, determine a scaling factor for each coil. This is the amount that each coil would need to be scaled by in order to satisfy the voltage limits (assuming diagonal $K_p$). 

    \begin{align*}
    \text{scale}[i] =& v_{max}[i] / \hat v[i] \;\;\; \text{ if }\;\;\; \hat v[i] \geq v_{max}[i],  \\
    & v_{min}[i] / \hat v[i] \;\;\; \text{ if }\;\;\; \hat v[i] \leq v_{min}[i], \\
    & 1.0  \;\;\;\; \text{otherwise.}
    \end{align*}

   \item Finally, compute the voltage requests with the original gain matrix and these scaled errors:

    \begin{equation}
        v = K_p (e_I \otimes \text{scale})
    \end{equation}

    \item The final voltages are not guaranteed to satisfy voltage limits with these steps (they will be very close if $K_p$ is diagonal-dominant), therefore there is an additional step of post-clipping the voltages. 
    
\end{itemize}

\subsectionbf{Vertical decoupling}\label{sec:notes_vert_decoupling}

In \cref{sec:static_plus_vertical} it was suggested that to improve vertical decoupling, all voltage commands for adjusting the net vertical position should be handled by the vertical controller. An effective method for achieving this within the IBSC framework is to add an extra plasma response mode corresponding to net vertical shifting of the equilibrium. 

For example, if the original response map is: 

\begin{equation}
    \delta y = T \delta I
\end{equation}

Now adding a response mode corresponding to a z-shift, gives an augmented response matrix $\hat T := \begin{bmatrix} \pdv{y}{z} & T \end{bmatrix}$. For measured shaping errors $e$, the IBSC correction is:

\begin{equation}
    \begin{bmatrix}
        \delta z \\ \delta I_c^{targ} 
    \end{bmatrix} = \hat T^\dagger e
\end{equation}

In other words, the IBSC inversion now additionally provides a $z$ term that represents the amount to shift the $z-$reference target used by the vertical controller. If there is any inconsistencies between the target z position and the target shape, this will get resolved via this mechanism. Additional vertical decoupling suggestions such as slowing down vertically anti-symmetric shape control actuation still apply.

\pagebreak
\section{Tutorial - NSTX-U static IBSC controller design}\label{app:nstxu_tutorial}

\subsectionbf{Overview}

This tutorial gives an example of designing the magnetic controller for NSTX-U using with inversion-based shape control. For this example, we make the following design decisions: 

\begin{itemize}
    \item Following the design options in \cref{tab:ibsc_design_choices}, we use a \textbf{QP-constrained controller}, based on \textbf{static} mappings of the \textbf{vacuum flux response}. 
    
    \item We will use the isoflux control method, as oppposed to gap control (see \cref{sec:gap_vs_isoflux}). 

    \item We will control a large number of shape control points (20) as well as the field at upper and lower x-points. 
    
    \item We neglect some of the hardware-specific implementation and observer details, such as sensor placement and characteristics. 
\end{itemize}

\subsectionbf{General Characteristics}

NSTX-U is a spherical tokamak designed (before recent hardware upgrades) to reach high elongation $\kappa > 2$, moderate plasma current $I_p \sim 1$MA, and pulse lengths of about 1 second. NSTX-U features low-inductance copper coils that can be driven very quickly. However, with high plasma elongation and vertical growth rate, the necessary control timescales are also fast. Additionally the vessel current interactions are significant.

A cross-section of the NSTX-U tokamak is shown in \cref{fig:geo_nstxu2}. NSTX-U features a long, slender ohmic coil that is used to drive flux for the plasma current and is a limiting factor on pulse length. The upper and lower PF1A/PF2A/PF3 coils can each be driven independently while the PF5 coils are driven together in series. Together with the OH coil this gives 8 independent actuators. This is not a high degree of freedom, and in general it is understood that spherical tokamaks do not have significant shaping capabilities. 

\begin{figure}[H]
    \centering
    \includegraphics[width=0.3\linewidth]{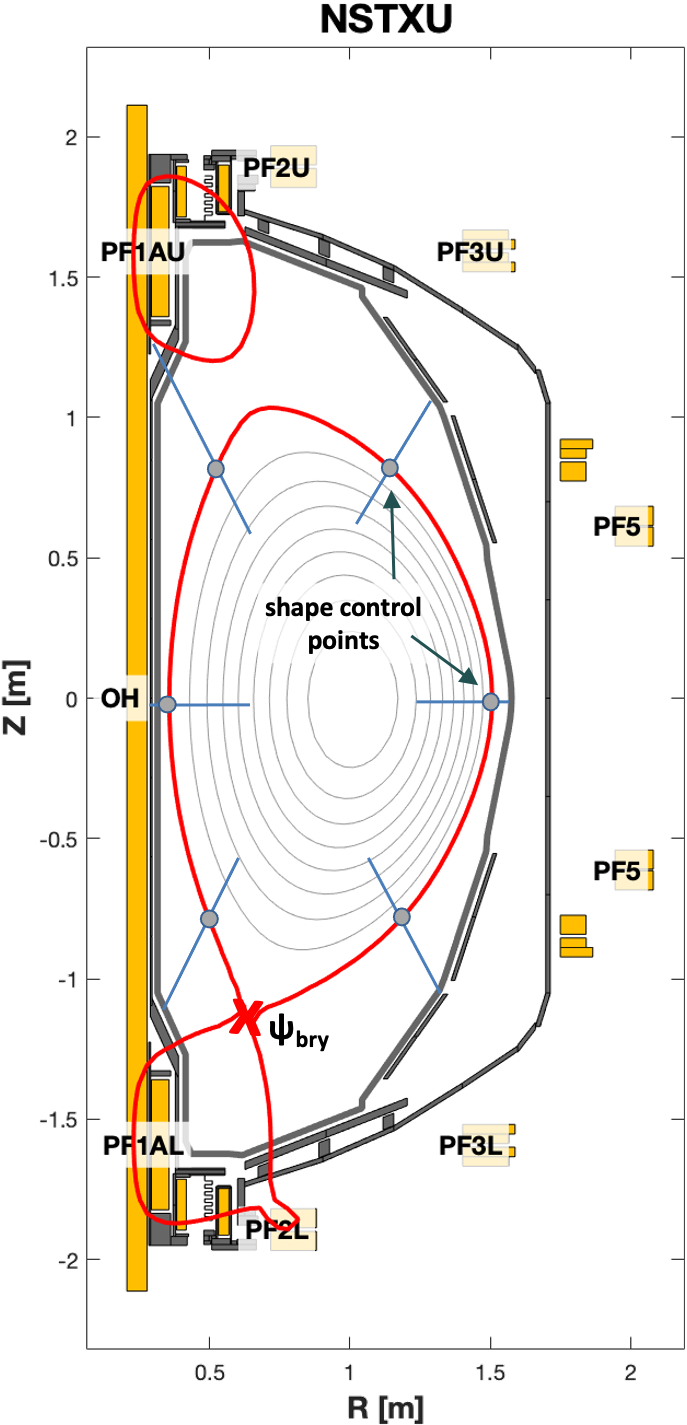}
    \caption{Geometry of the NSTX-U tokamak showing plasma equilibrium and 6 shape control points.}
    \label{fig:geo_nstxu2}
\end{figure}

\begin{table}[H] 
    \begin{center}
        \includegraphics[width=9cm]{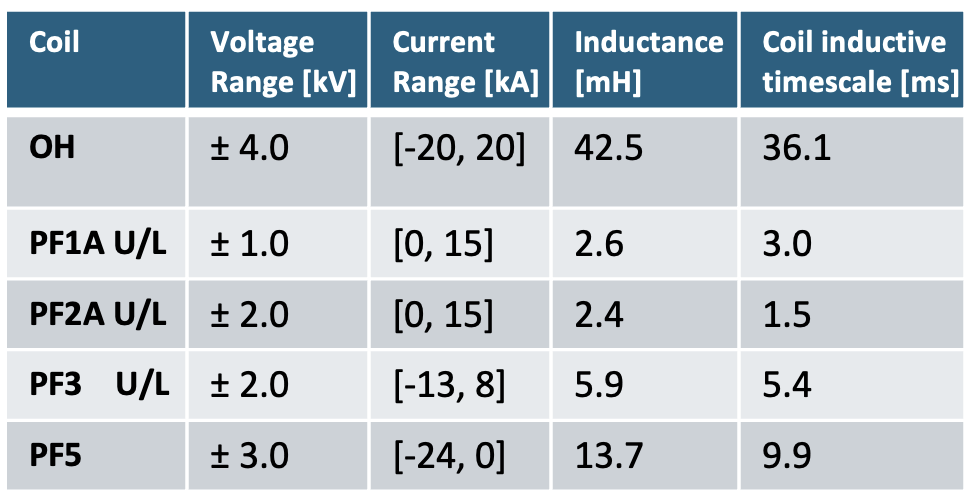}
    \end{center}
    \caption{Geometry and hardware parameters and associated timescales for NSTX-U.}
    \label{tab:nstxu_timescales}
\end{table}

To get a general sense of the control characteristics, we evaluate the vessel shielding and coil inductive timescales (\cref{eq:control_timescales}, repeated here for convenience): 

\begin{equation}
\begin{aligned}     
     \tau_{vessel \; shielding} &= 1 / \; \text{max(real(eig(} -\vec M_{vv}^{-1} \vec R_v \text{)))}  &= 25\text{ms}\\
     \tau_{coil \; inductive} &=  (L \Delta I_{typical}) / V_{max} &= 2-40\text{ms} \\    
     \tau_{vertical \; growth \; rate} &= \gamma^{-1}  = 1 / \; \text{max(real(eig(}\vec A \text{)))} &= 12\text{ms}
\end{aligned}
\end{equation}

The vessel shield time is found to be $25$ms. The coil inductive timescales are shown in \cref{tab:nstxu_timescales} and range from 2-40ms. The fact that the coil times are faster than the vessel shield time indicates that we are not voltage-limited by the coils. Instead control is likely to be limited by vessel shielding effects. We are unlikely to be able to control shaping features much faster than about 1.5-2x the vessel shield time and the induced vessel currents could impact control accuracy. 

For the particular equilibrium depicted in \cref{fig:geo_nstxu2}, the vertical growth time constant is found to be 12ms. To get a frame of reference, we perform a vacuum simulation of the midplane radial field response to several coils as shown in \cref{fig:nstxu_vacuum_br_response}. Applying 1kV of voltage to each coil, it takes about 5-10ms for the field to penetrate the vessel and reach 3Gs. This corresponds to a 1cm vertical shift based on the spatial variation of the vacuum field for this equilibrium. The point of this exercise is not to fully define the vertical controller yet, but just to illustrate that we should reasonably expect to control this equilibrium given that the timescales line up with the field response time and power supply capabilities.  

\begin{figure}[H]
    \centering
    \includegraphics[width=0.5\linewidth]{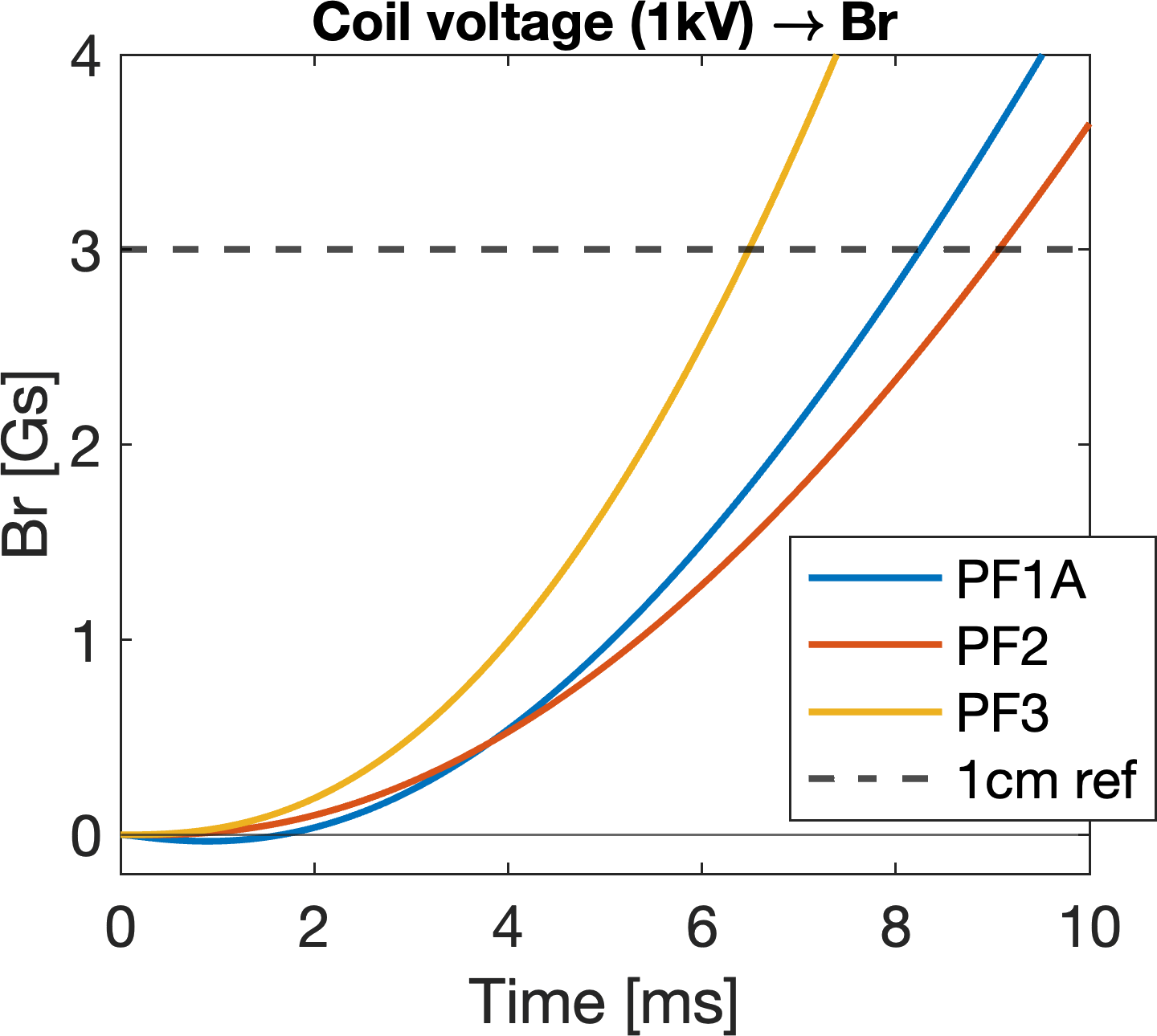}
    \caption{Vacuum response of the midplane radial field to a step change in power supply circuit voltage. The PF3 response is fastest, but PF1A and PF2 also respond on a relevant control timescale. For the given equilibrium, a 1cm shift in Z position corresponds to approximately 3Gs, based on the spatial gradient of the vacuum magnetic field for the equilibrium. The power supplies are able to supply relevant levels of field on the 5-10ms timescale.}
    \label{fig:nstxu_vacuum_br_response}
\end{figure}

\subsectionbf{Vertical control}\label{sec:nstxu_vert_control}

We now move onto the desiging the vertical controller. Following \cref{sec:notes_vert_control}, we will use a PD controller of the form:

\begin{equation}
    K_z = \hat u_z (k_p + k_d*s)
\end{equation}

In this example, we will design the controller around the equilibrium depicted in \cref{fig:geo_nstxu2}. This is a representative lower single-null equilibrium, with moderately high elongation $\kappa=1.9$, $I_p=600$kA, and $l_i=0.9$. 

We use the Toksys code with GSpert \cite{Welander2005} to linearize this equilibrium and obtain the state-space model. For the following, we will assume that the linearization code provides: 

\begin{equation}
    \begin{aligned}
        A,B &= \text{State-space dynamics matrices} \\
        \pdv{z_{cur}}{x}, \pdv{r_{cur}}{x}&= \text{Response of the current centroid position} \\ 
        \pdv{\psi_g}{x} &= \text{Response of the gridflux} 
    \end{aligned}
\end{equation}

Linearization codes such as GSpert may or may not provide the desired form of output matrix $C$ but this can be derived as in \cref{sec:output_linearization}. To scope out the initial controller design, we perform the following steps: 

\subsubsection{Select control vector direction}
We assign the vertical control vector $\hat u_z$ to be the antisymmetric combination of PF3U minus PF3L, with all other coils contributions set to zero. This follows the original NSTX-U controller design, though later we will consider other combinations. 

\subsubsection{Estimate order of magnitude for control gains}

There are many methods to obtain the desired control gains. For this example, we will describe a heuristic process based on correlating the vacuum field temporal response, with the vacuum field spatial variation, to get rough order-of-magnitude estimates for the vertical controller PD gains. The vacuum radial field for the equilibrium is obtained from greens functions via,

\begin{equation}
    B_{r, vac} = G_{Br,x} x
\end{equation}

We take an area-averaged gradient of the radial field, to obtain that, 

\begin{equation}
    \pdv{B_r,vac}{z} = 1.88 \times 10^{-2} \text{ T/m}
\end{equation}

This is roughly the amount of additional radial field that to be supplied per unit of vertical motion. We also obtain a step response of the radial field from applying voltage in the $\hat u_z$ direction, similar to the step response of \cref{fig:nstxu_vacuum_br_response}. This step response indicates that, after 12ms which is the vertical time constant, the ratio of field to voltage is: 

\begin{equation}
    \pdv{B_r,vac}{v}|_{t =12\text{ms}} = 2.0 \times 10^{-6} \text{ T/V}
\end{equation}

This is the amount of radial field supplied on the vertical instability timescale, per unit of power supply voltage. Let's assume that this amount of field is supplied equally by the proportional and derivative terms. For the proportional gain this would be:

\begin{equation}
    k_p = \left( \pdv{B_r,vac}{v}|_{\tau=1/84Hz} \right)^{-1} \left(\pdv{B_r,vac}{z} \right) = \frac{1.88 \times 10^{-2}\text{T/V}}{2.0 \times 10^{-6} \text{T/m}} = 8700 \text{ V/m}, 
\end{equation}

and for the derivative gain

\begin{equation}
    k_d = k_p / \gamma = \frac{8700 \text{V/m}}{84 \text{Hz}} = 104 \text{ V/(m/s)}
\end{equation}

Again, these parameter values are just estimates to get the correct order of magnitude for the controller gains, and must still be tuned for performance. 

\subsubsection{Evaluate metrics and tune}

We now will refine the PD controller gains to optimize various vertical control metrics and improve performance. For this example, we use a delay time of 0.1ms which is roughly the amount of power supply and measurement delay in the system. As described in \cref{sec:notes_vert_control} some of the key metrics are stability, phase margin, and $\Delta Z-$max. 

The following MATLAB code snippet demonstrates how to perform some of this analysis, using the PD gains estimated above:

\begin{figure}[H]\label{code:vert_control}
\caption{Matlab code for vertical control analysis.}
\begin{lstlisting}[language=Matlab]
% ============ Vertical Control Analysis ==============

% Model parameters
P = ss(A, B, dzdx, 0);      % A,B,dzdx provided by linearization code such as TokSys

% Controller parameters
kp = 8700;                  % [V/m] eqn 63
kd = 104;                   % [V/(m/s)] eqn 64 
delay_time = 1e-4;          % [s] time delay
tau_roll = 1e-6;            % [s] low-pass filter time for the derivative
u_z = [0 0 0 1 0 -1 0 0]';  % corresponds to the PF3U-PF3L direction

% Control law
s = tf('s');
K = u_z * (kp + kd*s / (tau_roll*s + 1));
delay = exp(-ps_delay_time * s);
K_with_delay = pade(K*delay, 2);   % Pade approx to make time delay realizable

% Loop transfer function
L = P * K_with_delay;

% Closed loop transfer function
CL = feedback(L,1); 

% Stability condition
is_stable = max(real(pole(CL))) < 0;

% Closed loop step response
step(CL)

% Nyquist plot to read phase margin
nyquist(L)
\end{lstlisting}
\end{figure}

\Cref{fig:nstxu_tut_vert0} shows the step response and nyquist figure obtained from running the code above, and we can see that the system is found to already be stable with the rough-estimate PD gains. However, the overall characteristics indicate poor tuning. The controller overshoot is very high $\sim80\%$ and the response is highly oscillatory. This is consistent with the Nyquist plot which has a small phase margin of 21$\degree$. After experimenting with the controller gains, we obtain an improved controller. The phase margin has increased to 34$^\degree$ and the overshoot reduced to 50\%. Per \cref{tab:vs_metrics} this is still acceptable operation but not considered highly robust.

\begin{figure}[H] 
    \begin{center}
        \includegraphics[width=0.95\linewidth]{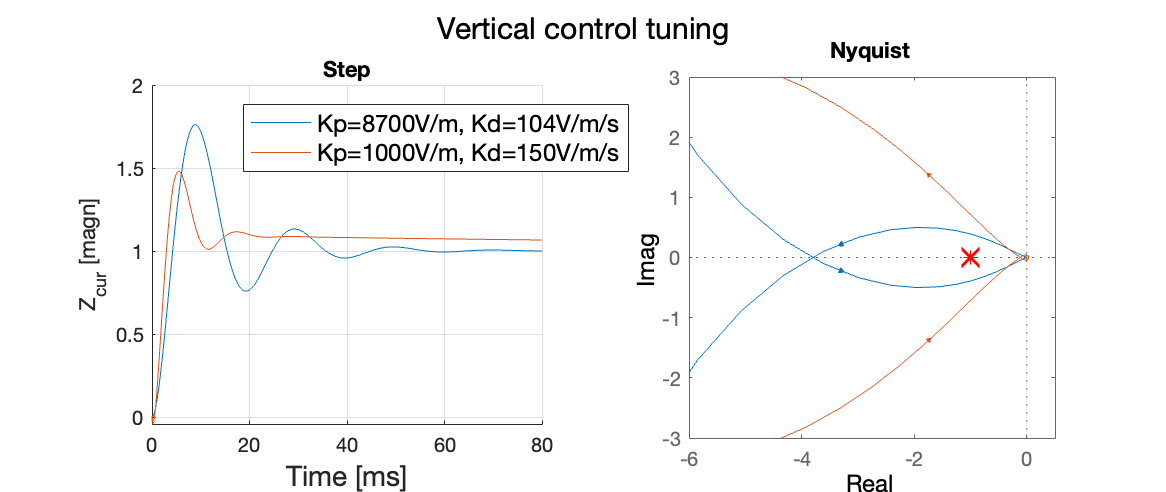}
    \end{center}
    \caption{Example of tuning the NSTX-U vertical proportional and derivative gains. The blue line is the initial controller obtained from order-of-magnitude estimates of the gains based on field responses. The red line is the controller after tuning the controller, and has less overshoot and increased phase margin.}
    \label{fig:nstxu_tut_vert0}
\end{figure}

While the phase margin indicates acceptable operation (in-line with operational experience of NSTX-U) a $\Delta Z_{max}$ analysis of this equilibrium indicates robust operation, with $\Delta Z_{max}$ values approaching 30cm or more. These values are obtained by specifying an initial condition as in the code snippet below, and performing a linear simulation as in block diagram \cref{fig:plant_ref_frame}. The reason for the optimistic $\Delta Z_{max}$ value is that the power supplies are very capable and not very constraining on the ability to stabilize an equilibrium. However this doesn't capture the performance sensitivity to things like modeling error, controller tuning, or reconstruction errors, for which phase margin is a better criterion. This illustrates the value in computing multiple metrics to get a complete picture of performance, as $\Delta Z_{max}$ alone would be too optimistic. 




\subsubsection{Additional controller performance}

The original vertical controller used only the PF3 upper and lower coils. However, \cref{fig:nstxu_vacuum_br_response} illustrates that the field penetration time for PF1A and PF2 is roughly as fast as PF3. Additionally, \cref{fig:nstxu_spatial_br} illustrates that including these coils can provide a more uniform field distribution than PF3 alone. Together, these suggest that there is opportunity to use additional coils to improve vertical control.

\begin{figure}[H]
    \centering
    \includegraphics[width=0.6\linewidth]{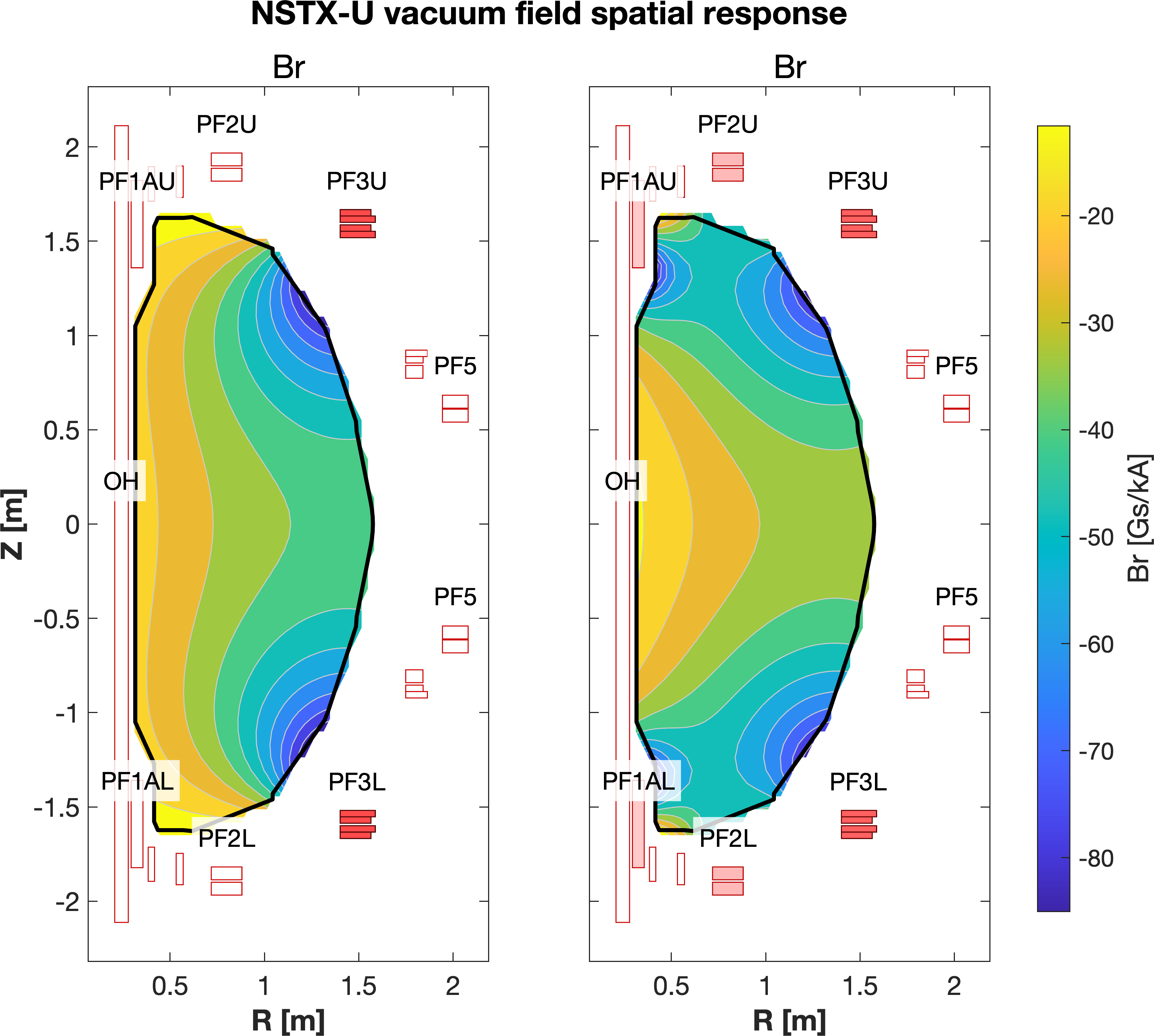}
    \caption{Radial magnetic field distribution created by \textbf{left)} PF3 coil pair only, \textbf{right)} PF3 coil pair with light usage of PF1A and PF2 coil pairs. This vector is obtained by attempting to optimize for a uniform radial magnetic field, and is slightly more uniform than the field produced on the left. Together with the dynamic response \cref{fig:nstxu_vacuum_br_response} this hints at being able to improve vertical control performance with the addition of PF1A and PF2 to the control vector.}
    \label{fig:nstxu_spatial_br}
\end{figure}

After some tuning, we find the best control vector direction is primarily in the PF3 direction, but with about 20-30\% use fraction on PF1A and PF2 as well. This is illustrated in \cref{fig:vs_timedelay}. For a representative equilibrium, we observe that overshoot is reduced by 15\% and the phase margin increases from 32$^\degree$ to 38$^\degree$. While these are not huge performance gains they come at nearly zero cost because it mostly just takes some software changes to implement. 

\Cref{fig:vs_timedelay} also illustrates the phenomenon that the performance is highly sensitive to any time delays in the system. Eliminating time delays does come at a high cost because it requires high performance hardware and computing systems. The time delay on NSTX-U was estimated to be well below 1ms and nearing 0.1ms \cite{Boyer2018}, which is a critical feature of meeting performance. At the more extreme delays near 1ms, the overshoot begins to exceed 100\% and the nyquist plot indicates low stability margin such that the controller would almost certainly be unstable in practice. This example illustrates that, especially for vertical control, the time delays should be explicitly modeled in order to avoid over-optimistic analysis. 

\begin{figure}[H] 
    \begin{center}
        \includegraphics[width=14cm]{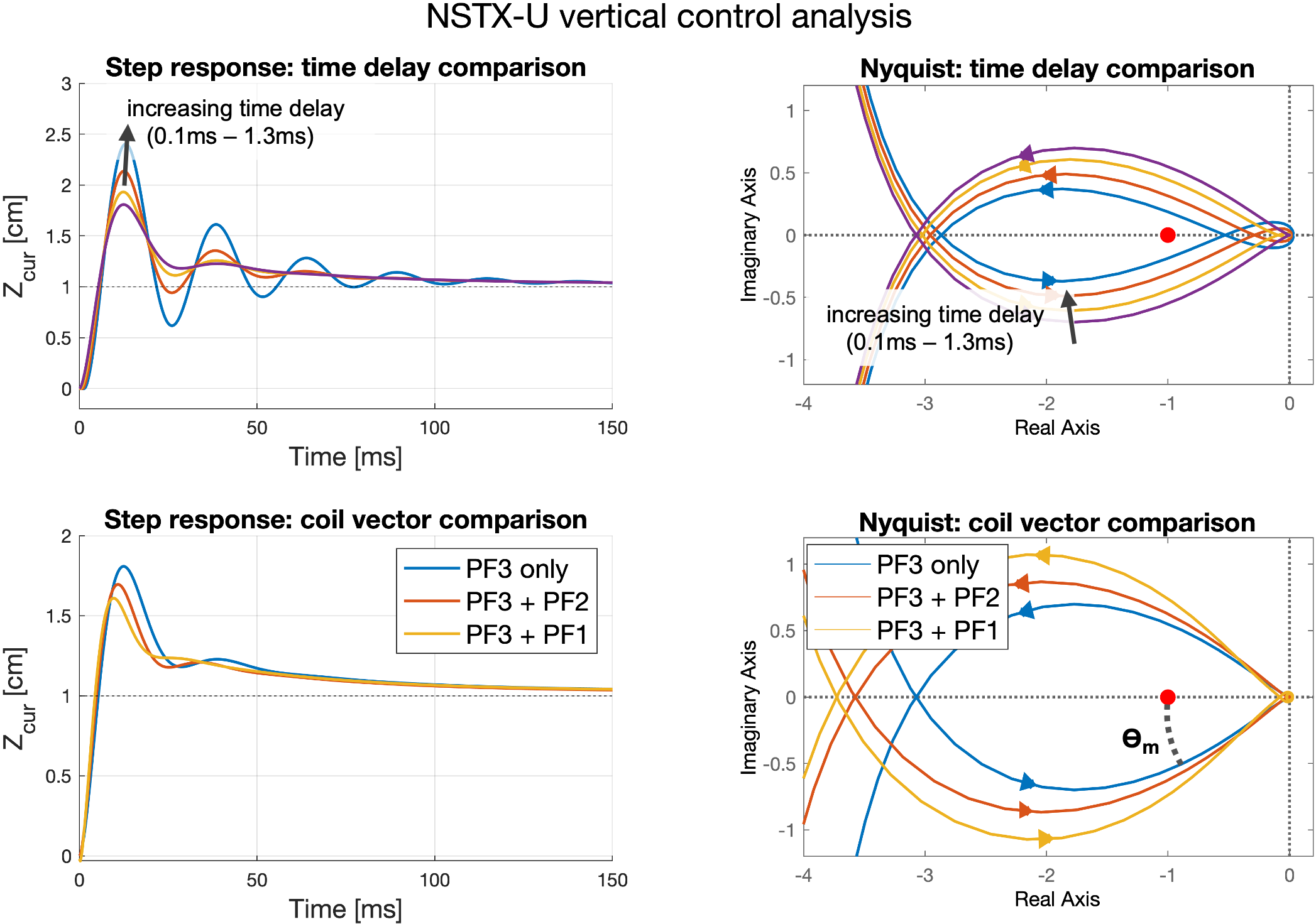}
    \end{center}
    \caption{Effect of various parameter and controller changes on the vertical controller performance. \textbf{Top)} The presence of any time delays in the system is detrimental to control. For a time delay of 1ms or more, the low phase margin or stability margin would likely lead to a controller that is unstable in practice. Because of the sensitivity to time delays, delays are a critical feature to explicitly model in any vertical control analysis. \textbf{bottom)} The simple change of adding PF1 and/or PF2 to the vertical controller can reduce the overshoot and increase control phase margin.}
    \label{fig:vs_timedelay}
\end{figure}

\subsectionbf{Radial and plasma current control}

\subsubsection{Radial control vector selection}

Having stabilized the plasma with vertical control, our next step in the process is to design radial and Ip control loops. Following \cref{sec:notes_r_ip_control}, we adopt the PID control laws of the form: 

\begin{equation}
    \begin{split}
    u_{r} &= \hat u_{r} ( k_P + k_I/s + k_D s) e_r \\
    u_{Ip} &= \hat u_{Ip} ( k_P + k_I/s + k_D s) e_{Ip}
\end{split}
\end{equation}

Our first step will to be evaluate the control vector directions. For radial control we desire a vector that supplies uniform vertical field without supplying vertical field or flux. One way to find such a vector is to solve:

\begin{equation}\label{eq:radial_control_vector}
    \begin{bmatrix} \vec 1 \\  \vec 0 \\  \vec 0 \end{bmatrix} = 
    \begin{bmatrix}
        W_{Bz} G_{Bz,I_c} \\ W_{Br} G_{Br,I_c} \\ W_\psi G_{\psi,I_c}
    \end{bmatrix} I_c,
\end{equation}

where $W_{<>}$ are weighting matrices and $G_{<>}$ are the greens functions for field and flux between the coils and target grid locations. For the grid locations, we use a large number of points spread out over the plasma domain. Taking a pseudoinverse of this system gives the coil current combinations and field response as shown in figures \cref{fig:radial_control_vector,fig:spatial_response_bz}.

\begin{figure}[H] 
    \begin{center}
        \includegraphics[width=7cm]{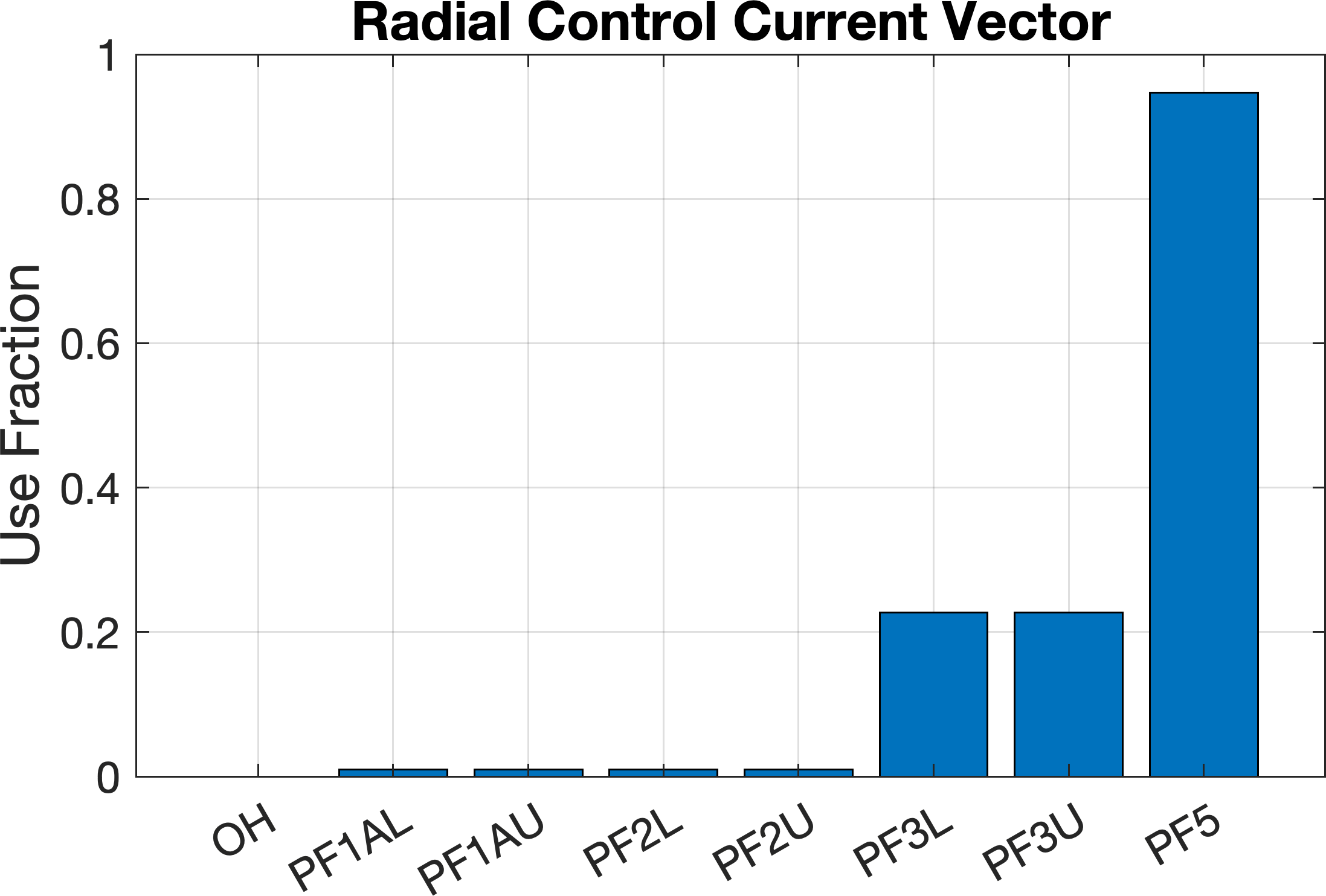}
    \end{center}
    \caption{Radial control vector obtained by optimizing for a vector combination that produces uniform vertical field without supply radial field or flux.}
    \label{fig:radial_control_vector}
\end{figure}

\begin{figure}[H] 
    \begin{center}
        \includegraphics[width=0.45\linewidth]{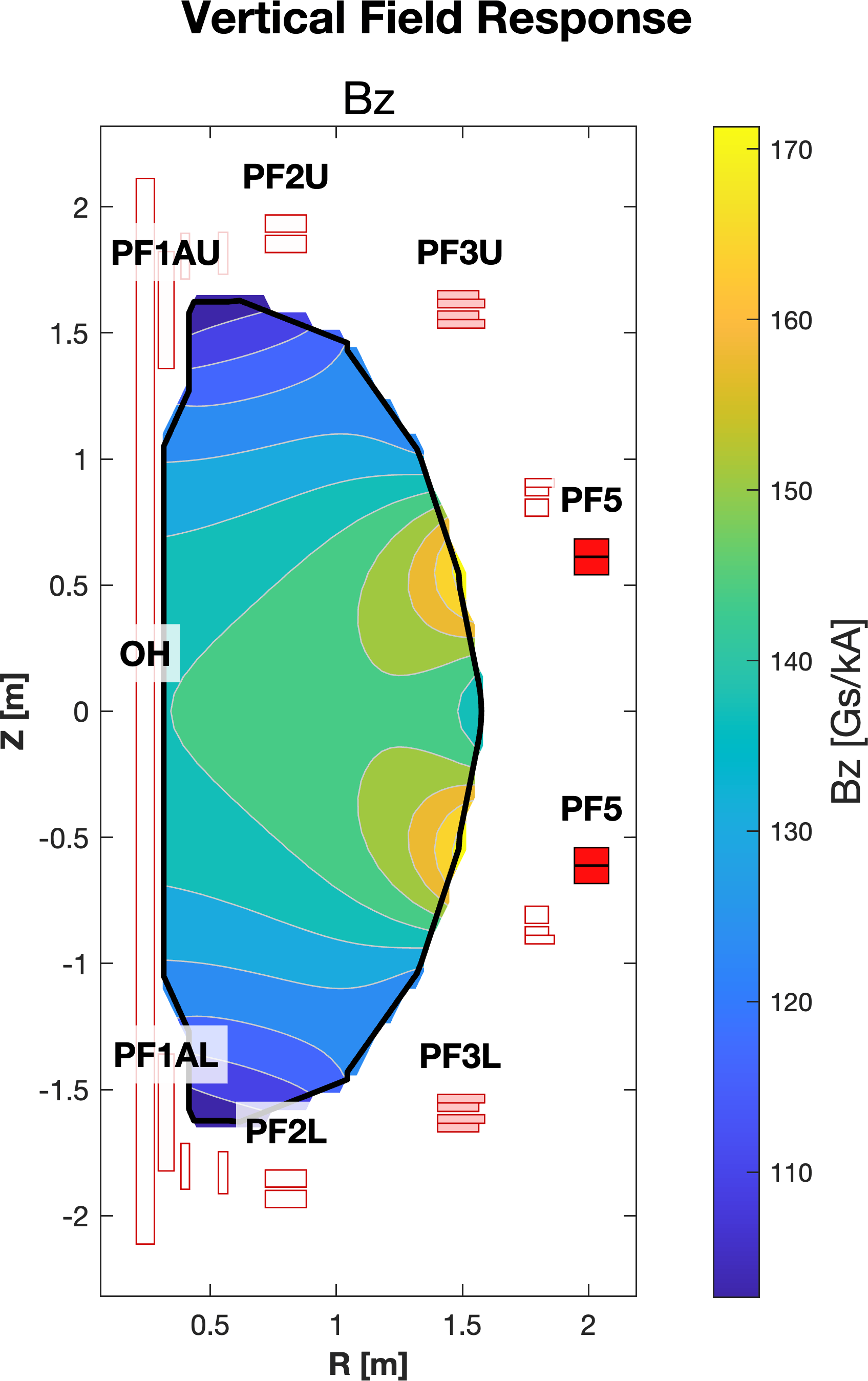}
    \end{center}
    \caption{Vertical magnetic field distribution produced by the radial control vector.}
    \label{fig:spatial_response_bz}
\end{figure}

The control vector direction consists primarily of PF5, but also about 25\% on the PF3 coils. This is the ratio of coil currents, which needs to be scaled by coil inductances for voltage control. From \cref{tab:nstxu_timescales}, the PF3 coil inductance is about 40\% of the PF5 inductance, so the final vector has $f=0.95$ usage on PF5 and $f=0.4*0.25=0.1$ fractional usage for PF3. 

\subsubsection{Plasma current control vector selection}

We repeat a similar exercise to find the Ip control vector, this time using the relationship,

\begin{equation}\label{eq:ip_control_vector}
    \begin{bmatrix} \vec 0 \\  \vec 0 \\  \vec 1 \end{bmatrix} = 
    \begin{bmatrix}
        W_{Bz} G_{Bz,I_c} \\ W_{Br} G_{Br,I_c} \\ W_\psi G_{\psi,I_c}
    \end{bmatrix} I_c,
\end{equation}

which provides a vector that supplies flux for the current drive while minimizing the magnitude of field that is produced over the plasma area. \Cref{fig:spatial_response_psi} illustrates this effect. The first plot illustrates 100\% usage of just the ohmic coil where it is observed that the flux contours are curved from the fringing field. In \cref{fig:spatial_response_psi}b the current control vector from optimization contains light usage of PF3 (10\%) and PF5 (3\%). This straightens out the flux contours over the majority of the plasma. There is no perfect solution though, as there is still a radial gradient in the flux instead of a perfectly flat contour, indicating some shaping perturbations will accompany the current drive. (Note that shaping effects from ohmic flux consumption were also observed in the GSPulse pulse designs \cite{Wai2025}) . 

\begin{figure}[H] 
    \begin{center}
        \includegraphics[width=11cm]{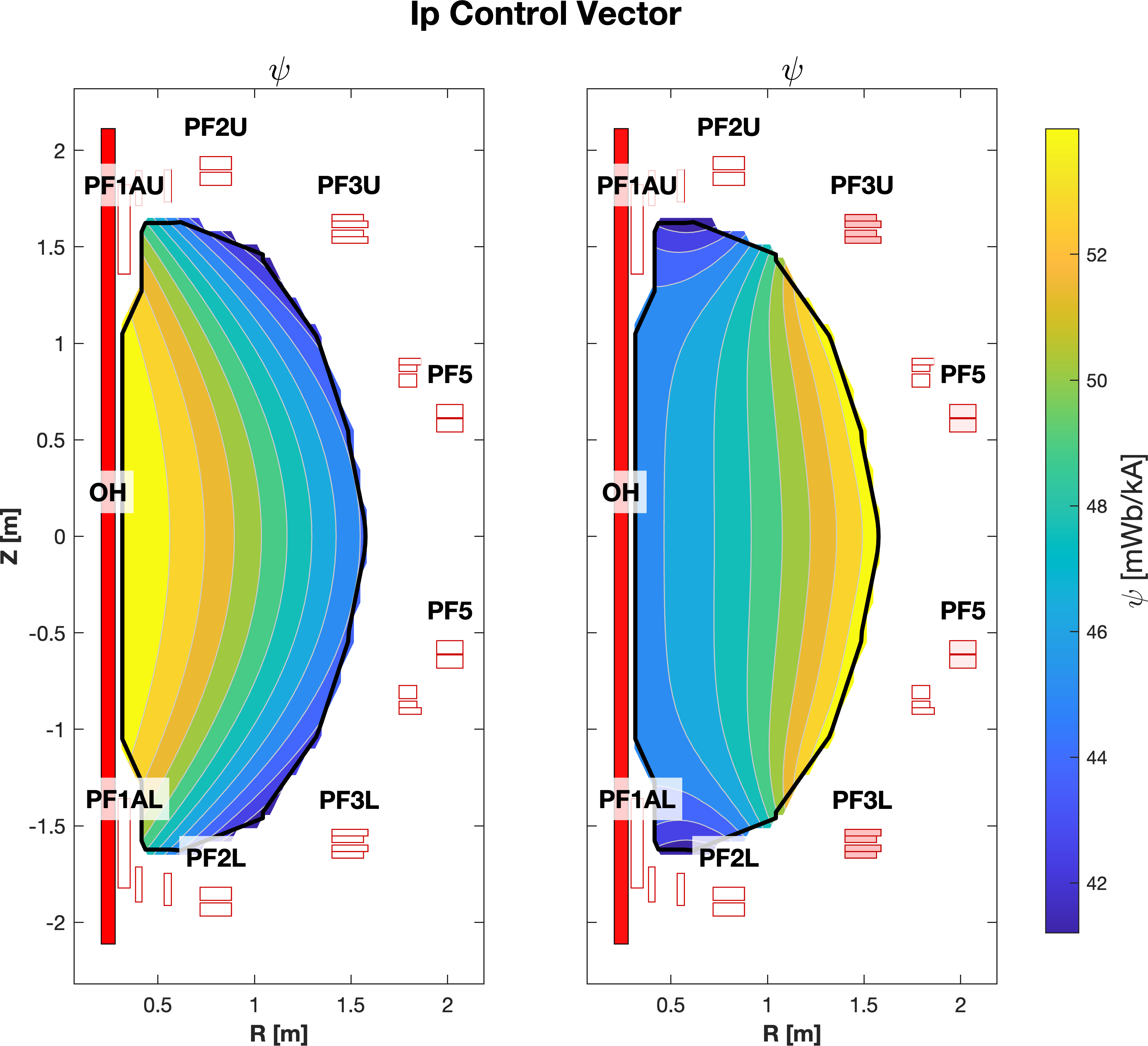}
    \end{center}
    \caption{Flux distribution produced by the proposed Ip control vectors, \textbf{left)} using the OH coil only and \textbf{right)} using the OH coil and light usage of PF3 and PF5. The flux contours in the optimized vector are more uniform in the second case. This is also consistent with GSPulse feedforward trajectory results \cite{Wai2025} that required continued evolution of the PF3 current throughout the pulse in order to maintain shape during the OH ramp.}
    \label{fig:spatial_response_psi}
\end{figure}

\subsubsection{Radial and Ip control tuning}

For NSTX-U radial control, we decide that it is most physically important to control the radial outer gap (instead of radial current centroid). Also, because we are using the isoflux method we will control in flux units. This means that our definition for the radial response is,

\begin{equation}
    \pdv{r}{x} := \pdv{\psi_{(R_{OMP}, Z_{OMP})}}{x} - \pdv{\psi_{(R_{BDEF}, Z_{BDEF})}}{x}
\end{equation}

which can be computed following the description in \ref{sec:output_linearization}. One complication is that the radial and Ip control can have some cross-talk (e.g., the identified control vectors both contained PF3 and PF5 usage), and so it is useful to tune these systems simultaneously. The following code snippet demonstrates how to close the loop for the radial and Ip controllers simultaneously and generate responses for analysis. 

\begin{figure}
\caption{Matlab code for combined vertical, radial, and Ip control analysis.}
\begin{lstlisting}[language=Matlab]
% ============ Combined Vertical/Radial/Ip Control Analysis ==============

% Plant description
dzrpdx = [dzdx; drdx; dIpdx];    % Vertical, radial, and Ip responses
P = ss(A, B, dzrpdx, 0);

% Controllers
s = tf('s'); 
K_z = u_z * (kp_z + ki_z/s + kd_z*s / (tau_roll*s + 1));   % Designed PID controllers
K_r = u_r * (kp_r + ki_r/s + kd_r*s / (tau_roll*s + 1));
K_p = u_p * (kp_p + ki_p/s + kd_p*s / (tau_roll*s + 1));
K = [K_z, K_r, K_p];   

% Power supply description
delay = exp(-ps_delay_time * s);   
K_with_delay = pade(K*delay, 2);  

% Analysis
L = P * K_with_delay;       % Open loop gain for Z, R, Ip
CL = feedback(L, eye(3));   % Closed loop system for Z, R, Ip
nyquist(L)                  % Nyquist plots for the 3x3 system
step(CL)                    % Step response plots for the 3x3 system
\end{lstlisting}
\end{figure}

For radial and Ip control, we individually increase $k_P$ to get the desired dynamic response (rise time of 20ms) and then set $k_I$ so that the steady-state error is zero. The step response for these individually controllers corresponds to the blue lines \cref{fig:nstxu_tut_step_zrp0}a/b. While the performance looks good when either the Ip or radial control loops are closed, when both loops are closed the step response shown in red is very poor with large oscillations and overshoot due to the coupling between controllers. This illustrates the importance of tuning the controllers simultaneously, since Ip control acts like a disturbance for the radial controller, and vice versa. It turns out that, while these controllers had good tracking capability, they have poor disturbance rejection. 

This oscillation can be tuned out of the system using the derivative term to increase damping, as shown in \cref{fig:nstxu_tut_step_zrp1}. The derivative term makes each system more robust against disturbances and cross-coupling. \Cref{fig:nstxu_tut_nyquist_zrp} supports this analysis, as we can see that the phase margin has increased from 50$^\degree$ to a healthy 92$^\degree$. 

\begin{figure}[H] 
    \begin{center}
        \includegraphics[width=0.9\linewidth]{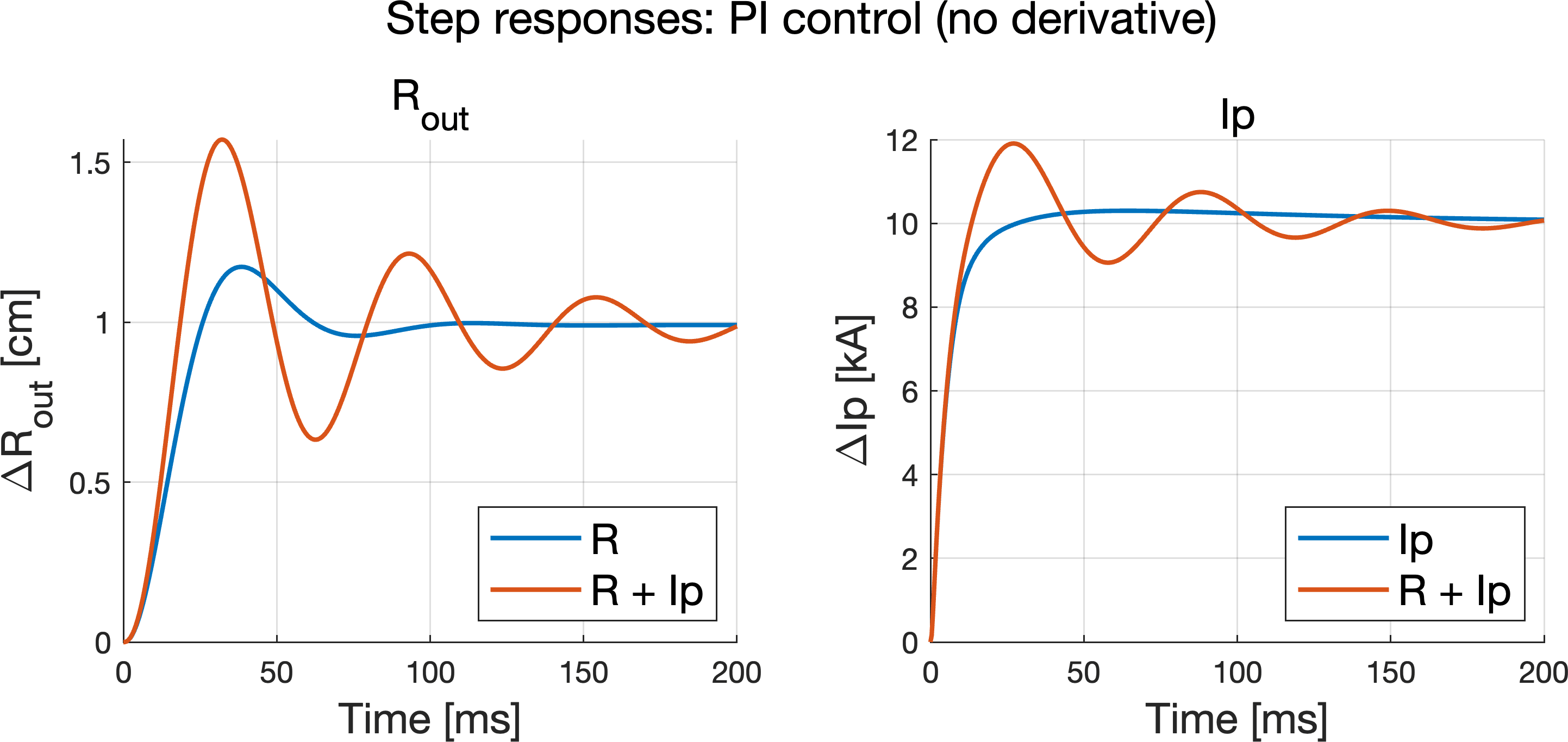}
    \end{center}
    \caption{Step responses of the radial and Ip controllers. Radial and Ip control generally have strong cross-coupling, so it is important to tune these controllers in a combined system. The blue line represents controllers tuned and tested with only a single (R or Ip) loop closed, whereas the red line is the response for the same controller in the combined system.}
    \label{fig:nstxu_tut_step_zrp0}
\end{figure}

\begin{figure}[H] 
    \begin{center}
        \includegraphics[width=0.9\linewidth]{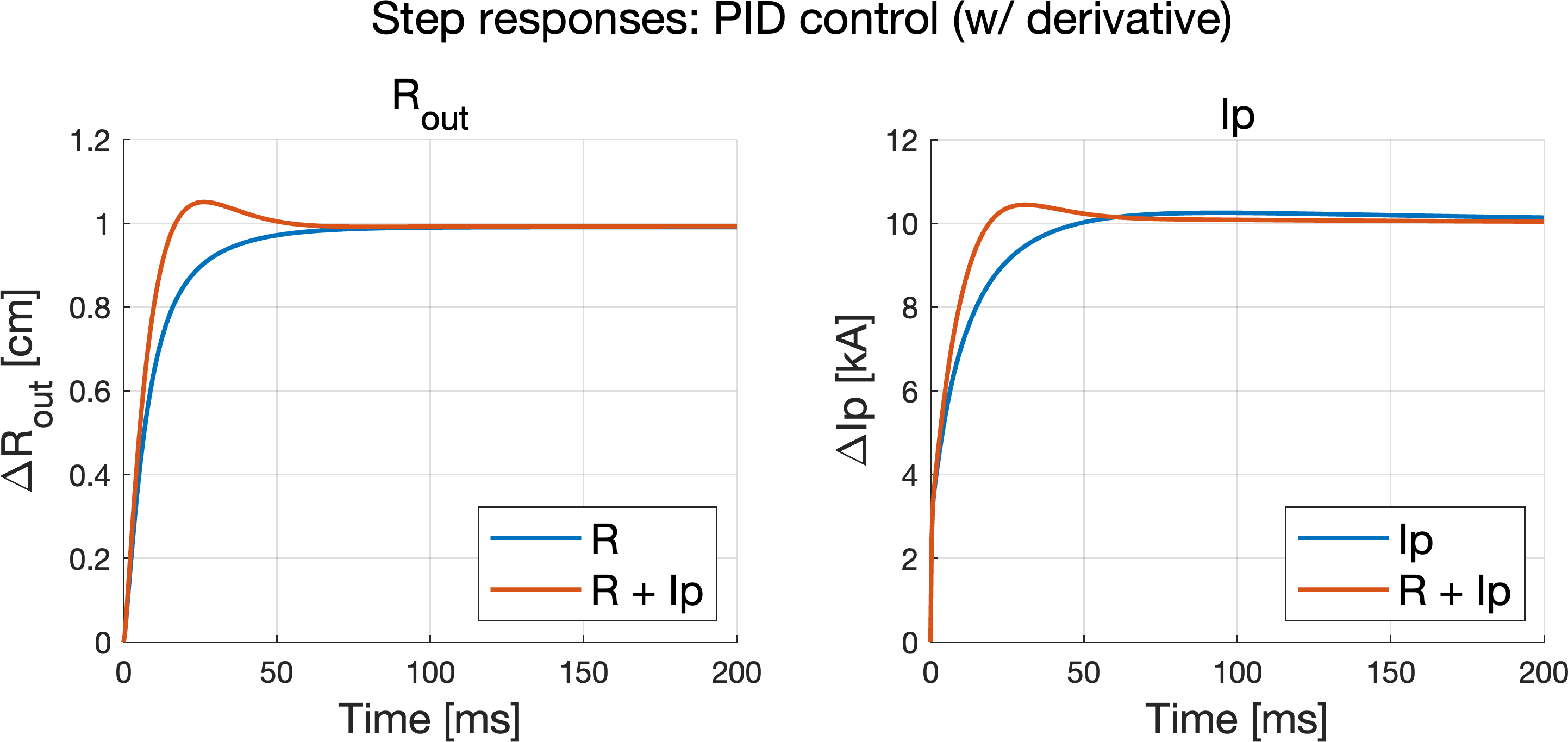}
    \end{center}
    \caption{Adding derivative gains, to increase damping, can reduce the cross-coupling across radial and Ip channels. There is much less performance degradation when testing in the combined system (red) compared to with individual control loops closed (blue), as compared to the degradation in \cref{fig:nstxu_tut_step_zrp0} }
    \label{fig:nstxu_tut_step_zrp1}
\end{figure}

\begin{figure}[H] 
    \begin{center}
        \includegraphics[width=0.6\linewidth]{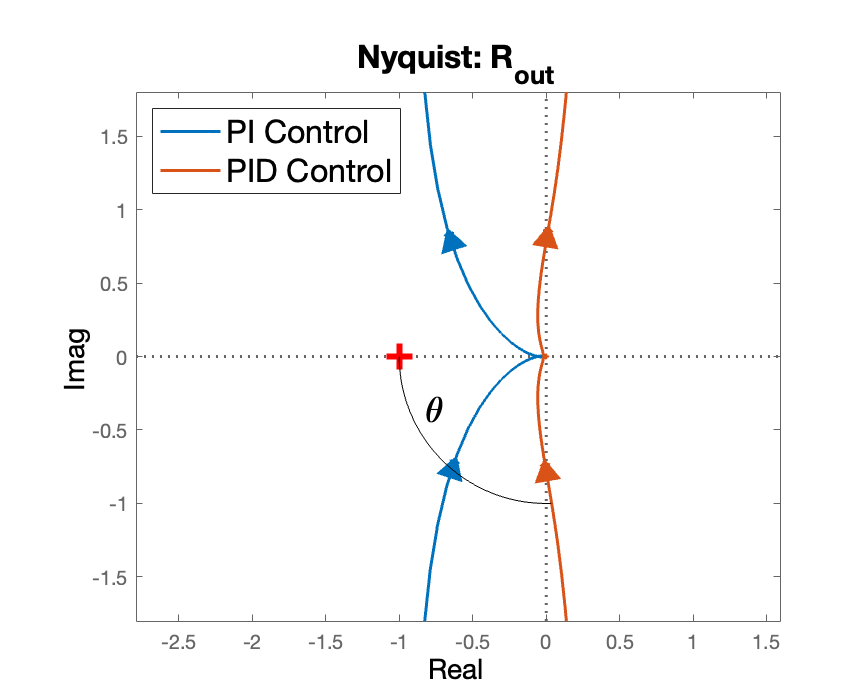}
    \end{center}
    \caption{Adding derivative gains increases the phase margin significantly from acceptable to robust levels of $>90^\degree$.}
    \label{fig:nstxu_tut_nyquist_zrp}
\end{figure}

\subsectionbf{Coil current control}\label{sec:nstxu_coil_control}

At this stage it is useful to design the coil current controller. Inversion of our static shape map will provide target coil currents, and then these target coil currents will be passed to a decoupling coil current controller to obtain power supply voltage commands. While the more sophisticated ALIGN technique can provide better decoupling, a simple and adequate approach is to scale the coil-to-coil mutual inductance matrices. We use the control law:

\begin{equation}
    K_{coil} = 120 * M_{cc}
\end{equation}

This is a MIMO proportional controller. In vacuum simulations, this gives reasonable performance for reference tracking, and the step response shown in \cref{fig:current_step_response2} shows a rise time of 25ms. There is a little bit of cross-coupling between the OH coil and PF1A which is expected given the relative size and position of these coils. The coupling between all the other coil combinations is significantly less. 

\begin{figure}[H]
    \centering
    \includegraphics[width=0.75\linewidth]{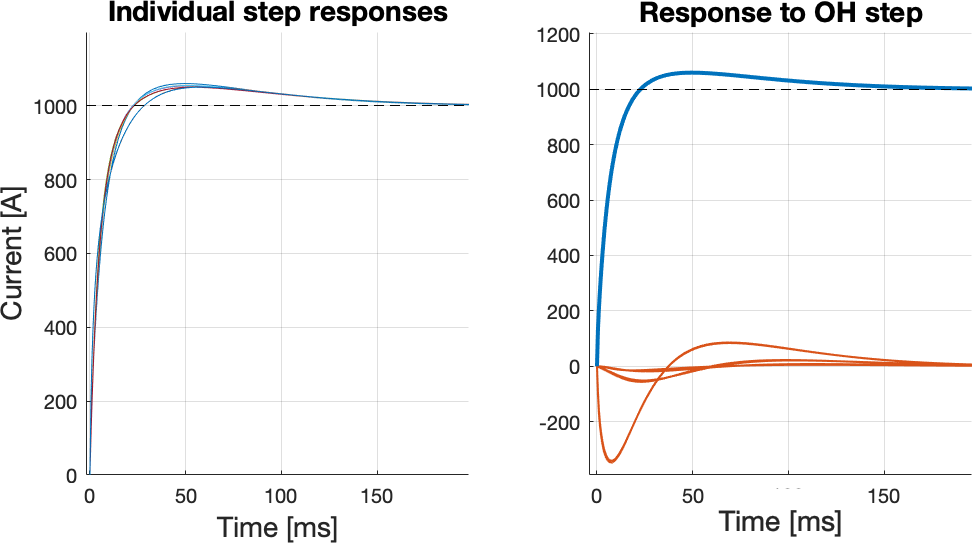}
    \caption{Step response for the coil current controller based on the coil-to-coil mutual inductances.}
    \label{fig:current_step_response2}
\end{figure}

\subsectionbf{Shape control via QP optimization of the static vacuum map}


We now move onto designing the shape controller. This tutorial uses the QP-controller to invert static maps based on the vacuum flux linearization. For this example, we choose an equilibrium and obtain its linearization and state-space model using GSpert. 

We define our controlled variables as,

\begin{equation}
    y = \begin{bmatrix} \psi_{cp} - \psi_x \\ \pdv{\psi_{xp}}{r} \\ \pdv{\psi_{xp}}{z} \end{bmatrix} 
\end{equation}

which are flux differences of control points versus the x-point, and the flux gradient at the x-point. This is a good choice of isoflux shaping variables for a lower single null diverted plasma, although for the full pulse we extend this to also control flux versus the touch point as well as flux gradient at both upper and lower x-points. Following \cref{sec:output_linearization} we obtain the output linearization $C_{0,vac}$ derived from the vacuum mutual inductance maps. 

The static map $T$ is taken from the columns of $C_{0,vac}$ that correspond to the coil currents.

\begin{equation}
    T = C_{vac}  \begin{bmatrix} I_{ncoils} \\ 0 \end{bmatrix}
\end{equation}

As a test case, we define the target shape perturbation shown in \cref{fig:nstxu_shape_target_tuning}, which is a perturbation to the plasma elongation and requires shifting the x-point upwards. We set up the quadratic program \cref{eq:shape_inv_QP} repeated here for convenience:


At this stage it becomes important to tune the weights $W_y$ and $W_{Ic}$ for the controller. These weights are used to tradeoff between matching shape targets and minimizing coil current perturbations. These weights can be roughly tuned at this stage by solving the QP for the coil currents, and then visualizing what the flux surfaces look like for these currents. After some playing around, we set the weights such that 1kA of coil current corresponds to 3mWb of isoflux shaping error. This tuning procedure is shown in \cref{fig:nstxu_shape_target_tuning}. For these weights we can see that predicted boundary generally aligns with the target control points, although there are some small differences especially near the upper control points and near the x-point. The currents that achieve this are fairly reasonable, requiring a few kA of distributed across the PF coil set. 

\begin{figure}[H]
    \centering
    \includegraphics[width=0.8\linewidth]{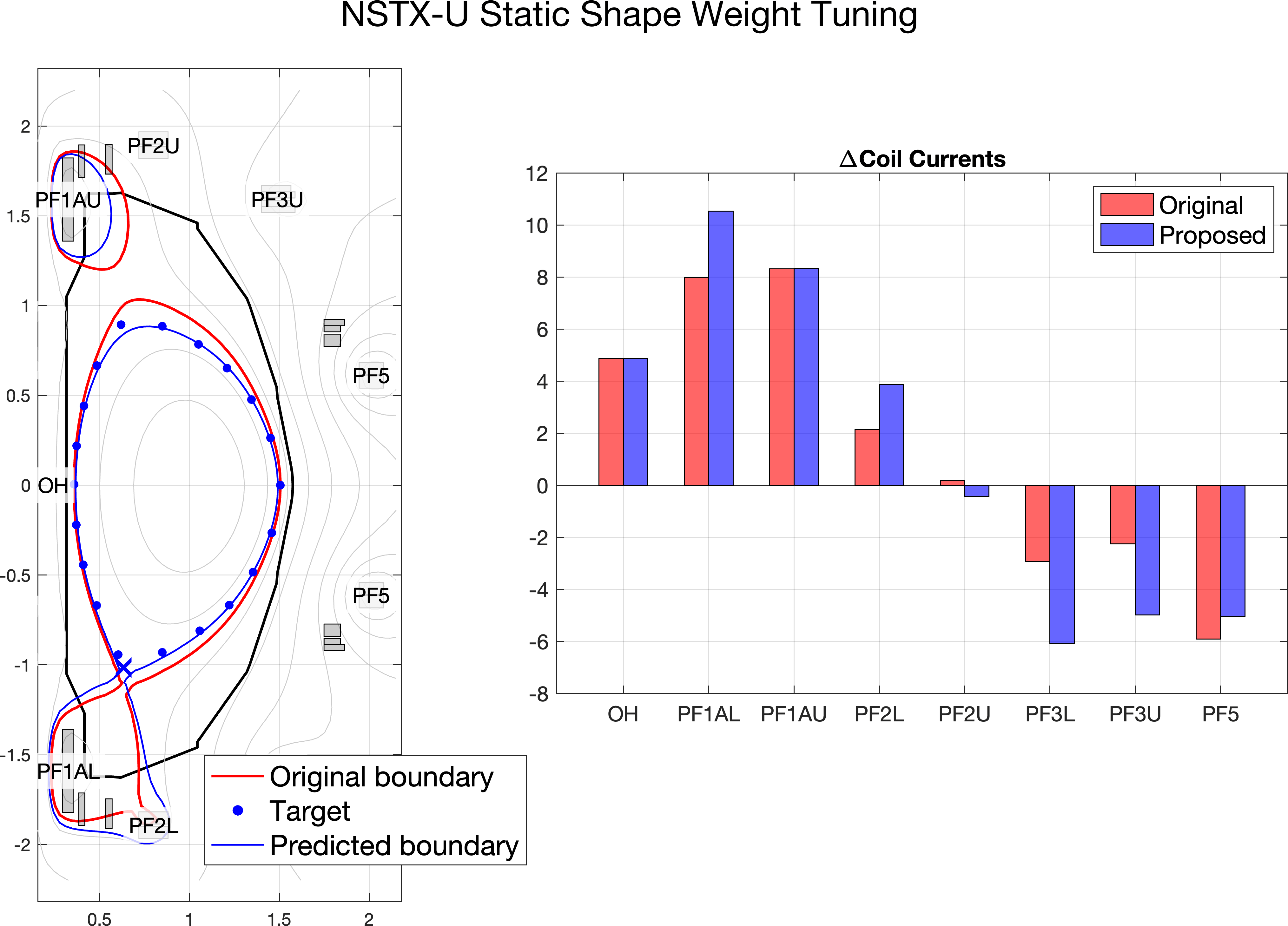}
    \caption{Illustration of the shape controller weight tuning procedure. The weights are adjusted until we are satisfied in the tradeoff balance between achieving the desired shape perturbation (left) versus the size of coil current perturbation needed to effect the shape change (right).}
    \label{fig:nstxu_shape_target_tuning}
\end{figure}

This is a good starting point for the controller although at this stage we have not applied any vertical decoupling yet. This particular shape perturbation was chosen because the elongation change and vertical shift for the target shape interact with the vertical control. We do achieve the large step change in upper gap position to meet the target elongation. However, during this process the plasma suffers lots of oscillations in its shaping which are observed in the current centroid z-position measurement and even the radial outer gap. 

\begin{figure}[H] 
    \begin{center}
        \includegraphics[width=7cm]{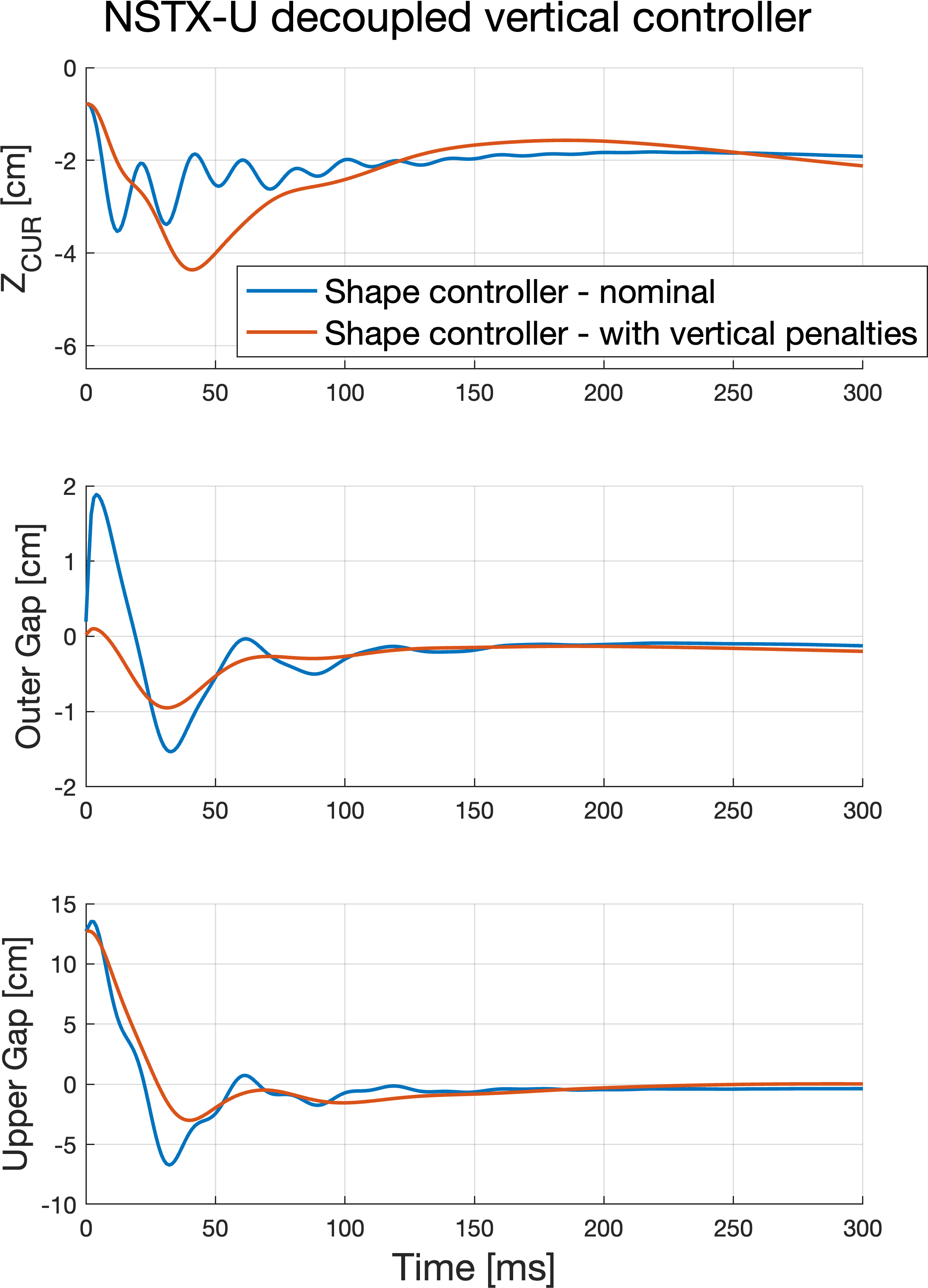}
    \end{center}
    \caption{Comparison of shape controller performance with and without applying vertical actuation penalties. The target shape perturbation is the one shown in \cref{fig:nstxu_shape_target_tuning} that both decreases the plasma elongation and slightly shifts the net vertical position. This is a perturbation that is expected to have cross-coupling between shape and vertical actuation. When vertical actuation penalties are applied, the controller-induced oscillations are removed and a much smoother trajectory is obtained.}
    \label{fig:nstxu_shape_vertical_decoupled2}
\end{figure}

As discussed in \cref{sec:static_plus_vertical}, for static map control designs it is important to explicitly add vertical decoupling in order to avoid interaction with the RHP zero. The primary solution is to slow down shape control actuation in the vertical control direction. One way to achieve this is via the $W_{Ic}$ weighting matrix. In this case, we add off-diagonal entries to $W_{Ic}$ that penalize up-down antisymmetric current combinations (e.g. penalize PFA1U - PFA1L, PF2U - PF2L, etc). This allows us to retain high gain in symmetric directions like radial shifts or compressing/elongating fields, but slows down antisymmetric directions. 

The other action that can be taken to improve vertical decoupling is to add a mode into the static map that corresponds to $\delta z$ shifts (see \cref{sec:notes_vert_decoupling}). The QP inversion will then output not just target coil currents, but a target $\delta z$ shift that can be passed to the vertical controller. This design ensures consistency that between the target shape and the target z-position, such that the vertical controller and shape controller will not be trying to achieve competing objectives. After performing both of these decoupling actions, we achieve the control performance shown in red in \cref{fig:nstxu_shape_vertical_decoupled2} which shows much less coupling interactions and better performance overall and was used to achieve the results of \cref{sec:nstxu_results}.

\section*{Acknowledgements}
This material is based upon work supported by the U.S. Department of Energy, Office of Science, Office of Fusion Energy Sciences, using the DIII-D National Fusion Facility, a DOE Office of Science user facility, under Awards DE-FC02-04ER54698, DE-AC02-09CH11466, DE-SC0015878, and DE-AC52-07NA27344. This work was supported by Commonwealth Fusion Systems.

\section*{Disclaimer}
This report is prepared as an account of work sponsored by an agency of the United States Government. Neither the United States Government nor any agency thereof, nor any of their employees, makes any warranty, express or implied, or assumes any legal liability or responsibility for the accuracy, completeness, or usefulness of any information, apparatus, product, or process disclosed, or represents that its use would not infringe privately owned rights. Reference herein to any specific commercial product, process, or service by trade name, trademark, manufacturer, or otherwise, does not necessarily constitute or imply its endorsement, recommendation, or favoring by the United States Government or any agency thereof. The views and opinions of authors expressed herein do not necessarily state or reflect those of the United States Government or any agency thereof.

\bibliography{bib}
\end{document}